\documentclass[onecolumn,nobibnotes,nofootinbib]{revtex4}
\usepackage{color}
\usepackage{amssymb}
\usepackage{amsmath}
\usepackage{epsfig}
\usepackage{graphicx}
\usepackage{eucal}

\newcommand{\be}{\begin{equation}}
\newcommand{\ee}{\end{equation}}
\newcommand{\beeq}{\begin{eqnarray}} 
\newcommand{\eeeq}{\end{eqnarray}} 

\def\Qdash{\overline{Q}} 

\newcommand{\uvec}[1]{\underline{#1}}
\begin{document}

\title{\bf Exact kinematics in the small $x$ evolution of the color dipole and gluon  cascade}

 \author{Leszek Motyka}

   \affiliation{ II Institute for Theoretical Physics, University of Hamburg,
Luruper Chaussee 149, D-22761, Germany\\
Institute of Physics, Jagellonian University
Reymonta 4, 30-059 Krak\'ow, Poland}

\author{Anna M. Sta\'sto}

\affiliation{Physics Department, Penn State University, University Park, 16802 PA, U.S.A.\\
RIKEN center, Brookhaven National Laboratory, Upton, 11973 NY, U.S.A.\\
Institute of Nuclear Physics, Polish Academy of Science, ul. Radzikowskiego 152, Krak\'ow, Poland}

 \vskip 2mm      

\vspace*{1cm}

\vskip1cm      
\begin{abstract}
The problem of kinematic effects in gluon and color dipole cascades  is addressed in the large $N_c$ limit of $SU(N_c)$ Yang--Mills theory. We investigate the  tree level multi-gluon components of the gluon light cone wave functions in the light cone gauge keeping the exact kinematics of the gluon emissions. We focus on the components with all helicities identical to the helicity of the incoming gluon.
The recurrence relations for the gluon wave functions are derived. 
In the case when the virtuality of the incoming gluon is neglected the exact form of the multi-gluon wave function is obtained. Furthermore, we propose an approximate scheme to treat the kinematic effects in the color dipole evolution kernel. The new kernel entangles longitudinal and transverse degrees of freedom and leads to a reduced diffusion in the impact parameter. When evaluated in the next-to-leading logarithmic (NLL) accuracy, the kernel reproduces the correct form of the double logarithmic terms of the dipole size ratios present in the exact NLL dipole kernel. Finally, we analyze the scattering of the incoming gluon light cone components off a gluon target and the fragmentation of the scattered state into the final state. The equivalence of the resulting amplitudes and the maximally-helicity-violating amplitudes is demonstrated in the special case when the target gluon is far in rapidity from the evolved gluon wave function.
\end{abstract}
\maketitle

\section{Introduction}

In this paper we investigate the possibility of using information available in the exact tree level QCD amplitudes in the large $N_c$ limit to improve the evolution equations of QCD. Following the Mueller's color dipole model approach \cite{AlMueller,Chen:1995pa} we choose the framework of the light cone perturbation theory (LCPT)~\cite{Weinberg:1966jm,Bjorken:1970ah,Lepage:1980fj,Brodsky:1997de} and adopt the light cone gauge. The attractive feature of the color dipole model is its probabilistic interpretation in terms of the multi-dipole densities. This allows to extract the total cross-section for a scattering of two small sized onia (quark-antiquark systems) at a large energy, $\sqrt{s}$, from the tree level amplitudes in the soft-gluon limit. The color dipole model provides a resummation scheme of the leading logarithmic (LL) contributions in  powers of $\alpha_s \log s$. It is remarkable, that the analysis of the tree level amplitudes in the color dipole framework  reproduces the results of the leading logarithmic Balitsky-Fadin-Kuraev-Lipatov (BFKL) evolution equation \cite{BFKL}, that incorporates quantum loops. This suggests that a more accurate treatment of the color dipole (or gluon) cascade at the tree level but beyond the soft gluon limit, may provide a lot of insight into higher order corrections to the BFKL equation.

The basic object in our analysis is the light cone wave function of a 
 gluon, that may be decomposed into multi-gluon components. The whole analysis of this paper is carried out in the large $N_c$ limit. As the simplest starting point we choose the configuration in which both the initial state gluon and all the gluons in the wave function carry the positive helicity. We aim at obtaining an exact description of the gluon cascade, so we keep the complete information about the gluon kinematic variables along the cascade. This implies that both the form of gluon splitting vertex and the energy denominators of the light cone perturbation theory are treated
exactly. This is different from the soft limit where the approximate forms were used.

As the first result, we find an exact form of a tree level multi-gluon components of the real gluon wave function in the chosen helicity sector. This is achieved by solving recurrence relations for the multi-gluon components of the gluon light cone wave function in the momentum space.  The natural variables in this formulation are closely related to the spinor products which are used in the construction of the maximally-helicity-violating amplitudes (MHV), see e.g.\ \cite{Parke:1986gb,Mangano:1990by}. 
This wave function can be also analyzed in the transverse coordinate space, after a suitable Fourier transform is performed.
Curiously enough, the obtained wave function in the coordinate space coincides with the one obtained in the soft gluon limit. This coincidence, however, is expected to hold only for the particular choice of helicities, that has been assumed.

In general, the description of the gluon cascade within LCPT is non-local, the energy denominators involve momenta of all gluons, and that makes the resummation difficult in a general case. In order to obtain a more practical improvement of the dipole model kernel, we propose a local approximation to the $1\to 2$ gluon splitting kernel in which the value of energy denominator is obtained from the invariant mass of the produced gluon pair. This leads to a new color dipole kernel. The form of the kernel in the soft gluon limit 
$\frac{dz}{z} \, d^2 \underline{x}_2  \frac{x_{01}^2}{x_{02}^2 x_{21}^2}$ 
gets modified to 
\be
\frac{dz}{z}\, z \;
  \frac{d^2 \underline{x}_2}{x_{01}^2}\, \left[K_1^2\left(\,{x_{02} \over x_{01}} \, \sqrt{z}\right)+K_1^2\left({x_{12} \over x_{01}} \, \sqrt{z}\right)-2\,K_1\left(\,{x_{02} \over x_{01}} \, \sqrt{z}\right)K_1\left({x_{12} \over x_{01}} \, \sqrt{z}\right)\frac{\underline{x}_{02}\cdot\underline{x}_{12}}{x_{02}x_{12}}\right] \; \;  ,
\label{eq:modkernel_0}
\ee
 where $\underline{x}_{01}$ is the parent dipole vector, $\underline{x}_{02}$, $\underline{x}_{21}$, are the daughter dipole vectors, $z$ is the fraction of the longitudinal momentum of the softer daughter gluon, and $K_1$ is the modified Bessel function. The modified kernel becomes equivalent to the leading logarithmic kernel for $zx^2_{02} \ll x^2_{01}$ and $zx^2_{21} \ll x^2_{01}$ but differs significantly otherwise. In more detail, the production of larger dipole sizes is exponentially suppressed above the cutoff size which depends on the longitudinal momentum fraction of the soft gluon. 
The modified kernel introduces corrections at all orders beyond the leading logarithmic approximation. In particular, we show that the modified kernel reproduces an important part of the NLL kernel, namely the double logarithmic terms.  
Interestingly enough, these are the dominant terms that violate the two dimensional conformal symmetry of the BFKL-dipole kernel both in ${\cal N}=4$ Supersymmetric Yang--Mills (SYM) theory and in QCD at NLL approximation \cite{Balitsky:2008zz,Fadin:2007xy,Fadin:2007de,Fadin:2007ee,Fadin:2006ha}. The suppression of the large dipole emission implies also a suppression of the diffusion in the impact parameter. 

Finally, we consider the production amplitude of $n$ gluons in a scattering process of two gluons. We impose additional conditions  that the produced state is arbitrarily far away in rapidity from the target gluon, and that both the incoming and all the outgoing gluons carry positive helicities. Using the LCPT we reproduce formally the Parke-Taylor form of the MHV amplitudes \cite{Parke:1986gb} in this limit. In the intermediate steps of the proof we analyze in detail the fragmentation amplitude of $m$ virtual gluons into $n\geq m$ real gluons. We find that an apparent entanglement of the fragmentation process of different gluons in LCPT can be unwound at the tree level, and in fact, fragmentation amplitude of each of the gluons is independent. We find an exact form of the tree level amplitude for a gluon fragmentation, again assuming that the helicity of the incoming and outgoing gluons are positive. We observe, that the $n$-gluon component of the incoming real gluon wave function is dual to the fragmentation amplitude of a gluon into $n$~real gluons i.e. the latter has the same form as the former under replacement of the gluon transverse positions by the gluon transverse (light cone) velocities.

The structure of this paper is as follows: in the next section we present the overview of the color dipole model and the next-to-leading order corrections to BFKL. In Sec.~\ref{sec:exactkinematics} we set up a general formalism for computing the multi-gluon tree-level wave function  with exact kinematics in the light cone perturbation theory and in the light cone gauge. We derive the recurrence relations for the wave functions with different number of external gluons in momentum space and in transverse coordinate space. In Sec.~\ref{sec:approx}, we derive the modified kernel for the dipole evolution in transverse coordinate space. We show that when expanded to the next-to-leading order in the strong coupling constant the double logarithmic terms in the next-to-leading calculation are recovered. We discuss the implications for the impact parameter dependence of the cross section. In Sec.~\ref{sec:mhv}
we construct the scattering amplitudes for external on-shell partons and demonstrate the general factorization property of the fragmentation in the light cone perturbation theory. The equivalence with the MHV amplitudes is also demonstrated in the high energy limit.  In the last section we briefly summarize our results. The Appendix contains the most technical part of the calculations.

\section{Overview of color dipole evolution in  QCD}

In a conventional approach to the high energy limit one computes the amplitude for scattering of  two highly energetic particles which interact via exchange of gluons in the $t$-channel.
   The emitted gluons in this scattering process are then resummed in the multi-Regge kinematics.  The amplitude turns out to be governed by the exchange of  a Reggeon with the quantum numbers equivalent to that of the  vacuum: the  Pomeron.   Its intercept is evaluated from the solution to the Balitsky-Fadin-Kuraev-Lipatov (BFKL) \cite{BFKL} evolution equation at small $x$. This equation describes the change of the amplitude with increasing rapidity between the scattering initial particles.
In the groundbreaking paper \cite{AlMueller} Mueller demonstrated that the small-$x$ cascade of gluons can be alternatively described as the evolution of the hadron wave function. The subsequent emissions of the soft gluons in  the hadron wave function can be computed   in the light cone perturbation theory. By soft gluons we mean here gluons whose longitudinal momentum fractions are much smaller than that of the incoming particles, the transverse momenta of the gluons are not restricted. In the large $N_c$ limit it was demonstrated that the correct degrees of freedom at high energy are $q\bar{q}$ dipoles. Furthermore, by taking the Fourier transform with respect to the transverse momenta it was shown that the soft gluon emissions factorize in the transverse coordinate space.
The resulting evolution equation for the scattering amplitude   was then formulated by considering the interaction of the hadron wave function with the target.  It can be written in the following form
\begin{equation}
\frac{\partial N_Y(\underline{x}_{01})}{\partial Y} \; = \; \bar{\alpha}_s \, \int \frac{d^2 \underline{x}_{02}}{2\pi} \frac{\underline{x}_{01}^2}{\underline{x}_{12}^2\underline{x}_{02}^2} \, \left[N_Y(\underline{x}_0,\underline{x}_2) +N_Y(\underline{x}_1,\underline{x}_2)-N_Y(\underline{x}_0,\underline{x}_1)   \right] \; .
\label{eq:muellerdipolell}
\end{equation}
Here, $N_Y$ is the scattering amplitude for a dipole on the target, $Y$ is the rapidity for the process and 
$\underline{x}_{0},\underline{x}_1,\underline{x}_2$ are the two-dimensional vectors  which specify the position of the color dipoles. To obtain the solution to this equation one needs to specify an amplitude at the initial rapidity $N_{Y_0}$.
The solution to this equation for the dipole scattering amplitude turned out to be  equivalent to the one from the original BFKL equation,
at least in  the case of the inclusive quantities \cite{Chen:1995pa}, like the total cross section.

 The dipole evolution proved to be a powerful framework for investigating the unitarity corrections \cite{Salam:1995uy,Mueller:1996te} in the scattering processes. By taking into account multiple rescatterings of the gluon components of the evolved hadron wave function onto  the nuclei the non-linear evolution equation was derived \cite{Kovchegov:1999yj,KOV2}. It differs from (\ref{eq:muellerdipolell}) by the presence of an additional  nonlinear term
 $$
-N_Y(\underline{x}_0,\underline{x}_2) \, N_Y(\underline{x}_1,\underline{x}_2) \; ,$$
which tames the growth of the amplitude with the energy and leads to the gluon saturation.  
  The same equation was derived in the context of the high energy operator expansion with Wilson lines \cite{Balitsky:1995ub,BAL2,BAL3,Balitsky:2001mr,Balitsky:2004rr,Balitsky:2006pf,Balitsky:2005we} and in the Color Glass Condensate framework  \cite{MCLERVEN,CGC1,JAL}, see also \cite{Iancu:2003xm} for a nice review. It is called the Balitsky-Kovchegov equation.
It is important to note
that the kernels both in the linear and non-linear equations are exactly the same (to the NLL level in the dipole approach).
\subsection{Next-to-leading corrections and  the kinematical effects}
Similarly to the $t$-channel approach the dipole evolution kernel receives large corrections at 
next-to-leading order. The next-to-leading order kernel in the momentum space was first evaluated in \cite{Fadin:1996nw,Fadin:1998py,Camici:1997ij,Ciafaloni:1998gs} in the forward case, and later in \cite{Fadin:2005zj,Fadin:2004zq} in the non-forward case.  The dipole kernel at next-to-leading accuracy was then computed in \cite{Balitsky:2006wa,Balitsky:2008zz}
 via direct calculation and also in \cite{Fadin:2007xy,Fadin:2007de,Fadin:2007ee,Fadin:2006ha} by performing the  Fourier transform of the existing result for the non-forward BFKL kernel in the momentum space. The running coupling corrections to the dipole kernel were also evaluated in \cite{Kovchegov:2006vj}. 
 There are different sources of the next-to-leading corrections and 
 it is quite well known that the major part of them which is  common to QCD and ${\cal N}=4$ Super Yang Mills theory comes from the kinematics, see for example \cite{Andersson:1995jt,Andersson:1995ju,Kwiecinski:1996td}. The slow convergence of the small $x$ series can be argued intuitively as follows. In the standard perturbative collinear approach one identifies the large scale $\mu^2$ which characterizes the short distance part of the process with strong interactions. The  limit of large $\mu^2$ implies automatically the smallness of the strong coupling constant, which allows to seek the solution to the hard scattering cross section and to  the splitting function in the form of the perturbative series. This series  should be convergent at least in the asymptotic sense. In the high energy limit the large parameter is the total available energy $\sqrt{s}$, the strong coupling however is not naturally small in this limit, and can take large values too. Therefore  there is  no guarantee that the expansion in powers of the strong coupling in the high energy limit will be strongly convergent.  Indeed, it is observed that the small $x$ expansion is poorly convergent, see for example \cite{Kwiecinski:1996td,Salam:1998tj,Ross:1998xw}.
 By assuming the  multi-Regge kinematics one makes  strong approximations  onto the phase space of the produced particles. As a result  the perturbative methods which rely on the expansion in terms of the coupling are simply inefficacious in correcting these kinematic approximations at higher orders.
 
   Even before the next-to-leading correction to the BFKL kernel was computed, it was shown that the constraints from the more careful treatment of the kinematics  (while being formally of the higher order) are numerically very important and  significantly  reduce the growth of the gluon density \cite{Andersson:1995jt,Andersson:1995ju,Kwiecinski:1996td}. Taking this effect into account leads to the predictions which are in a good agreement with phenomenology \cite{Kwiecinski:1997ee,Avsar:2005iz,Avsar:2006jy}. It was first demonstrated by Salam  \cite{Salam:1998tj,Salam:1999cn} that the kinematic effects manifest themselves at next-to-leading order in the form  of the double logarithms of ratios of transverse momenta, and correspond to the higher order poles in the Mellin space conjugated to the transverse momenta. 
The formalism of the collinear resummation developed later \cite{Ciafaloni:1998iv,Ciafaloni:1999yw,Ciafaloni:1999au,Ciafaloni:2003ek,Ciafaloni:2003rd,Ciafaloni:2007gf} was shown to stabilize the small $x$ series by imposing onto the kernel  the kinematic constraints and the momentum sum rule. 
The small $x$ expansion is also stabilized by the appropriate implementation of the running coupling effects, in the case of QCD. Another approach to match the DGLAP and BFKL resummations was constructed in parallel in Refs.~\cite{Altarelli:1999vw,Altarelli:2000mh,Altarelli:2001ap,ABF_improved,Altarelli:2003hk,Altarelli:2005ni} and similar results were found. Finally, the same problem was addressed in yet another scheme in Ref.\ \cite{THORNE,ThWh06}.

It is therefore quite urgent  to address the problem of the better treatment of the kinematics in the context of the dipole evolution. As the next-to-leading corrections take into account only part of the kinematic corrections it is necessary to investigate the origin and effect of the kinematical constraints on the dipole evolution. Such effects necessarily go beyond the  fixed order calculation and are part of the all-order (in powers of $\alpha_s^n \ln^m s$) resummation.

\section{Exact kinematics in the multi-gluon wave function}
\label{sec:exactkinematics}

\subsection{Setup of the problem}
\label{sec:exA}

In this section we shall derive the exact form of the $n$-gluon components of the real gluon wave function, $\Psi_n, \;\;n=1,2,\ldots$, using the LCPT for a special choice of helicities. Thus, we shall assume that the incoming gluon has the positive helicity and analyze the case when all the gluons in the wave function have all positive helicities as well. Such choice of the helicity state is dictated by an expected simplicity of the result. This choice also prepares  the ground for the comparison with the Parke-Taylor amplitudes that we shall make in Sec.~\ref{sec:mhv}. We understand the computation presented in this section as a demonstration of a potentially useful calculational technique and the first step towards the exact determination of the gluon wave function with arbitrary helicities at the tree level.

 \begin{figure}[ht]
\centerline{\epsfig{file=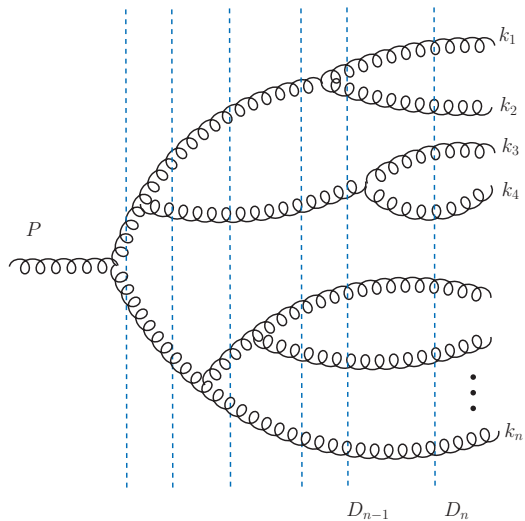,width=0.4\textwidth}}
\caption{The multi-gluon wave function. The vertical dashed lines symbolize different intermediate states where we need to evaluate the energy denominators. It is understood that the wave function scatters finally on some target.}
\label{fig:onium_n}
\end{figure}

To be specific, we start with a gluon with four momentum $P$, and the color index $a$, that develops a virtual fluctuation into states containing $n$ gluons with momenta $(k_1,\ldots k_n)$ and color indices $(a_1,a_2,\ldots,a_n)$, correspondingly
see Fig.~\ref{fig:onium_n}. For the initial virtual gluon with virtuality $-Q^2$ and a vanishing transverse momentum one has in the light cone variables,
\[
P^{\mu} = (P^+,  -\frac{Q^2}{2 P^+},\, \uvec{0})\; ,
\]
with $P^{\pm}=\frac{1}{\sqrt{2}}(P^0\pm P^3)$.
After $n-1$  splittings the wave function is shown in  Fig.~\ref{fig:onium_n}.  The last  $n$ gluons have  the corresponding momenta labeled $k_1,\dots,k_n$, as in  Fig~\ref{fig:onium_n}. Each of this momenta has components $k^{\mu}_i=(z_i P^{+},k_i^-,\underline{k}_i)$, with $z_i$ being the fraction of the initial $P^+$ momentum which is carried by the gluon labeled by $i$ and $\uvec{k}_i$ being the transverse component of the gluon momentum.
The rules of the light cone perturbation theory \cite{Weinberg:1966jm,Bjorken:1970ah,Lepage:1980fj,Brodsky:1997de}  require to evaluate the energy denominators for each of the intermediate states in Fig.~\ref{fig:onium_n}.  The energy denominator for the $n$ gluons in the intermediate state  is defined to be a difference between the light cone energies of the initial and the  intermediate state in question. For the wave function shown in Fig.~\ref{fig:onium_n} we assume  the last intermediate state is with $n$ gluons as depicted. The corresponding energy denominator for this state  reads

\begin{equation}
\overline{D}_n= P^{-}-\sum_{i=1}^n k_i^-=- \frac{1}{2P^+}\left(Q^2+ {\uvec{k}_1^2 \over z_1} + 
{\uvec{k}_2^2 \over  z_2} + \ldots +
 {\uvec{k}_n^2 \over z_n}\right) = - \frac{1}{2P^+} D_n\; ,
\label{eq:denominator_n}
\end{equation}
where we have used $$k_i^- = \frac{\underline{k}_i^2}{2 z_i P^+} \; ,$$ and we introduced the auxiliary notation for the (rescaled) denominator $D_n$.

In what follows, we shall focus on the color ordered multi-gluon amplitudes, that is the amplitudes decomposed in the basis of color tensors, 
$\,T^{a_1 a_2 \ldots a_n} \,=\, \mathrm{tr}\, (t^{a_1}   t^{a_2} \ldots t^{a_n})\,$, where $a_1,a_2, \ldots, a_n$ are the color indices of the gluons. 
Up to the leading terms in the large $N_c$ approximation, these tensors are orthogonal, see e.g.\  \cite{Mangano:1990by} for a detailed discussion. This decomposition has a transparent graphical representation in 't~Hooft's double line notation (see e.g.\ \cite{Manohar:1998xv} for a review). The $n$-gluon amplitudes in the basis 
$\,T^{a_1 a_2 \ldots a_n}\,$ are then represented by planar diagrams built of the effective colored fermion lines. This, in turn, leads to the color dipole representation proposed by Mueller \cite{AlMueller}, in which the color coefficient in the triple gluon vertex is equal to unity. In what follows, we shall use the $n$-gluon components of the incoming particle light cone wave function $\Psi_n(k_{1},k_2,\ldots, k_{n})$, assuming implicitly, that the corresponding color tensor is given by the planar diagram in the double line notation, that corresponds to the definite ordering $(a_1,a_2,\ldots,a_n)$ of the gluon color indices.

\subsection{Recursion relations}

We focus on components of the wave function in which all the gluons have positive 
helicities.  We work in the light-cone gauge, $\eta\cdot A = 0$, with vector
$\eta = (0,1,\uvec{0})$ in the light-cone coordinates. This choice of gauge defines 
the polarization four-vectors of the gluon with four-momentum $k$
\be
\epsilon^{(\pm)} = \epsilon_{\perp} ^{(\pm)} + 
{\uvec{\epsilon}^{(\pm)}\cdot \uvec{k} \over \eta\cdot k}\, \eta\; ,
\ee
where
$\epsilon_{\perp} ^{(\pm)} = (0,0,\uvec{\epsilon} ^{(\pm)})$, and
the transverse vector  is defined by 
$\uvec{\epsilon}^{(\pm)} = \mp {1\over \sqrt{2}}(1,\pm i)$.
The explicit projections of the triple and quartic gluon vertices show on this
helicity states show that~\cite{Brodsky:1997de}:
\begin{enumerate}
\item The helicity transitions $(-\to ++)$ and $(+\to --)$ are forbidden for the triple gluon vertex.
\item The $1\to3$ gluon transition given by the four-gluon vertex leads to mixed helicity composition of the three outgoing gluons.  
\end{enumerate}
These properties indicate, that the state after splitting always contains at least one  gluon with the helicity of the parent. Thus, if we require the final state to be composed from gluons with positive helicities, there can be no gluons with negative helicities at any intermediate step. So, the four gluon-vertex does not contribute to the splitting process that we consider. In fact, the whole branching process is driven by a $(+ \to ++)$ helicity projection of the triple gluon vertex. 

We define the projection of  the triple gluon vertex on the $(\pm)$ polarization
states, assuming that all the gluons are outgoing, 
\be
\tilde V_{\lambda_1 \lambda_2 \lambda_3 } ^{a_1 a_2 a_3} (k_1,k_2,k_3) 
= \epsilon^{(\lambda_1)\, \mu_1} \epsilon^{(\lambda_2)\, \mu_2 }   
\epsilon^{(\lambda_3)\, \mu_3}   
V_{\mu_1 \mu_2 \mu_3 } ^{a_1 a_2 a_3} (k_1,k_2,k_3) \; ,
\ee
where $a_1,a_2,a_3$ are of course color indices.
As we already stated, vertex $\tilde V_{\lambda_1 \lambda_2 \lambda_3} ^{a_1 a_2 a_3} (k_1,k_2,k_3) $ vanishes if all the {\em outgoing gluon helicities} are the same, $\lambda_1 = \lambda_2 = \lambda_3$.
When one helicity, say $\lambda_1=-1$, is different than the others
$\lambda_2 = \lambda_3=+1$, the vertex takes the following form in the 
light-cone variables,
\be
\tilde V_{-++} ^{a_1 a_2 a_3} (k_1,k_2,k_3) 
= g \,f^{a_1 a_2 a_3} \, z_1 \,
\uvec{\epsilon}^{(-)} \cdot
\left( {\uvec{k}_2 \over z_2} - {\uvec{k}_3 \over z_3}\right) \; ,
\label{eq:vhel}
\ee 
where the $\delta$-functions related to the conservation of the transverse and longitudinal 
`$+$' components of the momentum are implicit.
Here $g$ is the strong coupling constant and $f^{a_1 a_2 a_3}$ is the structure constant for the $SU(N_c)$ color group. 
For the case of interest, of $(+\to ++)$ transition, the amplitude is described 
by (\ref{eq:vhel}), with $z_1$ being the fraction of the $+$ component  of the 
momentum of  the incoming gluon. Note that all  dependence of the vertex on the transverse momenta of the daughter gluons is completely  absorbed into a variable
\be
\uvec{v}_{23} \equiv \left( {\uvec{k}_2 \over z_2} - {\uvec{k}_3 \over z_3}\right).
\label{eq:v23}
\ee
This variable is a relative transverse light cone velocity of the two gluons.  
Interestingly enough, the same variable is present when we consider the change of the energy denominator due to
the splitting. 
In a general situation, when the gluon with momentum $k_1$~belongs to a virtual gluon
cascade, the energy denominator before the
splitting of gluon~$1$ can be written as
\be
D_n = D_{n/1} + {\uvec{k}_1 ^2 \over z_1} \; ,
\ee
where $D_{n/1}=Q^2+\sum_{i>1} \frac{\uvec{k}_i^2}{z_i}$ does not contain the energy  of gluon $1$. Note that, we are using here the definition (\ref{eq:denominator_n}) for the energy denominator, which is different by the sign and with the $P^+$ dependence factored out.
After the gluon splits into two gluons with momenta $k_2$ and $k_3$ we have 
\be
D_{n+1} = D_{n/1} + {\uvec{k}_2 ^2 \over z_2} + {\uvec{k}_3 ^2 \over z_3} \; .
\ee  
 In the light cone perturbation theory the transverse and the $+$ components of the longitudinal momenta are conserved in the vertices therefore we have that  that $\uvec{k}_1 = \uvec{k}_2  +  \uvec{k}_3$ and
 $z_1 = z_2 + z_3$. Using this fact  one can express the change of the energy 
denominator as,
\be
D_{n+1} - D_n = {z_2 z_3 \over z_2 + z_3} \, 
\left( {\uvec{k}_2 \over z_2} - {\uvec{k}_3 \over z_3}\right)^2 \; .
\label{eq:dendifference}
\ee
It is convenient to introduce a variable that depends on  the longitudinal degrees of
freedom 
\be
\xi_{23} \equiv {z_2 z_3 \over z_2 + z_3} \; .
\ee 
We therefore see that the change in the denominator (\ref{eq:dendifference}) due to the splitting  is expressed through the variable $v_{23}$ as well
\be
D_{n+1} - D_n = \xi_{23}\, \uvec{v}_{23}^2 \; .
\label{eq:dendifference2}
\ee
In the light-cone formulation of the QCD, the intermediate line
that carries longitudinal momentum fraction $z_i$ is multiplied
by $1/\sqrt{z_i}$. It is therefore convenient to follow the convention by \cite{Brodsky:1997de}  and 
absorb such factors into all the gluon lines incoming and outgoing 
from the triple gluon vertex. Thus, we shall use
\be
\bar V_{\lambda_1 \lambda_2 \lambda_3} ^{a_1 a_2 a_3} (k_1,k_2,k_3) = 
{1\over \sqrt {z_1 z_2 z_3}} \, \tilde V_{\lambda_1 \lambda_2 \lambda_3} ^{a_1 a_2 a_3} (k_1,k_2,k_3)
= g f^{a_1 a_2 a_3} \,
{\uvec{\epsilon}^{(-)} \uvec{v}_{23} \over  \sqrt{\xi_{23}}} \; .
\ee
Following the discussion of the color structure of the amplitudes in Sec.~\ref{sec:exA}, in what follows we drop the structure constants  $f^{a_1 a_2 a_3}$.

Now, we can collect the vertex and the energy denominator together to get the effect of the gluon splitting on the virtual multi-gluon state wave function. We shall introduce the following notation. Let $\Psi_{n-1}(k_{1},k_2,\ldots, k_{i\,i+1}, \ldots, k_{n})$ (with $k_{i\,i+1}\equiv k_i+k_{i+1}$) be the $(n-1)$-gluon wave function in momentum space before the splitting of gluon with momentum $k_{i\, i+1}$, and  
$\Psi_{n} (k_1,k_2,\ldots, k_i,k_{i+1}, \ldots, k_{n})$ the wave function after 
splitting of this gluon. Then the splitting of the gluon with momentum $k_{i\, i+1}$ acts as follows,
\be
\Psi_{n} (k_1,\ldots, k_i, k_{i+1}, \ldots, k_{n}) = 
{g \over \sqrt{\xi_{i\, i+1}}} 
{\uvec{\epsilon}^{(-)} \uvec{v}_{i\, i+1} \over  D_{n-1} + \xi_{i\, i+1}\,
\uvec{v}_{i\, i+1}^2}\, \Psi_{n-1} (k_1, \ldots, k_{i\,i+1},\ldots, k_{n})\; ,
\label{eq:vxisplit}
\ee 
where the color degrees of freedom are treated in the way described in 
Sec.~\ref{sec:exA}. 
  
The formula (\ref{eq:vxisplit})
is the recurrence prescription for obtaining the wave function with $n$  virtual gluons from the wave function with $n-1$ gluons. Of course to obtain the full recurrence formula one needs to sum over the different possibilities of the splittings which will give us the following result
\be
\Psi_{n} (k_1,k_2,\ldots, k_{n}) = 
{g \over D_{n}}\, \sum_{i=2}^{n} 
{\uvec{\epsilon}^{(-)} \uvec{v}_{i-1 \, i} \over  \sqrt{\xi_{i-1 \, i}} }\, \Psi_{n-1} (k_1,\ldots ,k_{i-1 \, i},\ldots, k_{n})\; ,
\label{eq:recurrence1}
\ee
where $D_{n}$ is the denominator for the last intermediate state with $n$ gluons. 
The form of the splitting given by (\ref{eq:vxisplit}) and (\ref{eq:recurrence1}) looks highly 
symmetric, and all dependence on momenta of daughter gluons $i$~and~$i-1$ 
is embedded into two variables: $\xi_{i-1 \, i}$ and $\uvec{v}_{i-1 \, i}$. 
In this calculation we have kept the exact kinematics and therefore the energy denominator includes the full dependence on the momenta in the whole cascade. This is understandable, as  in the case of the exact kinematics the new wave function
has to carry the full information about the old wave function. We will come back to this point later in Sec.~III  when we discuss the high energy limit.

\subsection{Relation to helicity amplitudes and the collinear limit in the on-shell case}

It turns out that the variables  $\underline{v}_{jk}$ 
that we used to construct the wave functions in the previous subsection are related to the variables 
used in the framework of helicity amplitudes, see \cite{Mangano:1990by}
for a nice review. Namely, for given pair of on-shell momenta $k_i$ and $k_j$ we have that 
\be
\langle ij \rangle = \sqrt{z_i z_j} \; \uvec{\epsilon}^{(+)} 
\cdot \left( {\uvec{k}_i \over z_i} -  {\uvec{k}_j \over z_j} \right)\, ,
\qquad 
[ij] =  \sqrt{z_i z_j} \; \uvec{\epsilon}^{(-)}\cdot 
\left( {\uvec{k}_i \over z_i} -  {\uvec{k}_j \over z_j} \right)\, ,
\label{eq:ij}
\ee
where the symbols $[ij],\langle ij \rangle$ are the spinor products defined by
\be
\langle i | j \rangle = \langle i- | j+ \rangle\, ,  \; \; \; [ij] = \langle i+ | j- \rangle \; .
\label{eq:ijdef1}
\ee
The  chiral projections of the spinors for massless particles  are defined as
\be
|i \pm \rangle \; = \; \psi_{\pm}(k_i) \; = \; \frac{1}{2}(1\pm\gamma_5)\psi(k_i) \; \; , \;\;\;\;\langle \pm i| \; = \; \overline{\psi_{\pm}(k_i)} \; ,
\label{eq:ijdef2}
\ee
for a given momentum $k_i$.
The spinor products are complex square roots of the total energy mass squared for the pair of gluons $(i,j)$
\be
\langle \,ij\, \rangle [\,ij\,] = (k_i + k_j)^2 ,
\label{eq:ijs}
\ee
and they also satisfy $\langle ij \rangle = [ij]^*$.
Using the above definitions (\ref{eq:ij},\ref{eq:ijs}) we have that
$$
\langle \,ij\, \rangle [\,ij\,]=  z_i z_j \, \left( {\uvec{k}_i \over z_i} -  {\uvec{k}_j \over z_j}  \right)^2 \; .
$$
which is real and positive for the on-shell gluon momenta.
Finally, combining (\ref{eq:v23}) and (\ref{eq:ij}) we obtain
\be
\langle\, ij\,\rangle = 
\sqrt{z_i z_j} \; \uvec{\epsilon}^{(+)}\cdot \uvec{v}_{ij}\, ,
\qquad 
[\,ij\,] = \sqrt{z_i z_j} \; \uvec{\epsilon}^{(-)}\cdot \uvec{v}_{ij} \, ,
\ee
and the dependence on the transverse momenta in the light cone wave function
can be expressed by $\langle\,ij\,\rangle$ and $[\,ij\,]$.

Using these expressions we can check the collinear limit for the on-shell case.
The Eq.~(\ref{eq:vxisplit}) is part  of the recursion relation (\ref{eq:recurrence1}) for the off-shell
multi-gluon wave function. It actually describes the situation in which the gluon  with momentum $k_{12}$ splits into two daughter gluons, with momenta 
$k_1$ and $k_2$ respectively. It is interesting to investigate the 
collinear limit of the on-shell amplitude, which should get factorized. 
To get the on-shell amplitude one needs to drop the non-local denominator  $D_{n-1}$ in (\ref{eq:vxisplit}).
The factorizable limit for gluons $1$ and $2$ is then
\begin{multline}
 \Psi_{n} ^{(1||2)}(k_1,k_2,\ldots, k_{n}) = 
{g  \over \sqrt{\xi_{12}}} 
{\uvec{\epsilon}^{(-)} \uvec{v}_{12} \over   \xi_{12}\,
\uvec{v}_{12}^2}\, \Psi_{n-1} (k_1+k_2,k_3,\ldots, k_{n})  = \\
=\frac{1}{\sqrt{\xi_{12}}}\frac{g}{\sqrt{z(1-z)}}\frac{ [12]}{s_{12}}\Psi_{n-1} (k_1+k_2,k_3,\ldots, k_{n})
\label{eq:vxisplitcollinear}
\end{multline}
where $z=z_1/(z_1+z_2)$ and $s_{12}=(k_1+k_2)^2$. 
Relation (\ref{eq:vxisplitcollinear}) is, modulo $1/\sqrt{\xi_{12}}$ 
coefficient and the sign convention,  exactly  the factorization relation
on the collinear poles for the kinematical parts of the dual amplitudes
as shown  in \cite{Mangano:1990by}. It is interesting to note that the only thing that we have done here
is to identify the gluons with momenta $k_1,k_2$ as originating from the splitting of the gluon with momentum $k_{12}$ and therefore we have selected only one splitting out of $n-2$ possible combinations.\\

\subsection{Multi-gluon wave function in the coordinate representation}

It is interesting to investigate the form of the recursion relation (\ref{eq:vxisplit}) in
the transverse coordinate representation. By this we mean performing the Fourier transform with respect to the transverse components of the momenta, just like in the original dipole approach \cite{AlMueller}.
In this section we will consider a special case of the wave function where in the initial state we have only  one gluon. In the original approach the initial state was the quark-antiquark pair. We will discuss the latter case in the next section where we will derive the modified dipole kernel. 
In this section we want to demonstrate calculational techniques which allow us to resum the consecutive splittings in the wave function.
We therefore assume that initially we have the just one gluon, which has momentum $k_1$.
 The initial wave function in the transverse space is then written as follows
\be
{\Phi}_{1}(1)\equiv{\Phi}_{1}(z_1,\underline{r}_{1})=\int \frac{d^2 \underline{k}_1}{(2\pi)^2}\, e^{i \underline{k}_1 \cdot \underline{r}_{1}} \, \Psi_1(z_1,\uvec{k}_1) \; ,
\ee
where $\underline{r}_1$ is the coordinate  in the transverse space. In the rest of the paper the symbol $\Psi$ will always denote the wave function in the momentum space, whereas $\Phi$ will denote the wave function in the coordinate space.
For the wave function with $n$ gluons we define
\begin{multline}
\Phi_{n}(1,\ldots,n) \; \equiv  \;
\Phi_{n} (z_1,\uvec{r}_1;z_2,\uvec{r}_2; \ldots, z_n,\uvec{r}_n) \\
\; = \; \int {d^2 \uvec{k}_1 \over (2\pi)^2}
{d^2 \uvec{k}_2 \over (2\pi)^2} \ldots 
{d^2 \uvec{k}_n \over (2\pi)^2} \, 
\exp(i\uvec{k}_1 \cdot \uvec{r}_1 + i\uvec{k}_2 \cdot\uvec{r}_2 +\, \ldots 
\,+i\uvec{k}_n \cdot\uvec{r}_n) \; 
\Psi_{n} (k_1, \ldots, k_n),
\end{multline}
and similarly for $\Phi_{n-1}$ and $\Psi_{n-1}$.
For the purpose of the subsequent calculation it will be convenient to  change the  variables for the transverse momenta from $(\uvec{k}_1,\uvec{k}_2)$ to
$\uvec{K}_{12}=\uvec{k}_1+\uvec{k}_2$ and 
$\uvec{\kappa}_{12}=\xi_{12}\uvec{v}_{12}$, and the longitudinal variables from
$z_1,z_2$ to $z_{12} = z_1+z_2$, and $\xi_{12}$. These are actually the c.m.s.\ variables for gluons 1~and~2. In addition  one has:
$\,d^2 \uvec{k}_1 \, d^2 \uvec{k}_2 \, = \, 
d^2 \uvec{K}_{12}\, d^2 \uvec{\kappa}_{12}\,$, 
and $z_1 z_2 = \xi_{12} z_{12}$. The scalar products can be rewritten as follows,
$\,\uvec{k}_1 \uvec{r}_1 + \uvec{k}_2 \uvec{r}_2 =
\uvec{K}_{12} \uvec{R}_{12} + \uvec{\kappa}_{12} \uvec{r}_{12}$, where
$\uvec{R}_{12} = {z_1 \uvec{r}_1 + z_2 \uvec{r}_2 \over z_1 + z_2}$, and
$\uvec{r}_{12} = \uvec{r}_2 - \uvec{r}_1$.
Next, we represent the energy denominator,
$$\frac{1}{D_{n}} \,=\, \left[\,{\uvec{\kappa}_{12}^2 \over \xi _{12}} + {\uvec{K}^2_{12} \over z_{12}} 
+ {\uvec{k}_3 ^2 \over  z_3 } + \, \ldots \, +{\uvec{k}_n^2 \over z_n} +Q^2\,
\right]^{-1} \; , $$  as
\be
{1\over D_{n}} = \int_0 ^\infty d\tau \, \exp (-\tau  D_{n}) \; ,
\label{eq:den_tau}
\ee
where recall that 
$Q^2$ is the  virtuality of the incoming particle.
The above formula for the denominator (\ref{eq:den_tau}) may be regarded as the Hamiltonian representation with the Euclidean time $\tau$. Using this representation we arrive at
\begin{multline}
\Phi_{n}(1,2,\ldots,n) \; =  \; 
\int {d^2 \uvec{\kappa}_{12} \over (2\pi)^2}
{d^2 \uvec{K}_{12} \over (2\pi)^2} \ldots 
{d^2 \uvec{k}_n \over (2\pi)^2} \, 
\exp(i\uvec{\kappa}_{12} \cdot \uvec{r}_{12} + i\uvec{K}_{12} \cdot\uvec{R}_{12} 
+ \, \ldots \, + i\uvec{k}_n \cdot\uvec{r}_n) \; 
{\uvec{\epsilon}^{(-)}\uvec{v}_{12} \over \sqrt{\xi_{12}}} 
\\
\times \; 
\int_0 ^\infty d\tau e^{-\tau D_{n}}\, 
\Psi_{n-1} (K_{12},k_3,\ldots,k_n) \\
= -i\,{\uvec{\epsilon}^{(-)} \cdot \uvec{\partial}_{r_{12}}
\over \xi_{12}\sqrt{\xi_{12}}}\;
\int {d^2 \uvec{\kappa}_{12} \over (2\pi)^2}
{d^2 \uvec{K}_{12} \over (2\pi)^2} \ldots 
{d^2 \uvec{k}_n \over (2\pi)^2} \, 
\exp(i\uvec{\kappa}_{12} \cdot \uvec{r}_{12} + i\uvec{K}_{12} \cdot \uvec{R}_{12} 
+ \, \ldots \,+  i\uvec{k}_n \cdot \uvec{r}_n) \; \\
\times \;
\int_0 ^{\infty} d\tau \, \exp\left[-\tau \left( 
 {\uvec{\kappa}_{12}^2 \over \xi_{12}}
+ {\uvec{K}_{12}^2 \over z_{12}}
+ \ldots 
+ {\uvec{k}_n^2 \over z_n} + Q^2 \right)\right]  \\
\times \; 
\int d^2 \uvec{r}'_1 \ldots   d^2 \uvec{r}'_{n-1} \,  
\exp(-i(\uvec{k}_{1}+\uvec{k}_{2}) \cdot \uvec{r}'_1 - i\uvec{k}_{3} \cdot \uvec{r}'_2 
- \,  \ldots  \,-i\uvec{k}_n \cdot \uvec{r}'_{n-1})\, \Phi_{n-1} (1',\ldots,(n-1)')\; ,
\end{multline}
where we have used the recursion formula (\ref{eq:vxisplit}) and the inverse Fourier transform for the
$n$ gluon wave function.
Note, that the primed $z$ values are the following: $z'_1 = z_1+z_2$, $z'_2 = z_3$, \ldots, $z' _{n-1} = z_{n}$.
The  integrals over all transverse momenta are Gaussian and  can be easily performed, which results in 
\begin{multline}
\Phi_{n}(1,\ldots,n) \; =  \; 
 -i\,{\uvec{\epsilon}^{(-)} \cdot \uvec{\partial}_{r_{12}}
\over \xi_{12}\sqrt{\xi_{12}}}\; 
\xi_{12} z_{12} z_3 \ldots z_n \;
\int_0 ^{\infty} d\tau \, \left({1\over 4\pi\tau}\right)^{n}
\int d^2 \uvec{r}'_1 \ldots   d^2 \uvec{r}'_{n-1} \,  
 \\
\times \; 
\exp\left[-{1\over 4\tau} \left[\xi_{12}\uvec{r}_{12}^2 +
z_{12}(\uvec{r}'_1 -\uvec{R}_{12})^2 + 
z_3 (\uvec{r}'_2 - \uvec{r}_3)^2 + \ldots + 
z_n (\uvec{r}'_{n-1} - \uvec{r}_n)^2 \right] -Q^2\tau \right]
\; \Phi_{n-1}(1',2',\ldots,(n-1)') \; .
\end{multline}
 The integral over the `time' $\tau$ can be performed to give the expression with the modified Bessel function $K_0$
\begin{equation}
\Phi_{n}(1,\ldots,n) \; =  \; = \;  i\,{\uvec{\epsilon}^{(-)} \cdot \uvec{r}_{12}
\over \sqrt{\xi_{12}}}\; z_1 z_2 \ldots z_n \;
\int d^2 \uvec{r}'_1 \ldots  d^2 \uvec{r}'_{n-1} \;
\Phi_{n-1}(1',2',\ldots,(n-1)')\;
 \; \left( -{1\over \pi}{\partial \over \partial A} \right)^{n} 
\; 
 K_0 \left(\sqrt{Q^2 A}\right)  \; ,
\label{eq:multibessel}
\end{equation}
where 
\be
A\equiv\xi_{12}\uvec{r}_{12}^2 +
z_{12}(\uvec{r}'_1 -\uvec{R}_{12})^2 + 
z_3 (\uvec{r}'_2 - \uvec{r}_3)^2 + \ldots + 
z_n (\uvec{r}'_{n-1} - \uvec{r}_n)^2\,
\, .
\label{eq:A}
\ee
The integral kernel 
$\;\left( -{1\over \pi}{\partial \over \partial A} \right)^{n} 
\; K_0 \left(\sqrt{Q^2 A}\right)\;$ can be rewritten into a more elegant form using the relations between the modified Bessel functions and their derivatives, see for example \cite{GradRyzh}
$$
\left(\frac{d}{x\,dx}\right)^m [x^{-n}K_{n}(x)] = (-1)^m x^{-n-m}\, K_{n+m}(x) \; ,
$$
which gives 
$$
\left( -{1\over \pi}{\partial \over \partial A} \right)^{n} 
\; K_0 \left(\sqrt{Q^2 A}\right) \; = \; \frac{1}{(2\pi)^{n}} \, \left( \frac{Q^2}{A}\right)^{\frac{n}{2}} \, K_{n}(\sqrt{Q^2 A}) \; .
$$
Using the above relations we can recast the recurrence relation (\ref{eq:multibessel}) into
\begin{equation}
\Phi_{n}(1,\ldots,n) \; =  \;  i\,{\uvec{\epsilon}^{(-)} \cdot \uvec{r}_{12}
\over \sqrt{\xi_{12}}}\;  z_1 z_2 \ldots z_n 
\int \frac{d^2 \uvec{r}'_1 \ldots  d^2 \uvec{r}'_{n-1}}{(2\pi)^{n}}  \left( \frac{Q^2}{A}\right)^{\frac{n}{2}}  K_{n}(\sqrt{Q^2 A}) \;
\Phi_{n-1}(1',2',\ldots,(n-1)')\; ,
\label{eq:multibessel2}
\end{equation}
with $A$ defined above (\ref{eq:A}). 
The formula (\ref{eq:multibessel}) is the prescription for the off-shell tree level wave function with exact kinematics. Since it  depends on the coordinates of all $n-1$
gluons through the variable $A$ defined above, it is quite complicated. Also, it should be kept in mind that this is just formula for one particular splitting, one needs to sum over all possible splittings like in (\ref{eq:recurrence1}). We will show nevertheless that in the case where the incoming particle is on-shell there are significant simplifications, which allow to resum the multiple gluon splittings.
 The crucial difference with respect to the leading logarithmic approximation with the Regge kinematics is the appearance of the modified Bessel functions $K_{n}$ which contain the information about the gluon splitting. In the original approach the splitting of the gluon leads to the expression which is just a power in the transverse coordinates. This translates into the power-like behavior of the splitting kernel in the dipole equation. Here, because the kinematics is kept exact the functional dependence is governed by the Bessel functions, which for large values of their arguments  have exponential behavior  asymptotically. This will result in a qualitative difference when investigating the impact parameter dependence of the scattering amplitude. We will come back to this problem and discuss it in more detail at the end of Sec.~\ref{sec:approx}.
 
\subsection{Resumming the multi-gluon wave function in the case of the on-shell incoming gluon}

We consider here the multi-gluon wave function
that originates from subsequent splittings
of an on-shell incoming gluon with helicity~$+$.  One can also alternatively think about it as the incoming particle with a large momentum $P^+$ such that $P^- = -\frac{Q^2}{2P^+}$ is very small, at least as compared with the particles in the wave function. This will result in energy denominators which do not contain the initial $P^-$. We will assume that all gluons have  $+$~helicities which should be a situation in the   high energy limit, where the helicity flips are suppressed. We shall use 
the complex representation of the transverse vectors: 
$v_{ij} = \uvec\epsilon^{(+)}\cdot \uvec{v}_{ij}$,  
$v^*_{ij} = \uvec\epsilon^{(-)}\cdot \uvec{v}_{ij}$, and 
a useful notation, 
\be
{v}_{(i_1 i_2 \ldots i_p)(j_1 j_2 \ldots j_q)} = 
{{k}_{i_1} + {k}_{i_2} + \ldots + {k}_{i_p}
\over z_{i_1} + z_{i_2} + \ldots + z_{i_p}} -
{{k}_{j_1} + {k}_{j_2} + \ldots + {k}_{j_q}
\over z_{j_1}+ z_{j_2} + \ldots + z_{j_q}}\; ,
\ee
\be 
\xi_{(i_1 i_2 \ldots i_p)(j_1 j_2 \ldots j_q)} = 
{
(z_{i_1}+z_{i_2}+\ldots+z_{i_p})
(z_{j_1}+z_{j_2}+\ldots+z_{j_q})
\over 
z_{i_1}+ z_{i_2}+ \ldots+ z_{i_p} + z_{j_1} + z_{j_2} + \ldots + z_{j_q}} \;,
\ee
with notation $k_i \equiv \uvec{\epsilon}^{(+)}\cdot \uvec{k}_i$.
The global momentum conservation $\delta$-functions, 
$\Delta^{(n)} = \delta^{(2)} \left(\, \sum_{i=1} ^n \uvec{k}_i \,\right) \, 
\delta \left(\, 1 - \sum_{i=1} ^n z_i\, \right)$, that are present in 
all the expressions for the wave functions will not be displayed explicitly.
Thus, the incoming state has the wave function,
\be
\Psi_1(1) = 1.
\ee
In the following discussion we will consider color ordering in the amplitudes, therefore we will suppress color degrees of freedom.
In general, for a color-ordered amplitude, the gluon splitting acts on
the wave function as derived in  Eq.~(\ref{eq:recurrence1}) (for on-shell initial state though)
\[
-D_{n+1}\, \Psi_{n+1}(1,2,\ldots, n+1) \; = \; 
\]
\be
= \,g{v^*_{12} \over \sqrt{\xi_{12}}} \Psi_n (12,3,\ldots,n+1) \, +
 \,g{v^*_{23} \over \sqrt{\xi_{23}}} \Psi_n (1,23,\ldots,n+1) \, +
  \, \ldots \, +g{v^*_{n\, n+1} \over \sqrt{\xi_{n\,n+1}}} \Psi_n (1,2,\ldots,n\, n+1)\, ,
\label{eq:fullsplit}
\ee
with $D_{n+1}=\uvec{k}_1^2/z_1+\uvec{k}_2^2/z_2+\dots+\uvec{k}_{n+1}^2/z_{n+1}$.
We have introduced the notation $\Psi_n(1,\ldots,i-1 \, i,\ldots,n+1)$ where $ \; i-1 \, i \; $ means that it is the gluon with  the momentum $k_{i-1\, i}=k_{i-1}+k_i$.
After the first splitting one gets
\be
\Psi_2(1,2) \; =  \;
-g\, {1 \over \sqrt{\xi_{12}}}\,
{v^*_{12} \over \xi_{12} |v_{12}|^2} 
\; = \;  
-g\, {1\over\sqrt{\xi_{12}}}\, {1\over \xi_{12} v_{12}} \, ,
\ee
where we have taken that $P^-=0$.
According to (\ref{eq:fullsplit}), 
the next splitting leads from $\Psi_2(1,2)$ to $\Psi_3(1,2,3)$:
\[
-D_3 \Psi_3(1,2,3) \; = \;- g\, \left[ \,
{v^*_{12} \over \sqrt{\xi_{12}}}\Psi_2 (12,3) 
\, + \, {v^*_{23} \over \sqrt{\xi_{23}}} \Psi_2 (1,23) \, \right]
\]
\be
= \; g^2\, \left[ \, {v^*_{12} \over \sqrt{\xi_{12}\xi_{(12)3}}}
{1\over\xi_{(12)3}\, v_{(12)3}} 
\,+\,{v^*_{23} \over \sqrt{\xi_{23}\xi_{1(23)}}}
{1\over\xi_{1(23)}\, v_{1(23)}} \, \right] \, . 
\label{eq:wavefunction3}
\ee
This expression may be, after some simple algebra, simplified using 
$\xi_{12}\xi_{(12)3} \, = \, \xi_{23}\xi_{1(23)} \, = \, 
{z_1 z_2 z_3 \over z_1 + z_2 + z_3} \; = \; z_1 z_2 z_3$.
One obtains,
\be
\Psi_3(1,2,3) \; = \; g^2 \, {1 \over \sqrt{z_1 z_2 z_3}}\,
{1\over\xi_{(12)3}\xi_{1(23)}}\, {1\over v_{(12)3} \, v_{1(23)}}\; . 
\ee 
Note that, the energy denominator $D_3$ disappeared from the equation as it has canceled with the numerator when finding the common denominator for expression (\ref{eq:wavefunction3}).
The same procedure can be  iterated further. We shall give below the explicit 
form of the wave function obtained for 4 gluons and then present a
 generalization to an arbitrary $n$. 
Thus, for $n=4$ we found:
\be
\Psi_4(1,2,3,4) \; = \; -g^3\, 
{1 \over \sqrt{z_1 z_2 z_3 z_4}}\,
{1\over\xi_{(123)4}\, \xi_{(12)(34)}\, \xi_{1(234)}}\, 
{1\over v_{(123)4}\, v_{(12)(34)}\, v_{1(234) }}, 
\ee
and for a general integer $n>2$ one expects,
\[
\Psi_n(1,2,\ldots,n) \; = \; (-1)^{n-1}g^{n-1}\, {1 \over \sqrt{z_1 z_2 \ldots z_n}}\,
{1\over\xi_{(12\ldots n-1)n}\,\xi_{(12\ldots n-2)(n-1\,n )} \,
\ldots \, \xi_{1(2\ldots n)}}
\]
\be
\times\;
{1\over v_{(12\ldots n-1)n}\, v_{(12\ldots n-2)(n-1\,n )} \,
\ldots \, v_{1(2\ldots n)}} \; . 
\label{eq:psinfact}
\ee
This formula was explicitly verified for $n=2,3,4,5,6$.  The proof for arbitrary $n$ can be done by mathematical induction and proceeds as follows. 
We assume that the wave function $\Psi_{n}$ satisfies the above conjecture (\ref{eq:psinfact}).
Using (\ref{eq:fullsplit}) the wave function $\Psi_{n+1}$ has  then the form
\begin{multline}
\label{eq:psin}
-D_{n+1}\, \Psi_{n+1}(1,2,\ldots, n+1) \; = \; g\,\sum_{i=1}^{n} \, \frac{v^*_{i\, i+1}}{\sqrt{\xi_{i\,i+1}}}  \Psi_{n}(1,2,\ldots,(i \, i+1),
\dots ,n+1) \, = \\
(-1)^{n-1} g^n \,\sum_{i=1}^{n} \, \frac{v^*_{i\,i+1}}{\sqrt{\xi_{i\,i+1}}} \, \frac{1}{\sqrt{z_1z_2\dots(z_{i}+z_{i+1})\dots z_{n}}}\frac{1}{(\xi_{(12\ldots n)n+1}\,\xi_{(12\ldots n-1)(n\,n+1 )} \,
\ldots \, \xi_{1(2\ldots n+1)} )'} \times \\
\frac{1}{(v_{(12\ldots n)n+1}\, v_{(12\ldots n-1)(n\,n+1 )} \,
\ldots \, v_{1(2\ldots n+1)})'} \; .
\end{multline}
We have inserted the  symbol $'$ to denote the fact that the indices $i\,i+1$ have to be taken  together, or in other words
in each term of the sum in (\ref{eq:psin}) for a given $i$ the denominator does not have the term of the form  $ \xi_{(1\dots i)(i+1\dots n+1)} v_{(1\dots i)(i+1\dots n+1)}$. 
The expression under the square root is 
\be
\xi_{i\,i+1}\,z_1z_2\dots(z_{i}+z_{i+1})\dots z_{n}=\frac{z_i z_{i+1}}{z_i+z_{i+1}}\,z_1z_2\dots(z_{i}+z_{i+1})\dots z_{n}=
z_1z_2\dots z_{n+1} \; ,
\label{eq:zproduct}
\ee
because $i$ goes from $1$ to $n$, there are $n+1$ terms in the product (\ref{eq:zproduct}).
Now we write (\ref{eq:psin}) in a form with the common denominator for  all the terms which gives
\begin{multline}
\label{eq:psin2}
D_{n+1}\, \Psi_{n+1}(1,2,\ldots, n+1) \; = \; (-1)^{n} g^n\,{ \sum_{i=1}^{n} v_{i\,i+1}\xi_{(1\dots i)(i+1\dots n+1)} v_{(1\dots i)(i+1\dots n+1)}\over \sqrt{z_1 z_2 \ldots z_{n+1}}\,\xi_{(12\ldots n)n+1}\,\xi_{(12\ldots n-1)(n\,n+1 )} \,
\ldots \, \xi_{1(2\ldots n+1)}} \times \\
{1\over v_{(12\ldots n)n+1}\, v_{(12\ldots n-1)(n\,n+1 )} \,
\ldots \, v_{1(2\ldots n+1)}} \; .
\end{multline}
The denominator in the above equation is just what gives the wave function $\Psi_{n+1}$.  
An essential point is to show that
\begin{equation}
\sum_{i=1}^{n} v_{i\,i+1}\xi_{(1\dots i)(i+1\dots n+1)} v_{(1\dots i)(i+1\dots n+1)} =\sum_{j=1}^{n+1} \frac{\uvec{k}_j^2}{z_j} \equiv D_{n+1} \; ,
\label{eq:ndenominator}
\end{equation}
that is that the numerator on the right hand side in (\ref{eq:psin2}) cancels $D_{n+1}$ on the left hand side of this equation.
We notice that,
\be
v_{(1\ldots i)(i+1 \ldots n)} \xi _{(1\ldots i)(i+1 \ldots n)}
\; = \; \left({k_1 + \ldots + k_i \over z_1 + \ldots + z_i} +  
{k_1 + \ldots k_i \over 1 - (z_1 + \ldots + z_i)}\right)\,  
\xi _{(1\ldots i)(i+1 \ldots n)} \; = \; 
 \sum_{j=1}^{i} k_j\;.
 \label{eq:vxi_sumk}
\ee
Therefore the left hand side of (\ref{eq:ndenominator})  is 
\begin{multline}
\sum_{i=1}^{n} v_{i\,i+1}\xi_{(1\dots i)(i+1\dots n+1)} v_{(1\dots i)(i+1\dots n+1)} = \sum_{i=1}^{n} v_{i\,i+1} \sum_{j=1}^{i} k_j = 
 \sum_{i=1}^{n} \bigg(\frac{k_i}{z_i}-\frac{k_{i+1}}{z_{i+1}}\bigg) \sum_{j=1}^{i} k_j = \\
 =\sum_{i=1}^{n} \frac{k_i}{z_i} \sum_{j=1}^{i} k_j - \sum_{i=1}^{n} \frac{k_{i+1}}{z_{i+1}} \sum_{j=1}^{i} k_j 
 = \sum_{i=1}^{n} \frac{k_i^2}{z_i}+\sum_{i=2}^{n}\frac{k_i}{z_i}\sum_{j=1}^{i-1}k_j+\frac{k_{n+1}^2}{z_{n+1}}-\sum_{i=1}^{n-1}\frac{k_{i+1}}{z_{i+1}}\sum_{j=1}^{i}k_j \; .
\label{eq:proof_denom}
\end{multline}
The second and fourth term cancel, and the first and third
give
\begin{equation}
 \sum_{i=1}^{n+1} \frac{k_i^2}{z_i} = D_{n+1} \; .
\end{equation}
which is exactly the denominator, so we have proven (\ref{eq:ndenominator}).
Note that, we have also used 
$$
\sum_{i=1}^{n} k_i = - k_{n+1} \; ,
$$
i.e. the condition which is coming from the energy-momentum conservation.
This completes the proof, as $\Psi_{n+1}$ is exactly of the form (\ref{eq:psinfact}) for $n\rightarrow n+1$.

The resummed form (\ref{eq:psinfact}) looks relatively simple. 
Its dependence on transverse momenta of gluons factorized into 
reciprocals of  $v_{(12\ldots p)(p \ldots n)}$ with the splitting  index
$p$ takes all possible positions from $p=2$ to $p=n-1$.
The wave function given by (\ref{eq:psinfact}) may be also expressed in
terms of the $\langle\, ij \, \rangle$ symbols in the following way:
\[
\Psi_n(1,2,\ldots,n) \;=\;(-1)^{n-1}g^{n-1} \, {1 \over \sqrt{z_1 z_2 \ldots z_n}}\;
{1\over \sqrt{\xi_{(12\ldots n-1)n}\,\xi_{(12\ldots n-2)(n-1\,n )} \,
\ldots \, \xi_{1(2\ldots n)}}} \;
\]
\be
{1\over 
\langle(12\ldots n-1)n \rangle\, 
\langle(12\ldots n-2)(n-1\,n )\rangle \,
\ldots \, 
\langle 1(2\ldots n)\rangle}. 
\ee 

The wave function given by (\ref{eq:psinfact}) can be simplified further.

Denoting $k_{(1\ldots p)} \equiv k_1 + \ldots + k_p$ we may write:
\be
\Psi_n(1,2,\ldots,n) \;=\;(-1)^{n-1}g^{n-1}\, {1 \over \sqrt{z_1 z_2 \ldots z_n}}\;
{1\over k_{(1)} k_{(12)} \ldots k_{(12\ldots n-1)}}.
\label{eq:psinki}
\ee
One may go back to the real representation of the transverse vectors and 
perform the Fourier transform of (\ref{eq:psinki}),
\be
\Phi_n (z_1,\uvec{r}_1; \ldots ; z_n,\uvec{r}_n)\; = \;
\int {d^2 \uvec{k}_1 \over (2\pi)^2} \ldots {d^2 \uvec{k}_n \over (2\pi)^2}\; 
\exp(i \uvec{k}_1 \cdot \uvec{r}_1 + \ldots + i \uvec{k}_n \cdot \uvec{r}_n) \;
\Psi_n (z_1,\uvec{k}_1,\ldots,z_n,\uvec{k}_n).
\ee
It is convenient to change the momentum variables:
\be
\{\uvec{k}_1,\uvec{k}_2,\ldots \uvec{k}_n\} \to 
\{\uvec{k}_{(1)},\uvec{k}_{(12)},\ldots,\uvec{k}_{(12\ldots n)}\}.
\ee
The Jacobian of the transformation is trivial, and one
expresses the gluon momenta in the following way:
\be
\uvec{k}_1 = \uvec{k}_{(1)},\quad 
\uvec{k}_2 = \uvec{k}_{(12)} - \uvec{k}_{(1)}, \quad \ldots, \quad 
\uvec{k}_n = \uvec{k}_{(12\ldots n)} - \uvec{k}_{(12\ldots n-1)}.
\ee
The Fourier exponent takes the form, 
\be
\exp(i \uvec{k}_1 \cdot \uvec{r}_1 + \ldots + i \uvec{k}_n \cdot \uvec{r}_n) 
\; = \; \exp(i \uvec{k}_{(1)} \cdot (\uvec{r}_1-\uvec{r}_2) +  
i \uvec{k}_{(12)} \cdot (\uvec{r}_2-\uvec{r}_3) + \ldots 
i \uvec{k}_{(12\ldots n-1)} \cdot (\uvec{r}_{n-1}-\uvec{r}_n) + 
i \uvec{k}_{(12\ldots n)} \cdot \uvec{r}_n),
\ee
and the final answer for the wave function in position space reads \be
\Phi_n (z_1,\uvec{r}_1; \ldots ; z_n,\uvec{r}_n)\; = \; (-1)^{n-1} g^{n-1}
\delta\left( 1-\sum_{i=1} ^n z_i \right)\, {1\over \sqrt{z_1 z_2 \ldots z_n}}\;
{
\uvec{\epsilon}^{(-)}\uvec{r}_{12} \,  \uvec{\epsilon}^{(-)}\uvec{r}_{23} \,
\ldots \, \uvec{\epsilon}^{(-)}\uvec{r}_{n-1\,n} 
\over 
r^2 _{12} \, r^2 _{23} \, \ldots \,  r^2 _{n-1\,n} }.
\label{eq:onshelltrcoowf}
\ee
Surprisingly, this is precisely the structure of the gluon wave function 
in the LL~dipole model. However,  it is crucial for the derivation
that the incoming gluon is on-shell, and that we chose a very special
component of the wave function, with all positive gluon helicities.  
Note that the on-shellness (or quasi-onshellness) of the incoming particle means that in the light cone perturbation theory we neglect  in the energy denominators the energy  of this particular particle and keep only the energies of the subsequent emitted gluons.

\section{Dipole kernel with approximate kinematics}
\label{sec:approx}
\subsection{Dipole wave-function in the leading logarithmic approximation.}

In the previous section we have presented a general approach for the treatment of the light cone wave function while keeping the kinematics exact. We were able to construct the recurrence relation for the off-shell case and resum the wave function for the special case of the on-shell incoming particle and with all the gluons having the same helicities.
However, for the practical purposes, it would be desirable to construct the evolution equation in rapidity that would resum the gluon emissions just like in the original dipole approach developed by Mueller \cite{AlMueller}. We will show in this section that it is possible, though not all the exact kinematical effects are kept here. The modified equation should include at least part of the corrections due to the exact kinematics. 
These corrections will prove to be sufficient to recover the double logarithmic corrections  found  in the next-to-leading calculation.
 
Let us therefore recall the original construction of the dipole wave function at small $x$ with soft gluons which was first developed  by Mueller \cite{AlMueller}.  Unlike  the previous section, where we took the initial particle to be the gluon we will consider here the $q\bar{q}$ pair in the color singlet state, which is the onium. The  heavy onium wave function, which consists of the quark-antiquark pair without any additional gluons, is defined as
\be
\Psi^{(0)}(z_1,\underline{k}_1), \; \; \; \;  \;  {\rm where } \; \; \;  \; \;  z_1=k_1^+/P^+ \; . 
\ee
The initial momentum is $P$, the quark has four-momentum $k_1=(k_1^+,k_1^-,\underline{k}_1)$ and the antiquark has momentum $P-k_1$. In the transverse space the wave-function reads
\be
{\Phi}^{(0)}(z_1,\underline{x}_{01})=\int \frac{d^2 \underline{k}_1}{(2\pi)^2}\, e^{i \underline{k}_1 \cdot \underline{x}_{01}} \, \Psi^{(0)}(z_1,\underline{k}_1) \; ,
\ee
where $\underline{x}_{01}$ is the size of the dipole $01$ in the transverse space. Strictly speaking, this representation is a mixed one: coordinate space in the transverse components and momentum space in the longitudinal degrees of freedom. 

\begin{figure}[ht]
\centerline{\epsfig{file=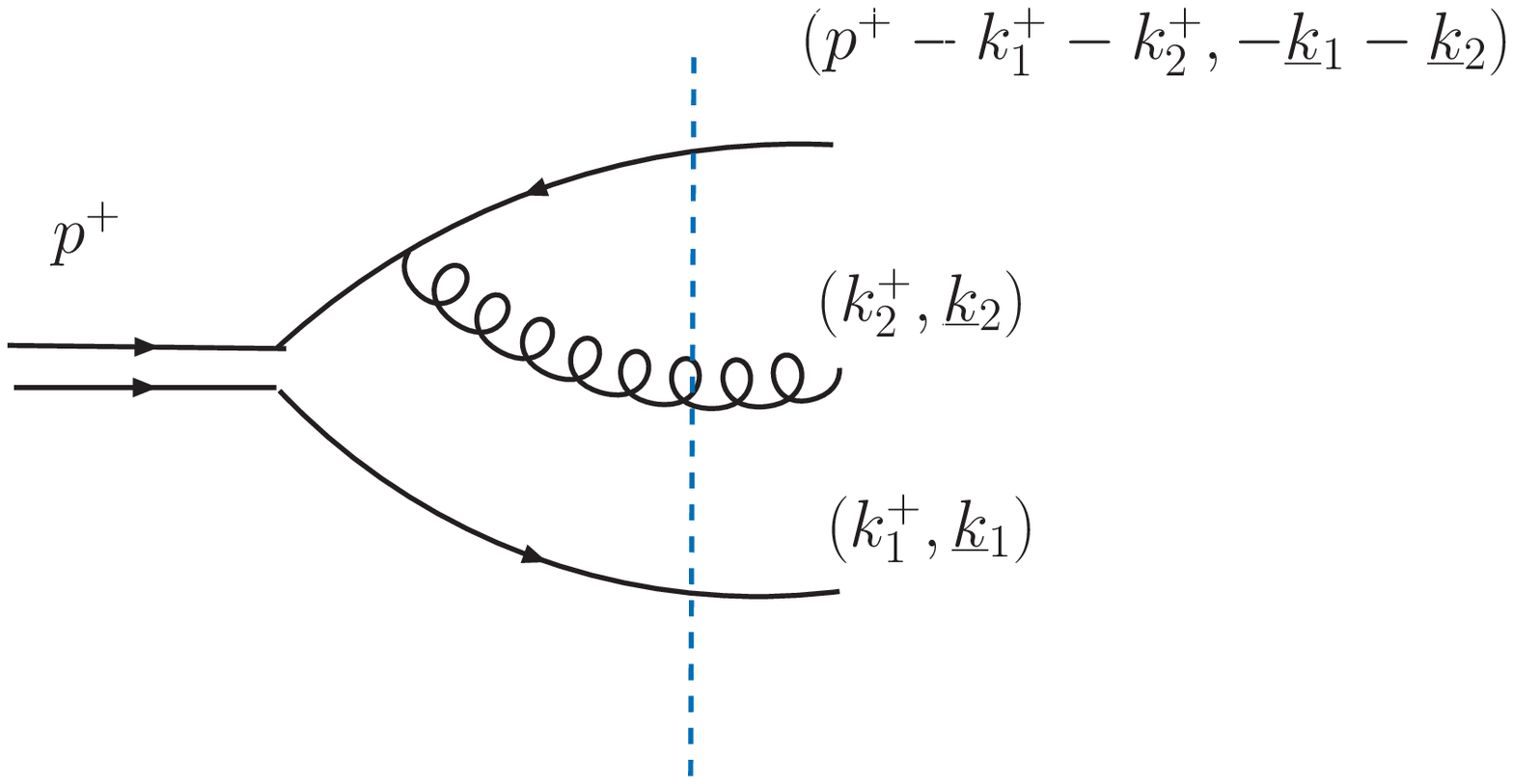,width=7.8cm}\hspace*{1cm}\epsfig{file=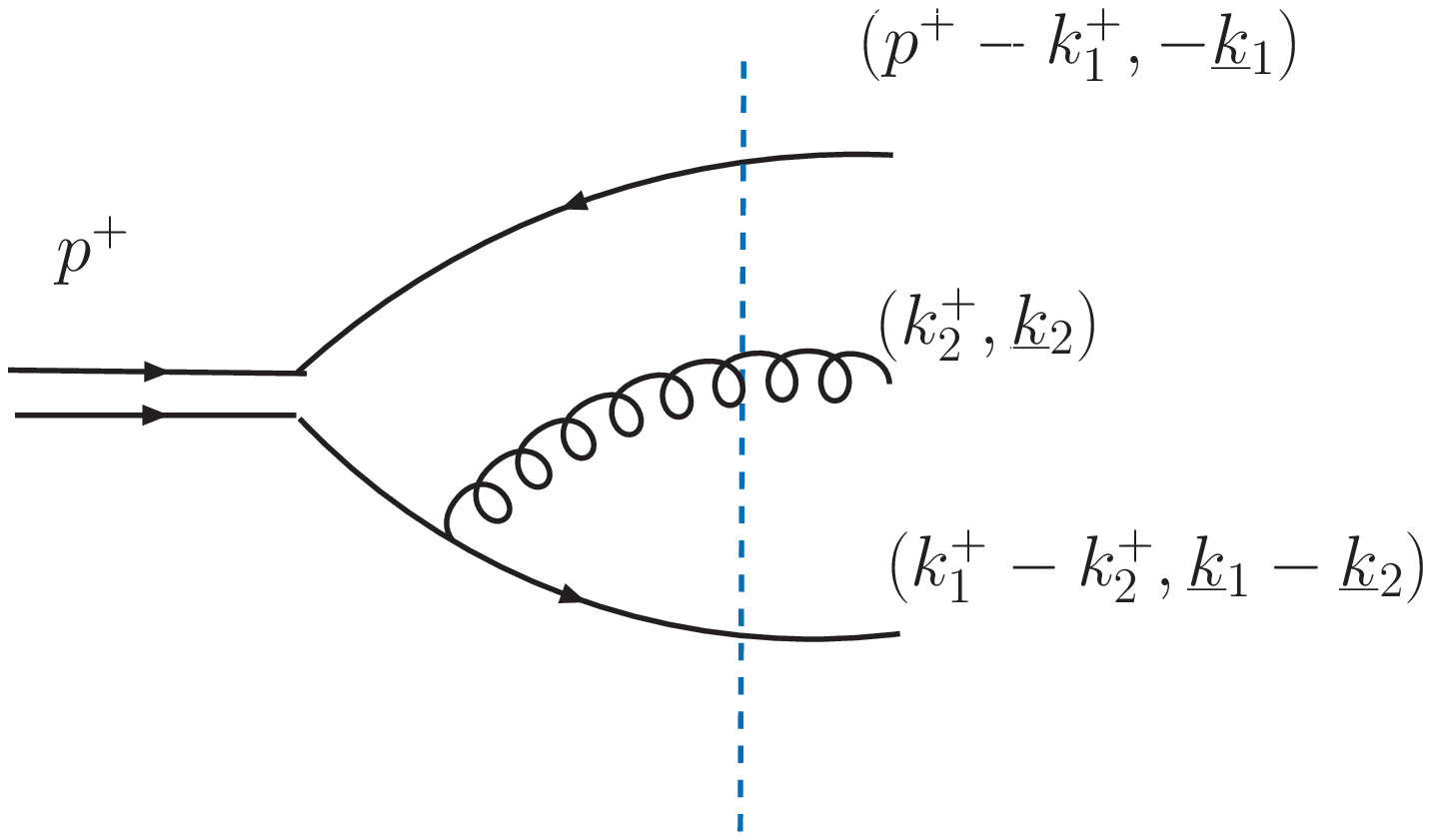,width=7cm}}
\caption{Dipole wave function with one soft gluon. Dashed line indicates the energy denominator for the intermediate state.}
\label{fig:graph1}
\end{figure}

Such representation was shown \cite{AlMueller} to be very convenient for the purpose of studying the high energy limit, as only in this limit there is a decoupling of the transverse and longitudinal degrees of freedom.
Let us investigate the emission of one gluon with momentum $k_2$, see Fig.~\ref{fig:graph1}. 
If we were to keep the kinematics exact we would arrive at the analog of the formula (\ref{eq:vxisplit})
for the case of one splitting, where say the gluon is emitted from the upper anti-quark line (Fig.~\ref{fig:graph1})
\be
\Psi^{(1)}_{\rm exact}(z_1,\underline{k}_1;z_2,\underline{k}_2)= \frac{g t^a \bar{u}(k_1)\gamma\cdot\epsilon_2 u(k_1+k_2)}{\bar{D}_1}  \Psi^{(0)}(z_1,\underline{k}_1)  \,  \; .
\label{eq:psi1exact}
\ee
Here the exact energy denominator has the form
\be
\bar{D}_1 = P^--[(P-k_1-k_2)^-+k_{1}^-+k_{2}^-] \; ,
\label{eq:den1}
\ee
and similarly for the second emission from the other quark line.
Following the procedure presented in the previous section one would 
obtain the recursion formulae which depend on the kinematics of the whole cascade. The simplification that arises in the high energy limit
is  due to the fact that the emission of the very soft daughter gluon gets factorized from the rest of the wave function. In other words there is no recoil due to the emission of the gluon and the rest of the whole cascade is frozen and is not affected by the splitting. This allows to factorize the emissions of the gluons and resum their emissions in the form of the evolution equation which is differential in rapidity.  We shall show that the leading logarithmic approach will be changed when the more exact kinematics is taken into account, but the resummation of the emission can be still recast in the form of the evolution equation.
To this aim let us recall the original assumptions made in \cite{AlMueller} that lead to the derivation of the dipole evolution equation in the leading logarithmic approximation.

The following assumptions are done in order to reproduce the leading logarithmic approximation, or the limit of the high energy
 \begin{itemize}
\item The emitted gluon is longitudinally soft: $k_2^+ \ll k_1^+$, 
 and one defines  $z_1 = k_1^+/P^+, z_2=k_2^+/P^+$ which are  the fractions of the longitudinal momenta. Obviously $z_2 \ll z_1$.
\item The coupling of the gluon to the quark (antiquark) is eikonal
$$g t^a \bar{u}(k_1)\gamma \cdot \epsilon_2 u(k_1+k_2) = 2 g t^a  k_1 \cdot \epsilon_2 \; , $$
where $t^a$ is the color matrix in the fundamental representation and $\epsilon_2$ is the polarization vector of the emitted gluon with momentum $k_2$.
\item Since the gluon is longitudinally soft $k_2^+ \ll k_1^+$ one can keep in graphs the leading term in the energy denominator (see Fig.~\ref{fig:graph1})
$$
\bar{D}_1=\frac{1}{P^--[(P-k_1-k_2)^-+k_{1}^-+k_{2}^-]} \simeq \frac{1}{k_{2}^-} \; , 
$$
because $k_{2}^-$ is the dominant term ($k_{2}^-=\underline{k}_{2}^2/2k_2^+$ and $k_2^+$ is very small, $k_2^+\ll k_1^+,P^+$). Here, $P$ is the initial momentum, see Fig.~\ref{fig:graph1}.

\end{itemize}

If one puts together all these assumptions  one arrives at the formula for the wave function with one emitted gluon
\be
\Psi^{(1)}(z_1,\underline{k}_1;z_2,\underline{k}_2)=2 g t^a \frac{\underline{\epsilon}_2 \cdot \underline{k}_2}{\underline{k}_2^2} [ \Psi^{(0)}(z_1,\underline{k}_1)-\Psi^{(0)}(z_1,\underline{k}_1+\underline{k}_2)]  \, \Theta(z_1-z_2) \; .
\label{eq:psi1mueller}
\ee
At this stage there is  only ordering in the longitudinal fractions $z_1 \gg z_2$ which is extended to the weak inequality $z_1 \ge z_2 $ in the subsequent calculation. 
The transverse momenta are unrestricted in this approximation. 
One can perform the Fourier transform of the above expression to obtain the formula for the wave function with one soft gluon in the transverse space
 \be
{\Phi}^{(1)}(z1,\underline{x}_{01};z_2,\underline{x}_{02}) \; = \; 
 \int \frac{d^2 \underline{k}_1}{(2\pi)^2}\frac{d^2 \underline{k}_2}{(2\pi)^2} \, e^{i \underline{k}_1 \cdot \underline{x}_{01}+i \underline{k}_2 \cdot \underline{x}_{02} }\, \Psi^{(1)}(z_1,\underline{k}_1;z_2,\underline{k}_2)\; ,  \ee
 which yields the explicit result
 \be
{ \Phi}^{(1)}(z_1,\underline{x}_{01};z_2,\underline{x}_{02}) \;=\; 
 -\frac{ig t^a}{\pi}\bigg(\frac{\underline{x}_{20}}{x_{20}^2}-\frac{\underline{x}_{21}}{x_{21}^2}\bigg)\cdot \underline{\epsilon}_2 \, \Psi^{(0)}(z_1,\underline{x}_{01})\; ,
 \label{eq:soft1}
 \ee
 where $x_{ij}^2\equiv \underline{x}_{ij}^2$.
We see the advantage of using the coordinate space representation because 
the soft gluons factorize in (\ref{eq:soft1}) as demonstrated  in \cite{AlMueller}. The modulus squared of the one gluon wave function has then explicitly the form
\be
\left|\Phi^{(1)}\right|^2(z_1,\underline{x}_{01}) \;= \; \int_{z_0}^{z_1} \frac{dz_2}{z_2}\int \frac{d^2\underline{x}_{02}}{2\pi} \, \frac{\uvec{x}_{01}^2}{\uvec{x}_{02}^2 \, \uvec{x}_{12}^2} \, \left|\Phi^{(0)}(z_1,\underline{x}_{01})\right|^2 \; .
\ee
In this form it is particularly transparent that in the soft gluon limit the transverse and longitudinal degrees of freedom decouple. Therefore 
the transverse space coordinate representation is especially useful. The wave function with one soft gluon is just the wave function without any soft gluons times the branching probability. The measure (or branching probability) in this case reads
\be
\label{eq:lldipolekernel}
\frac{d^2\underline{x}_{02} \, \uvec{x}_{01}^2}{\uvec{x}_{02}^2 \, \uvec{x}_{12}^2} \; , 
\ee
is the dipole splitting kernel in the leading logarithmic approximation which appears in the dipole evolution equation for the dipole scattering amplitude  and it was originally derived in \cite{AlMueller}. The same kernel appears in nonlinear version of the evolution equation with additional gluon rescatterings: the  Balitsky-Kovchegov equation \cite{Kovchegov:1999yj,BAL2}. It is straightforward to check that this kernel is invariant with respect to   the $2$-dimensional conformal transformations.
In deriving the equation (\ref{eq:psi1mueller}) one uses crucial assumption about the strong ordering in the energies
\be k_2^- \gg k_1^- \;  .
\label{energy_ordering}
\ee

In the original approach \cite{AlMueller} one uses the assumption of the  softness of the gluon with respect to the parent dipole and treats the transverse momenta as unordered which is equivalent to the Regge kinematics. This enables   to make the approximations as described above and 
factorize the soft gluon contribution.
However,  for the consistency of the calculations we should  keep
the inequality (\ref{energy_ordering}) (where $k_1^-=\frac{\underline{k}_1^2}{2k_1^+},  \; k_2^-=\frac{\underline{k}_2^2}{2k_2^+}$) exact which gives
\be
\frac{\underline{k}_2^2}{k_2^+} > \frac{\underline{k}_1^2}{k_1^+} \; , 
\label{kc_dipole}
\ee
with $k_1^+>k_2^+$. \\
If there are  more gluon emissions we will of course have 
\be
\dots \, \frac{\underline{k}_{i_4}^2}{k_{i_4}^+} > \frac{\underline{k}_{i_3}^2}{k_{i_3}^+}>\frac{\underline{k}_{i_2}^2}{k_{i_2}^+} > \frac{\underline{k}_{i_1}^2}{k_{i_1}^+}\; ,
\ee
and $$\dots  < k_{i_4}^+ < k_{i_3}^+<k_{i_2}^+<k_{i_1}^+ \; ,$$
where the indices $i_1,\ldots,i_4$ enumerate subsequent emissions along one branch of the gluon cascade.
 
As a result of  the ordering (\ref{kc_dipole})  the region in the transverse momenta $\underline{k}_2^2$ for the gluon emission is limited
\be 
\Theta(\underline{k}_2^2-\underline{k}_1^2) + \Theta(\underline{k}_1^2-\underline{k}_2^2)\Theta\left( \underline{k}_2^2- \underline{k}_1^{2} \frac{k_2^+}{k_1^+}\right) \; .
\label{kc_dipole1}
\ee
This means that there is  a constraint which restricts the  transverse momenta of the daughter gluon (labeled $2$).
which is the step function $\Theta(\underline{k}_2^2- \underline{k}_1^{2} \frac{k_2^+}{k_1^+})$.
The constraint (\ref{kc_dipole},\ref{kc_dipole1}) means that the momenta of the emitted gluons (those with $k_2$) are cutoff in the infrared.
  Note that, the ordering (\ref{kc_dipole}) is exactly the same as the one discussed in \cite{Dokshitzer:2005bf,DM,Marchesini:2006ax}, it is ordering in the fluctuation time.    One has to think about the cascade as developing from the hadron side and the onium as a model of a hadron. 
 
 Given the discussion above, the wave function of the onium with one soft gluon should be modified to include the constraint (\ref{kc_dipole}) and with this modification reads
 \be
 \Psi^{(1)}(z_1,\underline{k}_1;z_2,\underline{k}_2)=2 g t^a \frac{\underline{\epsilon}_2 \cdot \underline{k}_2}{\underline{k}_2^2} \, [ \Psi^{(0)}(z_1,\underline{k}_1)-\Psi^{(0)}(z_1,\underline{k}_1+\underline{k}_2)]  \, \Theta(z_1-z_2) \, \Theta\bigg( \frac{\underline{k}_2^2}{z_2}-\frac{\underline{k}_1^2}{z_1}\bigg) \; ,
 \label{psi1_kc}
 \ee
 where we used $z_1$ and $z_2$ instead of $k_1^+$ and $k_2^+$. 
 At this point it is no longer so trivial to perform the Fourier transform into the coordinate space
 since the transverse momenta are entangled now and the soft gluons do not factorize as before in coordinate space. In particular, there will be  a shift (or distortion) of the original dipole since we encounter integrals of the type (if we choose to make the $\underline{k}_1$ integral first in the following example )
 \be
 \int \frac{d^2 \underline{k}_1}{(2\pi)^2} \, e^{i \underline{k}_1 \cdot \underline{x}_{01}} \, \Psi^{(0)}(z_1,\underline{k}_1)  \,\Theta\bigg( \frac{\underline{k}_2^2}{z_2}-\frac{\underline{k}_1^2}{z_1}\bigg)\; .
 \ee
 
 Therefore we see that the kinematical constraint emerges in the light-cone perturbation theory from the more exact treatment of the energy denominators in the  graphs. In the $t$-channel formulation of the BFKL Pomeron, the analogous consistency constraint arises, when one takes into account the fact that the virtualities of the exchanged gluons are dominated by the transverse parts. This leads to the constraint on the transverse momenta of the emitted gluons, see for example \cite{Kwiecinski:1996td}. 
 
 \subsection{Modified energy denominators in the dipole evolution}
 
 In the previous section we have seen how the kinematical constraint emerges from the more careful treatment of the energy denominators.
 In this section we will demonstrate that one can include this effect into the 
  dipole wave function.  In particular we will arrive at a modified dipole evolution equation.
 Let us take the more exact version of the  energy denominator which includes the energy of the parent emitter
 $$
 \frac{1}{k_1^-+k_2^-}  \; .
 $$
 With this modification the formula (\ref{eq:psi1mueller}) for the dipole wave function in momentum space with one gluon becomes
\be
\Psi^{(1)}(z_1,\underline{k}_1; z_2,\underline{k}_2)=2 g t^a \frac{\underline{\epsilon}_2 \cdot \underline{k}_2}{\underline{k}_2^2+\underline{k}_1^2\frac{k_2^+}{k_1^+}} [ \Psi^{(0)}(z_1,\underline{k}_1)-\Psi^{(0)}(z_1,\underline{k}_1+\underline{k}_2)]  \, .
\label{eq:psi1mueller_mod}
\ee
We still keep the vertex to be eikonal, the only modifications are in the energy denominator. Let us define the scale
\be
\overline{Q}^2 \; \equiv \; \underline{k}_1^2 \, \frac{k_2^+}{k_1^+} = \underline{k}_1^2 \, z\; , \; \; z=\frac{z_2}{z_1}\; ,
\label{eq:momqbar}
\ee
 and perform the two-dimensional Fourier 
 transform of (\ref{eq:psi1mueller_mod}) to the coordinate space
 \be
{\Phi}^{(1)}(z_1,\underline{x}_{01}; z_2,\underline{x}_{02}) = 2 g t^a  \int \frac{d^2 \underline{k}_1}{(2\pi)^2} \frac{d^2 \underline{k}_2}{(2\pi)^2}  e^{i \underline{k}_1 \cdot \underline{x}_{01}+i\underline{k}_2 \cdot \underline{x}_{02} } [ \Psi^{(0)}(z_1,\underline{k}_1)-\Psi^{(0)}(z_1,\underline{k}_1+\underline{k}_2)] \frac{\underline{\epsilon}_2 \cdot \underline{k}_2}{\underline{k}_2^2+\overline{Q}^2} \; . 
 \ee
Let us take the first term  in the above formula
\begin{multline}
2 g t^a  \int \frac{d^2 \underline{k}_1}{(2\pi)^2} \frac{d^2 \underline{k}_2}{(2\pi)^2}  e^{i \underline{k}_1 \cdot \underline{x}_{01}+i\underline{k}_2 \cdot \underline{x}_{02} }  \Psi^{(0)}(\underline{k}_1,z_1) \frac{\underline{\epsilon}_2 \cdot \underline{k}_2}{\underline{k}_2^2+\overline{Q}^2}  = \, \\
= \, 2 g t^a  \int \frac{d^2 \underline{k}_1}{(2\pi)^2}  e^{i \underline{k}_1 \cdot \underline{x}_{01} }  \Psi^{(0)}(\underline{k}_1,z_1) \frac{i}{2\pi} \overline{Q} K_1(\overline{Q}x_{02}) \frac{\underline{\epsilon}_2 \cdot \underline{x}_{02}}{x_{02}} \\
= \, 2 g t^a  \int d^2 \underline{r} \, {\Phi}^{(0)}(\underline{r},z) \, \int \frac{d^2 \underline{k}_1}{(2\pi)^2}  e^{i \underline{k}_1 \cdot (\underline{x}_{01} -\underline{r})}   \frac{i}{2\pi} \overline{Q} K_1(\overline{Q}x_{02}) \frac{\underline{\epsilon}_2 \cdot \underline{x}_{02}}{x_{02}} \; .
\label{eq:fouriertransformk1}
 \end{multline}
 To get the last line, we have used the inverse Fourier transform to represent the wave function in the coordinate space ${\Phi}^{(0)}(z,\underline{r})$. 
 Since the expression with the modified Bessel function $\overline{Q} K_1(\overline{Q}x_{02})$  depends on $\underline{k}_1$ we  do not get a delta function $\delta^{(2)}(\underline{r}-\underline{x}_{01})$ after the integration over $\underline{k}_1$.
  This  means that the  parent dipole
 recoils or changes its size because of the emission of the daughter dipole. 
 If we insist that the recoil is small we can still make  an improvement over the original LL dipole formula by making the following approximations
 \be
 \overline{Q}_{01} \simeq \frac{1}{x_{01}}  \sqrt{\frac{k_2^+}{k_1^+}} = \frac{1}{x_{01}}   \sqrt{z} \; , 
 \label{eq:qapprox}
 \ee
 and after performing the $k_1$ integration which in this approximation gives  delta function  we finally get
 \be
 2 g t^a  \, {\Phi}^{(0)}(z,\underline{x}_{01}) \,    \frac{i}{2\pi} \overline{Q}_{01} K_1(\overline{Q}_{01}\, x_{02}) \frac{\underline{\epsilon}_2 \cdot \underline{x}_{02}}{x_{02}} \; .
 \label{eq:transversemod}
 \ee
 This is an  amplitude for the one gluon emission in the coordinate space improved by taking into account the next, subleading term in the energy denominator.
 Note, that it is a simplified and special  version of (\ref{eq:multibessel2}). The difference is that
 we do not keep all the terms in the energy denominators but only the ones which are related to the `parent' emitter and the last gluon in the emission. This allows to factorize the splitting from the rest of the cascade.
 Therefore in this case each emission will be governed by the same kernel (\ref{eq:transversemod})
 whereas in the exact case the splitting is governed by more complicated expression with the 
 Bessel function whose order changes with the number of the gluons in the wave function.
 
The last expression (\ref{eq:transversemod}) obviously reduces to the original LL dipole formula, compare (\ref{eq:soft1}),  by expanding the Bessel function $K_1$ for the small values of the argument
 \be
 \overline{Q}_{01} K_1\left(\overline{Q}_{01}\, x_{02}\right) = \frac{\sqrt{z}}{x_{01}} K_1\left(\frac{x_{02}}{x_{01}}\sqrt{z}\right) \; \simeq \; \frac{1}{x_{02}}\; , \; \; \;  {\text{for} } \; \; \frac{x_{02}}{x_{01}}\sqrt{z}\rightarrow 0 \; .
  \ee
 The expression (\ref{eq:transversemod}) becomes in this limit
 \be
 2 g t^a  \, {\Phi}^{(0)}(\underline{x}_{01},z) \,    \frac{i}{2\pi}  \frac{\underline{\epsilon}_2 \cdot \underline{x}_{02}}{x_{02}^2} \; , 
 \label{eq:transverse}
 \ee
 which is the original LL formula \cite{AlMueller} as expected.
 Therefore, (\ref{eq:transversemod}) is an improvement over the original formula (\ref{eq:transverse}).
 Note that, there is a   similarity of  (\ref{eq:transversemod}) with the form of the  dipole formula for the $F_2$ structure function
 \be    
\label{eq:f2dipole}     
F_{T,L}(x,Q^2)\, =\,      \frac{Q^2}{4\pi^2\alpha_{\rm em}}\,  \,
\int d^2{\bf r} \int_0^1 dz\; |\Psi_{T,L}^{\gamma^*, q\bar{q}}(r,z,Q^2)|^2\; \hat{\sigma}(r,x)\,,     
\ee   
where $\Psi_{T,L}^{\gamma^*, q\bar{q}}$ is the  wave function for the splitting of      
the virtual photon into a $q\bar{q}$ pair (dipole),      
and $\hat{\sigma}$    
is the imaginary part of the forward scattering amplitude   
of the $q\bar{q}$ dipole on the proton, called   
the dipole cross section, which    
describes  the interaction of the dipole with the proton.     
In addition, ${\bf{r}}$ is the transverse separation of the quarks in the   
$q\bar{q}$ pair,   and $z$ is the light-cone momentum fraction of the photon   
carried by the quark (or antiquark).   As usual, $-Q^2$ is the photon   
virtuality and $x$ is the Bjorken variable defined as $x=Q^2/s$ with $s$ the total  energy of the $\gamma^* p$ system.
The wave function of the virtual photon is given by the following equations: 
\beeq  
\label{eq:wavet}     
|\Psi_{T}^{\gamma^*, q\bar{q}}|^2 &=&     
\frac{3\,\alpha_{em}}{2\pi^2}\sum_f  e_f^2     
\left\{ [z^2+(1-z)^2] \Qdash_{f}^{2} K_{1}^{2}(\Qdash_f r)  +\, m_f^2\ K_{0}^{2}(\Qdash_f r)     
\right\}\,,     
\\ \nonumber     
\label{eq:wavel}     
|\Psi_{L}^{\gamma^*, q\bar{q}}|^2 &=&     
\frac{3\,\alpha_{em}}{2\pi^2} \sum_f e_f^2     
\left\{     
4 Q^2 z^2(1-z)^2 K_0^2(\Qdash_f r)      
\right\}\,,     
\eeeq     
where the sum is performed over quarks with flavor $f$,      
charge $e_f$, and     
\be     
\label{eq:qdash}     
\Qdash_{f}^{2}\, =\, z(1-z) Q^2+m_f^2 \; \; .    
\ee   
We see that whenever $z$ or $1-z$ are small the wave-function can be approximated by 
$$
 \bar{Q}_f^2 K_1^2(\bar{Q}_f r) \simeq \frac{1}{r^2} \; , 
$$
in a close analogy with  the dipole kernel in the LL approximation, compare the square of (\ref{eq:transverse}). 
We see that (\ref{eq:transversemod}) contains the kinematical constraint
in the transverse space which  is realized by the exponential tail of the Bessel function for large values of its argument
  \be
 \overline{Q}_{01} K_1(\overline{Q}_{01} x_{02}) = \frac{\sqrt{z}}{x_{01}} K_1\left(\frac{x_{02}}{x_{01}}\sqrt{z}\right) \; \simeq \;  \sqrt{\frac{\pi\sqrt{z}}{2 x_{02} x_{01}}} \exp\left(-\frac{x_{02}}{x_{01}}\sqrt{z}\right) \; , \;\;\;  \frac{x_{02}}{x_{01}}\sqrt{z}\rightarrow \infty \; ,
  \ee
 
We have  deliberately  chosen the notation of $\overline{Q}$ variable in (\ref{eq:momqbar}) to make contact with the analogous variable in the wave function of the photon $\overline{Q}^2\equiv Q^2z(1-z)$ (\ref{eq:wavet},\ref{eq:wavel}). 
 Of course we still have here small $z$, so that $(1-z)\simeq 1$.

The improved dipole kernel coming from squaring (\ref{eq:transversemod}) plus the term from the second graph, reads 
\begin{multline}
  d^2 \underline{x}_2 \bigg(\overline{Q}_{01} K_1(\overline{Q}_{01} x_{02}) \frac{\underline{\epsilon}_2 \cdot \underline{x}_{02}}{x_{02}} - \overline{Q}_{01} K_1(\overline{Q}_{01} x_{12}) \frac{\underline{\epsilon}_2 \cdot \underline{x}_{12}}{x_{12}} \bigg)^2= \\
  =\,d^2 \underline{x}_2 \, \overline{Q}^2_{01}\, \left[K_1^2(\overline{Q}_{01}\,x_{02})+K_1^2(\overline{Q}_{01}\,x_{12})-2K_1(\overline{Q}_{01}\,x_{02})K_1(\overline{Q}_{01}\,x_{12})\frac{\underline{x}_{02}\cdot\underline{x}_{12}}{x_{02}x_{12}}\right] \; ,
\label{eq:modkernel}
\end{multline}
with  \be\overline{Q}_{01}=\frac{1}{x_{01}}\sqrt{z} \; , \label{eq:q01sec}\ee compare Eq. (\ref{eq:qapprox}).
We will refer to it as the {\it quasilocal} case because in this approximation we are keeping only
terms in the energy denominators which refer to the daughter and the parent dipole, without
any other dipoles in the cascade. It is straightforward to verify that the (\ref{eq:modkernel}) simplifies
to  (\ref{eq:lldipolekernel}) when $z\rightarrow 0$. The kernel (\ref{eq:modkernel}) is very similar in form
to the one with the massive gluon, which also can be expressed in terms of the Bessel functions.
 Here, however the argument of the Bessel functions depends
on the $\overline{Q}_{01}$, and consequently on the longitudinal momentum $z$, see (\ref{eq:q01sec}). The transverse and longitudinal momenta
are not  separated any more, even though we can still use a single, closed integral equation for the evolution of the dipole amplitude
in the rapidity.  In the LL approximation the evolution depended only on the previous step in rapidity, with the branching
that was independent of the rapidity or $x$. The modified kernel (\ref{eq:modkernel})  contains the branchings which depend explicitly
on the longitudinal variable, and therefore on  all the steps in the evolution in rapidity. This is a qualitative difference as this means that there
is now a `memory' in the evolution of the system of dipoles. The probability of the emission of next dipoles depends on the evolution
variable (`time') $z$.
\subsection{Soft gluon emission with the hard gluon displacement}
 
The approximation that we made in deriving (\ref{eq:modkernel}) preserved an important feature of the soft gluon limit: the variation of the emitter gluon position in the process of soft gluon emission is neglected. Thus, transverse position of the harder of the two daughter gluons coincides with the transverse position of the parent gluon. A  more accurate treatment of the emission process shows, however, that it is not the case. In order to demonstrate is explicitly, we go back to expression (\ref{eq:fouriertransformk1}) with $\overline{Q}=\underline{k}_1\, \sqrt{z}$, 
 \be
 2 g t^a  \int d^2 \underline{r} \, {\Phi}^{(0)}(\underline{r},z) \, \int \frac{d^2 \underline{k}_1}{(2\pi)^2}  e^{i \underline{k}_1 \cdot (\underline{x}_{01} -\underline{r})}   \frac{i}{2\pi} \overline{Q} K_1(\overline{Q}x_{02}) \frac{\underline{\epsilon}_2 \cdot \underline{x}_{02}}{x_{02}} \; \; ,
 \ee
 and perform the integral over $d^2\uvec{k}_1$ to get, 
 \begin{multline}
 2 g t^a  \int d^2 \underline{r} \, {\Phi}^{(0)}(\underline{r},z) \, \int \frac{k_1 dk_1}{2\pi}J_0(k_1 |\underline{x}_{01}-\underline{r} |) \frac{i}{2\pi}k_1 \sqrt{z} K_1(k_1 \sqrt{z} x_{02}) \frac{\underline{\epsilon}_2 \cdot \underline{x}_{02}}{x_{02}} =\\
= 2 g t^a  \int d^2 \underline{r} \, {\Phi}^{(0)}(\underline{r},z) \,\frac{\underline{\epsilon}_2 \cdot \underline{x}_{02}}{x_{02}}  \frac{i}{2\pi} \sqrt{z} \int \frac{k_1^2 dk_1}{2\pi}J_0(k_1 |\underline{x}_{01}-\underline{r} |)  K_1(k_1 \sqrt{z} x_{02}) = \\
=2 g t^a  \int d^2 \underline{r} \, {\Phi}^{(0)}(\underline{r},z) \,\frac{\underline{\epsilon}_2 \cdot \underline{x}_{02}}{x_{02}^2}  \frac{i}{(2\pi)^2}  \frac{2 (\sqrt{z} x_{02})^2}{[(\underline{x}_{01}-\underline{r})^2+(\sqrt{z}x_{02})^2]^2} \; .
\label{eq:transverseshift}
 \end{multline}
The last term on the r.h.s.\ is a Cauchy type distribution in two dimensions which gives delta function $\delta^{(2)}(\underline{x}_{01}-\underline{r})$ when the scale parameter $\sqrt{z}\,x_{02}\rightarrow 0$.
In this limit one  reproduces expression (\ref{eq:transverse}). Last line of (\ref{eq:transverseshift}) gives more a precise expression in the situation when the change of the emitter gluon position is taken into account. 
Note, that the integration in (\ref{eq:transverseshift}) over the transverse vector $\uvec{r}$ occurs already at the amplitude level, so the evolution equation for the dipole scattering amplitude with this kernel would involve a cumbersome triple integration over transverse coordinates. Finally, let us conclude that in a general case of the gluon splitting, the transverse position of neither of the daughter gluons coincides with the position of the parent gluon.

\subsection{Modified dipole evolution kernel and the NLL BFKL kernel}

In the previous subsection we proposed an approximate scheme to go beyond the soft gluon limit in the description of the onium wave function. In this scheme one better accounts for the kinematical effects in the evolution of the gluon wave function. 
A significant part of the phase space for the gluon emissions is cut out. The effect is of course subleading  with respect to the leading logarithmic approximation or the  original dipole cascade. Nevertheless, the constraint introduces important corrections and cannot be neglected in the analysis of the color dipole / BFKL evolution beyond the LL approximation. In particular, we shall demonstrate explicitly that in the NLL approximation the modified dipole kernel (\ref{eq:modkernel}) saturates the double collinear logarithmic contributions that are found in the NLL BFKL kernel. As discussed in \cite{Salam:1998tj} these contributions violate the renormalization group constraints and should be resummed. The resummation of these spurious terms within the conventional $t$-channel formulation of the BFKL evolution in the momentum space is conveniently realized by imposing the so called {\em kinematical constraint} (or {\em the consistency constraint})  \cite{Andersson:1995jt,Andersson:1995ju,Kwiecinski:1996td}. The modified dipole kernel (\ref{eq:modkernel}) and the BFKL kernel with the kinematic constraint have similar origin and describe the same effect, a modification of the evolution kernel due to a more accurate treatment of the kinematical effects. In what follows, we shall discuss in detail the problem of the double collinear logarithms in the dipole representation.
  
In the exact next-to-leading calculation \cite{Balitsky:2008zz}, \cite{Fadin:2007xy,Fadin:2007de,Fadin:2007ee,Fadin:2006ha} the terms with the largest power of the collinear logarithms come from the following part of the NLL~BFKL kernel in the coordinate representation,

\be
{\cal K^{\rm NLO}_{\rm DL}} 
\otimes N_Y \; = \; -2\bar{\alpha}_s^2 \int  \frac{d^2 \underline{x}_{2}}{2\pi} \, \frac{x_{01}^2}{x_{02}^2 x_{12}^2} \log\left(\frac{x_{02}}{x_{01}}\right) \log\left(\frac{x_{12}}{x_{01}}\right)\,[N_Y(\uvec{x}_{0},\uvec{x}_2)+N_Y(\uvec{x}_{1},\uvec{x}_2)-N_Y(\uvec{x}_{0},\uvec{x}_1)] \; ,
\label{eq:doublelogexact}
\ee
where $N_Y(\uvec{x}_i,\uvec{x}_j)$ in the above equation is the scattering amplitude of the color dipole $(\uvec{x}_i,\uvec{x}_j)$ with the target and $Y=\ln 1/x$ is the rapidity between the incoming dipole and the target particle. 
It is important to note that these double logarithmic terms are the same in the QCD as in the conformally invariant ${\cal N}=4$ SYM theory~\cite{Fadin:2007xy}. Thus, we conjecture that this universality originates from universal kinematical effects in gluon emission.

\begin{figure}[ht]
\centerline{\epsfig{file=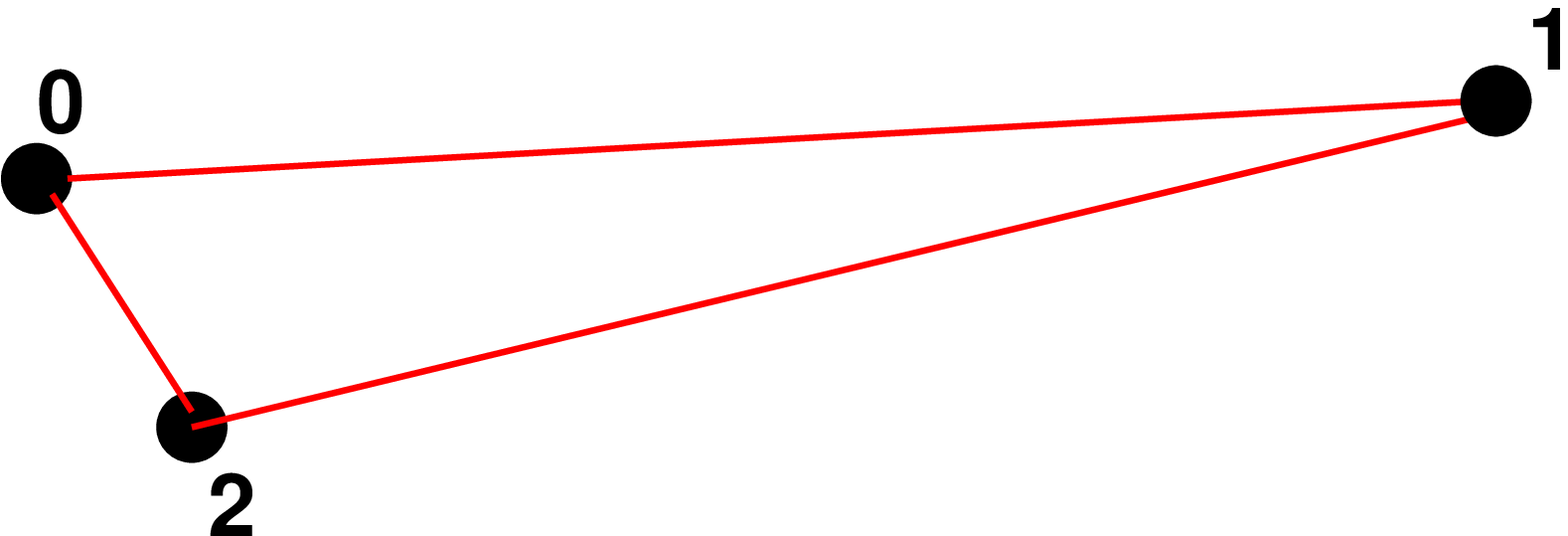,width=5cm}\hspace*{4cm}\epsfig{file=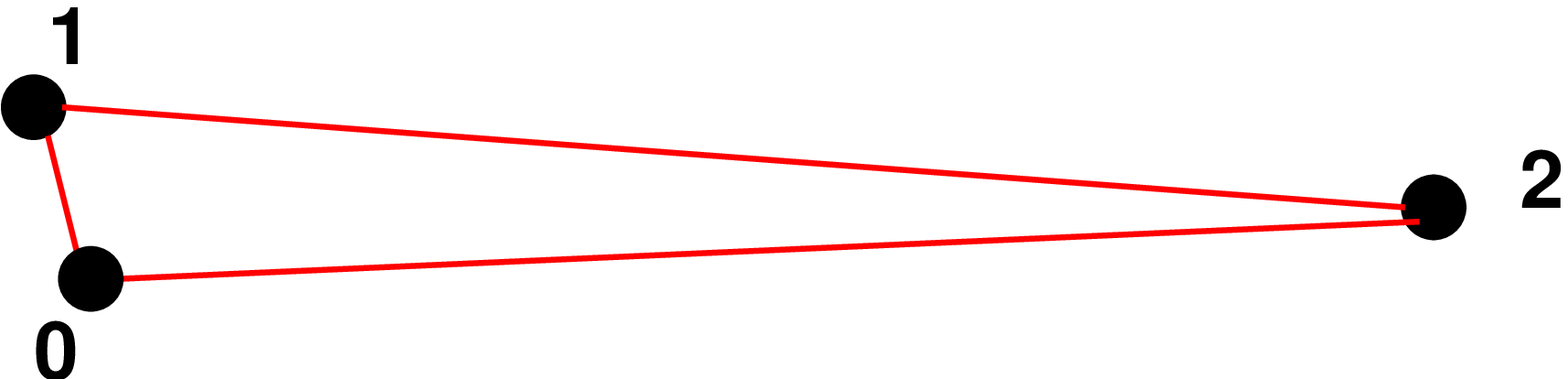,width=5cm}}
\caption{Dipole configurations in two collinear limits. Left plot: $x_{02} \ll x_{01}\;\ \mbox{and}\;\; x_{12} \sim x_{01}   $; right plot:  $x_{02} \gg x_{01} \;\;\mbox{and}\;\;x_{12} \sim x_{02} \; .$}
\label{fig:dipolecollinear}
\end{figure}

Let us compare the behavior of (\ref{eq:modkernel}) and  (\ref{eq:doublelogexact}) in the collinear and the anti-collinear limits. These are limits when the parent and daughter dipole sizes are strongly ordered. In the anti-collinear limit one of the daughter dipoles is much smaller than the parent dipole and we have, 
$$(\,x_{02} \ll x_{01}\;\; \mbox{and}\;\; x_{12} \sim x_{01}\,)  \;\;\; \mbox{or}\;\;\; (\, x_{12} \ll x_{01}\;\; \mbox{and}\;\; x_{02} \sim x_{01} \,) \; .$$
This situation is depicted on the left plot in Fig.~\ref{fig:dipolecollinear}.
In these configurations no super-leading logarithmic enhancement of NLL BFKL kernel occurs, in particular the part of the NLL kernel (\ref{eq:doublelogexact}) is suppressed as $$\log \frac{x_{12}}{x_{01}}\,\simeq\, \log \frac{x_{01}}{x_{01}}\, \to \,0  \; \; \; \;  {\rm  or} \; \; \; \;  \log \frac{x_{02}}{x_{01}}\,\simeq\, \log \frac{x_{01}}{x_{01}}\, \to \,0 \; .
$$
Thus, no  anti-collinear double logarithms are generated.
In the collinear limit both daughter dipoles are larger than the parent dipole. This configuration is shown in the right plot in Fig.~\ref{fig:dipolecollinear}. $$x_{02} \gg x_{01} \;\; \mbox{and}\;\; x_{12} \sim x_{02} \; .$$
In this case the most singular part (\ref{eq:doublelogexact}) of the next-to-leading kernel tends to 
$$ {\cal K^{\rm NLO}_{\rm DL}} \otimes N_Y \;  \; \rightarrow  \;-\frac{\bar{\alpha}_s^2}{\pi} \int  \frac{d^2 \underline{x}_{2} \, x_{01}^2}{x_{02}^4} \log^2\left(\frac{x_{02}}{x_{01}}\right) \,[\dots] \;,$$
where $[\dots]$ stands for the expression in the square brackets as in (\ref{eq:doublelogexact}). 

Note that kernel (\ref{eq:doublelogexact}) is large and negative in the regime where the produced dipoles $\underline{x}_{02}$  and $\underline{x}_{12}$ are very large compared to the dipole $x_{01}$. This is exactly the same regime where the modified kernel (\ref{eq:modkernel}) is exponentially cutting off the contributions of the large $(02),(12)$ dipoles. Clearly, in such configuration the phenomenologically improved kernel (\ref{eq:modkernel}) is better behaved than the NLL kernel. 

In the case of the kinematic constraint in the BFKL equation in momentum space,the double collinear logarithmic term is robust. These terms are entirely determined by the dependence of the effective cut-off scale for the transverse momentum on the longitudinal momentum fraction and does not depend on the details of the cut-off procedure. The same property should hold true in the position space. Thus, for the purpose of extraction of the contribution enhanced by the double collinear logarithms, it is sufficient to use a simplified dipole kernel
\be
{x_{01}^2 \over x^2_{02} x^2 _{12}} \, 
\theta(x^2_{01} - z x_{02}^2 ) \, \theta(x^2 _{01} - z x^2_{12})\,.
\ee  
This kernel introduces the cut-off on large dipole sizes consistent
with the effective cut-offs given by the more accurate expression  
(\ref{eq:modkernel}), but it is more convenient to analyze. 
The physical meaning of this kernel is simple: if the daughter dipole size is small enough, the deviations from the LL kernel of the kinematic origin are small, and if the dipole sizes are too large, the kinematical effects introduce a distortion, which leads to a suppression of the emission. This suppression, in particular, is visible in the Bessel improved kernel. 

The integral evolution equation for the dipole scattering amplitude with this kernel reads,
\be
N_{01}(x) = N^{(0)} _{01} + \overline{\alpha}_s
\int_x ^1 {dz \over z} \, \frac{ d^2 x_2}{2\pi} \; 
{x_{01}^2 \over x^2_{02} x^2 _{12}} \, 
\theta(z x^2_{01} -  x x_{02}^2 ) \, \theta(z x^2 _{01} - x x^2_{12})
\left[ N_{02}(z) + N_{21}(z) - N_{01}(z) \right].
\ee
We shall go to the differential form w.r.t.\ the longitudinal variable $x$.
One gets,
\[
-x {\partial \over \partial x} N_{01}(x) = \overline{\alpha}_s
\int \frac{ d^2 x_2}{2\pi} \;  
{x_{01}^2 \over x^2_{02} x^2 _{12}} \; \left\{ \,\rule{0ex}{3ex} 
\theta(x^2_{01} - x_{02}^2 ) \, \theta(x^2 _{01} - x^2_{12})
\left[ N_{02}(x) + N_{21}(x) - N_{01}(x) \right]
\right.
\]
\[
+ \frac{\overline{\alpha}_s}{2\pi} \int_x ^1 \, {dz \over z} \, 
\left[
\delta\left( x - z{x_{01}^2 \over x_{02}^2} \right) \, 
\theta (z x^2 _{01} - x x^2_{12}) + 
\delta\left( x - z{x_{01}^2 \over x_{12}^2} \right) \, 
\theta (z x^2 _{01} - x x^2_{02})\right] 
\]
\be
\times \; 
\left.
\left[ N_{02}(z) + N_{21}(z) - N_{01}(z) \right]
\rule{0ex}{3ex}\right\}.
\label{eq:sim_full}
\ee
Clearly, due to limitations in the $x_2$ integration imposed by the
$\theta$~functions, the first line does not introduce the double logarithms.
Let us focus on the part of the equation containing the integration over $z$.
This gives, after the $z$ integration
\[
\int d^2 x_2 \; 
{x_{01}^2 \over x^2_{02} x^2 _{12}} \, 
\left\{\,
\theta(x_{02}^2 - x_{01}^2) \theta(x_{02}^2 - x_{12}^2) \, \left[ 
   N_{02}\left( x \, {x^2_{02} \over x^2_{01} }\right) \,+\,
   N_{21}\left( x \, {x^2_{02} \over x^2_{01} }\right) \,-\,
   N_{01}\left( x \, {x^2_{02} \over x^2_{01} }\right)
\right] 
+
\right.
\]
\be
\left.
\theta(x_{12}^2 - x_{01}^2) \theta(x_{12}^2 - x_{02}^2) \left[ 
   N_{02}\left( x \, {x^2_{12} \over x^2_{01} }\right) \,+\,
   N_{21}\left( x \, {x^2_{12} \over x^2_{01} }\right) \,-\,
   N_{01}\left( x \, {x^2_{12} \over x^2_{01} }\right)
\right]
\right\}. 
\label{eq:sim_zint}
\ee
Being interested in the NLL accuracy we may expand the
dipole scattering amplitudes with the shifted $x$~argument, e.g.\ 
\be
N_{ij}(x \, x^2_{02} / x^2_{01} )
\ee
using a logarithmic variable $\log(x)$, 
\be
N_{ij}(x \, x^2_{02} / x^2_{01} )
\, \simeq \, 
N_{ij}(x) \,+\,x {\partial \over \partial x}\,N_{ij}(x) \,
\log \left( {x^2_{02} \over x^2_{01}} \right).
\ee
The terms $N_{ij}(x)$ (without the differentiation w.r.t.\ $x$) 
may be combined with the first line of (\ref{eq:sim_full}) to recover
the LL~dipole emission kernel. 
The emerging correction to the evolution equation takes the form,
\[
[\delta K \otimes N]_{01}(x) \; = \; 
{\bar \alpha_s \over 2\pi} \; \int d^2 x_2 \,
{x_{01}^2 \over x^2_{02} x^2 _{12}} \, 
\times
\]
\[
\left\{\, \theta(x_{02}^2 - x_{01}^2) \theta(x_{02}^2 - x_{12}^2)\, 
\log \left( {x^2_{02} \over x^2_{01}} \right)\,
\,x {\partial \over \partial x}\,
[\, N_{02}(x) \, + \, N_{21}(x) \, - N_{01}(x) \,] \;
+
\right.
\]
\be
\left.
\theta(x_{12}^2 - x_{01}^2) \theta(x_{12}^2 - x_{02}^2)\, 
\log \left( {x^2_{12} \over x^2_{01}} \right)\,
\,x {\partial \over \partial x}\,
[\, N_{02}(x) \, + \, N_{21}(x) \, - N_{01}(x) \,] \;
\right\}. 
\label{eq:sim_nll}
\ee
Next, we shall use the BFKL equation to eliminate the $x$-differential, 
$\;x {\partial/\partial x}\,N_{ij}(x)\;$,   
\be
x {\partial \over \partial x}\,N_{ij}(x) \, = \,
-{\bar\alpha_s \over 2\pi}\; 
\int d^2 x_k \, { x^2_{ij} \over x^2_{ik} x^2_{kj}} \;
\left[ N_{ik}(x) + N_{kj}(x) - N_{ij}(x) \right].
\ee

Let us evaluate the action of the complete non-leading part of the
expanded kernel on a probe function 
$N^{(\gamma)}_{ij}(x) = n(x) (x^2 _{ij})^{\gamma}$,
for $\gamma \to 1$. In this regime, the action of the LO order 
BFKL (dipole) kernel gives obviously, 
\be
K\otimes N^{(\gamma)}_{ij}(x) \simeq 
{\bar\alpha_s \over 1-\gamma} N^{(\gamma)}_{ij}(x).
\ee   
This expression is inserted in (\ref{eq:sim_nll}). 
First, we evaluate 
\be
-{\bar \alpha_s \over 2\pi}\;
\int d^2 x_2 \, {x_{01}^2 \over x_{02}^2 x_{21}^2} \,
\theta(x_{02}^2 - x_{01}^2) \theta(x_{02}^2 - x_{12}^2)\, 
\log\left( {x_{02} ^2 \over x_{01}^2} \right)\,
{\bar\alpha_s \over 1-\gamma}\, 
\left[ 
N^{(\gamma)}_{02}(x) + 
N^{(\gamma)}_{21}(x) -
N^{(\gamma)}_{01}(x) 
\right],
\ee
for $\gamma \to 1$.
The leading singularity at $\gamma=1$, is determined by 
the asymptotically large dipole sizes, $x_{02}^2 = r^2 \gg x_{01} ^2$
and $x_{12}^2 \simeq r^2 \gg x_{01} ^2$. 
Thus, the integral may be approximated by
\be
-{\bar \alpha_s \over 2\pi}\;
\int_{x^2_{01}} ^{\infty} {\pi \over 2}\, dr^2 \; {x^2_{01} \over r^4} \,
\log \left( {r^2 \over x^2 _{01}} \right) \,
{2\bar\alpha_s \over 1-\gamma} \, r^{2\gamma}
\; = \;
-x_{01}^{2\gamma} \, {\bar\alpha_s^2 \over (1-\gamma)^3},
\label{eq:halfdl}
\ee
where we took into account the effect of the constraint imposed
by the function $\theta(x_{02}^2 - x_{12}^2)$.
Since $\uvec{x}_{12} = \uvec{x}_{02} - \uvec{x}_{01}$, and 
$|\uvec{x}_{02}| \gg |\uvec{x}_{01}|$,
this $\theta$ function constrains $\uvec{x}_{12}$ to a half-plane for 
which the angle between $\uvec{x}_{02}$ and $\uvec{x}_{01}$ is smaller 
than $\pi / 2$. Therefore the angular part of the integration 
$d^2 x_2$ gives only $\pi$ and not $2\pi$.
The same contribution as the one computed in 
(\ref{eq:halfdl}) comes from the term proportional to 
$\,\theta(x_{12}^2 - x_{01}^2) \theta(x_{12}^2 - x_{02}^2)\,$ 
in (\ref{eq:sim_nll}), and the full answer reads,
\be
[\delta K\otimes N^{(\gamma)} ]_{01} (x) \; = \;
-{\bar\alpha_s^2 \over (1-\gamma)^3}\, N^{(\gamma)}_{01} (x)
\, + \, {\cal O}\left( {\bar\alpha_s^2 \over (1-\gamma)^2} \right).
\ee
This is precisely the same leading collinear singularity as the one found in 
the exact NLL kernel (\ref{eq:doublelogexact}), \cite{Fadin:2007xy,Fadin:2007de,Fadin:2007ee,Fadin:2006ha}.

Finally, let us comment on the issue of the conformal invariance of the NLL BFKL kernel. The part of NLL kernel given by (\ref{eq:doublelogexact}) is  scale invariant. This  means that the rescaling transformation  $\uvec{x_i} \rightarrow C \uvec{x}_i $ leaves this part unchanged for an arbitrary number $C$. It is, however, straightforward to verify that ${\cal K^{\rm NLO}_{\rm DL}}$ is not conformally invariant in two transverse dimensions, for instance, the action of the kernel changes under the inversion $\uvec{x}_i  \rightarrow \frac{\uvec{x}_i}{x_i^2}$. Interestingly enough, ${\cal K^{\rm NLO}_{\rm DL}}$ is the only part of the NLL BFKL kernel in ${\cal N}=4$ SYM theory that is non-conformal 
(of course in the QCD, already the scale invariance of the NLL BFKL/dipole kernel is broken by the running coupling effects). 
As mentioned before the modified dipole kernel (\ref{eq:modkernel}) is also non-conformal in two dimensions since the arguments of the Bessel-McDonald function  are exactly the same dipole size ratios $\frac{x_{02}}{x_{01}}\;,\,\frac{x_{12}}{x_{01}}$ as in the above next-to-leading contribution (\ref{eq:doublelogexact}). 
 
\subsection{Diffusion in impact parameter space}
 
 As is clear from the form of the modified kernel (\ref{eq:modkernel}), the corrections from the energy denominators imply large modifications of the diffusion properties in the impact parameter space. The modified Bessel functions  exponentially suppress the production of large size dipoles above a $z$-dependent characteristic size
 $$
 \overline{Q}_{01}^2 K_1^2(\overline{Q}_{01}x_{02})\, \simeq \, \frac{\pi}{2}\frac{\overline{Q}_{01}}{x_{02}} e^{-2x_{02}\overline{Q}_{01}}=  \frac{\pi}{2}\frac{\sqrt{z}}{x_{02}\,x_{01}} e^{-2\frac{x_{02}}{x_{01}}\sqrt{z}}\;,  \; \; \; \;  \overline{Q}_{01}x_{02} \rightarrow \infty . 
 $$
The effective cut-off size $\sim x_{01}/\sqrt{z}$ grows with the decreasing $z$.
This cutoff on the dipole size is analogous to the coherence effect in the cascade of the gluon emissions \cite{DM}.  The maximal opening angle prevents the gluons from being emitted into a certain kinematic regime. Here, the effect is to prevent the emission of very large dipoles  and as  a result the  diffusion in the impact parameter space is very much suppressed. 
 
 To illustrate the diffusion better we inspect the Balitsky-Kovchegov equation with the LL and the modified kernel respectively. We would like to stress however that the fact that we are using the modified dipole kernel in the non-linear equation is our assumption only, based on the analogy with the leading and next-to-leading logarithmic 
 calculations. This assumption is also motivated by the the fact that the dipole amplitude will still  have the limiting value equal to unity
 in the saturation regime. In principle, the modification to the dipole kernel implies also the corrections to the triple Pomeron vertex, \cite{Bartels:1992ym,Bartels:1993ih}. This is an interesting problem which deserves further analysis but it is outside the scope of the present paper. 
 
 Recall first  the leading-logarithmic analysis from \cite{KGBAS} where the power tails in impact parameter space were shown to emerge just after the first step in the evolution in rapidity of the BFKL and BK equations. We take  the initial condition for the amplitude to be exponentially decreasing in impact parameter 
\begin{equation}
N(y=\ln 1/x=0,x_{01}) \equiv N^{(0)}_{01}\, \sim \,  \exp(-\mu b_{01}) \;  , 
\nonumber
\end{equation}
 where the impact parameter is defined as $\underline{b}_{01}=\frac{1}{2}(\underline{x}_0+\underline{x}_1)$. Next, we perform one iteration of the dipole-BK equation with the LL kernel
\be
N_{01} (x)= N_{01}^{(0)}+\frac{\alpha_s N_c}{2 \pi^2}\int_x^1 \frac{dz}{z}\int d^2 \underline{x}_2 \frac{x_{01}^2}{x_{02}^2 x_{12}^2} \, [ N_{02}+N_{12}-N_{01}-N_{02}N_{12}] \;.
\label{eq:bk}
\ee
For a small dipole $x_{01}$ and at large value of the impact parameter $b_{01}$  the non-vanishing contribution to the amplitude comes from the configurations with large dipoles, i.e. such that one end is at $x_0$ and the other one is at $x_2$ located close to the center of the target where the field is strong. In that case
$N_{01}\sim 0$ because of small dipole at large impact parameter and $N_{02}\sim N_{12} \sim  1$ and $x_{02}\sim x_{12} \sim b_{01}$. The r.h.s of the LL BK equation (\ref{eq:bk}) is then
\be
\int d^2 \underline{x}_2 \frac{x_{01}^2}{x_{02}^2 x_{12}^2} \, [ N_{02}+N_{12}-N_{01}-N_{02}N_{12}] \simeq \frac{x_{01}^2}{b_{01}^4} \int_{\cal R} d^2 \underline{x}_2 \; ,
\ee
where ${\cal R}$ is the integration region where approximately $x_{02}\sim b_{01}$.
 Therefore in the case of the  LL equation the diffusion in the transverse space leads always to the power like tails in impact parameter even with the exponentially falling initial conditions.
 This leads to the violation of the Froissart bound \cite{Froissart} even in the presence of the saturation corrections, as was first pointed out in \cite{KW,Kovner:2001bh,Kovner:2002yt}. 
 
The situation changes when the modified kernel (\ref{eq:modkernel}) is considered. 
Here a similar analysis leads to 
\begin{multline}
\int_{z_0-\delta z}^{z_0}\frac{dz}{z}\int d^2 \underline{x}_2 \, \bar{Q}^2_{01}\left[K_1^2(\bar{Q}_{01} x_{02})+K_1^2(\bar{Q}_{01}x_{12})-2K_1(\bar{Q}_{01}x_{02})K_1(\bar{Q}_{01}x_{12})\frac{\underline{x}_{02}\cdot\underline{x}_{12}}{x_{02}x_{12}}\right] \times \\
\times\, [ N_{02}+N_{12}-N_{01}-N_{02}N_{12}]  \\
\simeq \int_{z_0-\delta z}^{z_0}\frac{dz}{z}\int d^2 \underline{x}_2 \,\left[\theta(x_{01}/\sqrt{z}C-x_{02})\frac{x_{01}^2}{x_{02}^2 x_{12}^2} \; + \right. \\ \left.
+\,\theta(x_{02}-x_{01}/\sqrt{z} C) \frac{\pi}{2}\frac{\sqrt{z} x_{01}}{x_{02}^3} \exp\left(-2 \frac{x_{02}}{x_{01}}\sqrt{z}\right)  \right] 
\,[N_{02}+N_{12}-N_{01}-N_{02}N_{12}] \, , \end{multline}
where $C\sim 1$, and where we made approximations in the second term that $x_{01} \ll x_{12} \sim x_{02}\sim b_{01}$ which gives the dominant contribution.
 The first term is the short range contribution and the second one is the  long range one. It is evident that the behavior of the scattering amplitude for small dipoles $x_{01}$ and at large values of the impact parameter $b_{01}$ is governed by 
 \be
 {N}(x_{01},b_{01}) \sim \exp\left(-2\frac{b_{01}}{x_{01}}\sqrt{z}\right) \; .
 \label{eq:ndipole}
 \ee
 Therefore there is an exponential but with the effective mass which becomes
 smaller as $\sqrt{z}/x_{01}$ decreases.
 One can actually see two limits in this behavior. We can take $b_{01},x_{01}$ fixed
 and so by decreasing $z$, or increasing the energy, the exponential tails become power like.
 One can also take $b_{01}$ and $z$ fixed, and change the dipole size $x_{01}$.
 For small dipole sizes $x_{01}\ll b_{01}\sqrt{z}$ the tails are exponential, but for larger
 ones they become again power like.
 It is interesting to note that the largest effect of the modification is for the dipoles with small sizes,
 even though the cutoff inside the integral equation is acting on the large dipole sizes. This is 
 result of the fact that the relevant parameter is $x_{02}/x_{01}\sqrt{z}$, i.e. is proportional to the ratio
 of the dipole sizes.
  The only way to eliminate the power-like tails 
 is to put in (essentially by hand) the fixed mass term,  which limits the range of the interactions \cite{Heisenberg}, for example 
 $$
 \overline{Q}^2_{01}\rightarrow \overline{Q}^2_{01}+m^2 \; \; .
 $$
 In this way one will get an amplitude which has exponential tails with fixed radius and this will
 lead to a behavior consistent with the Froissart bound  \cite{Froissart} (modulo normalization).
 
\section{MHV scattering amplitudes from the light-cone wave function}
\label{sec:mhv}
\subsection{Parke-Taylor amplitudes}
In the previous sections we have analyzed the (tree level) wave function with arbitrary number of gluons in the light cone formulation. This corresponds to the initial state evolution, that precedes the scattering.
In this section we take into account also the final state evolution and evaluate the production amplitude of $n$-gluons in 2~gluon scattering.
The exact tree level amplitudes with an arbitrary number of the external on-shell gluons are known. These are Parke-Taylor amplitudes \cite{Parke:1986gb} (see \cite{Mangano:1990by} for a comprehensive review) and can be recast in the following form
\be
{\cal M}_n \;=\; \sum_{\{1,\dots,n\}} {\rm tr}(t^{a_1}t^{a_2}\dots t^{a_n}) \; m(p_1,\epsilon_1;p_2,\epsilon_2;\ldots;p_n,\epsilon_n) \; ,
\label{eq:ptsum}
\ee
where $a_1,a_2,\ldots,a_n$, $p_1,p_2,\ldots,p_n$ and $\epsilon_1,\epsilon_2,\ldots,\epsilon_n$ are the color indices, momenta and the helicities of the external $n$ gluons. Matrices $t^a$ are in the fundamental representation of the color $SU(N_c)$ group. The sum in (\ref{eq:ptsum}) is over the $(n-1)!$ non-cyclic permutations of the set $\{0,1,\dots,n\}$.
\begin{figure}[ht]
\centerline{\epsfig{file=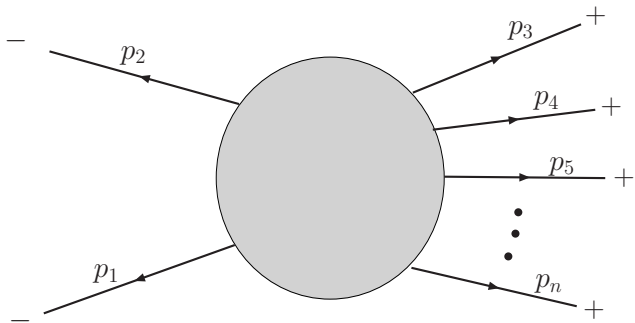,width=0.5\textwidth}}
\caption{The $n$-gluon production amplitude with $(-,-,+,\dots,+)$ helicity configuration. All the gluons are taken to be outgoing.}
\label{fig:mhv}
\end{figure}
 The kinematical parts of the amplitude denoted by $m(1,2,\ldots,n)\equiv m(p_1,\epsilon_1;p_2,\epsilon_2;\ldots;p_n,\epsilon_n)$ are color independent and gauge invariant. These objects  have a number of important properties
\begin{enumerate}
\item $m(1,2,\ldots,n)$ are invariant under cyclic permutations of  the set $\{0,1,\dots,n\}$.
\item $m(n,n-1,\ldots,1) = (-1)^n\, m(1,2,\ldots,n)$ (reversal symmetry).
\item $m(1,2,\ldots,n)+m(2,1,\ldots,n)+m(2,3,1,\ldots,n)+\ldots+m(2,3,\ldots,n,1)=0$ (dual Ward identity).
\item Incoherence to leading number of colors
$$
\sum_{\rm colors} |{\cal M}_n|^2 \, = \, N_c^{n-2}(N_c^2-1) \, \sum_{\{1,\dots,n\}} \left\{ |m(1,2,\ldots,n)|^2 + {\cal O}(N_c^2)   \right\} \; .
$$
\end{enumerate}
It can be shown that the amplitudes where all the gluons have the same helicities, or only one is different from the others are vanishing (we assume that all the gluons are outgoing) 
$$
m(\pm,\pm,\dots,\pm)=m(\mp,\pm,\pm,\dots,\pm)=0\; .
$$
The non-vanishing amplitude for the configuration $(-,-,+,\dots,+)$ depicted in Fig.~\ref{fig:mhv} at the tree level is given by the formula
\be
m(1^-,2^-,3^+,\ldots,n^+) \, = \, i g^{n-2}\, \frac{{\langle 1 2 \rangle}^4}{\langle1 2  \rangle \langle2 3   \rangle\ldots \langle n\!-\!2\; n\!-\!1   \rangle\langle  n\!-\!1\;n  \rangle \langle n 1 \rangle} \; ,
\label{eq:ptmhv}
\ee
where the spinor products are defined by Eqs.~(\ref{eq:ijdef1}),(\ref{eq:ijdef2}).

The comparison of the BFKL amplitudes with the Parke-Taylor expressions was also performed some time ago in \cite{DelDuca:1995zy,DelDuca:1993pp} but only in the limit of the multi-Regge kinematics. Here, we shall re-derive the Parke-Taylor amplitudes within the color dipole picture in a much less restrictive kinematical limit. In order to compare the amplitudes obtained within the light cone perturbation theory formalism with the MHV  amplitudes we need to consider the scattering process of the evolved wave function onto the target. We will simplify the problem by analyzing the case where the evolved projectile gluon scatters on a single target gluon which is separated from the virtual gluons in the projectile by a large rapidity interval. The exchange between the projectile and the target will be treated in the high energy limit.  In this limit the interaction between the projectile and the target is mediated by an instantaneous part of the gluon propagator in the light cone gauge defined as
$$
D_{\rm inst}^{\mu\nu}  = \frac{\eta^{\mu} \eta^{\nu}}{(k^+)^2}\; ,
$$
see for example \cite{Lepage:1980fj}. We restrict the kinematics of the exchange, but still, the internal structure of the projectile gluon field is accurately represented. In principle, the technique applied could be used to evaluate the scattering amplitudes without any kinematical restrictions, but an analysis of the completely general case would be much more complicated. 
 
\begin{figure}[ht]
\centerline{\epsfig{file=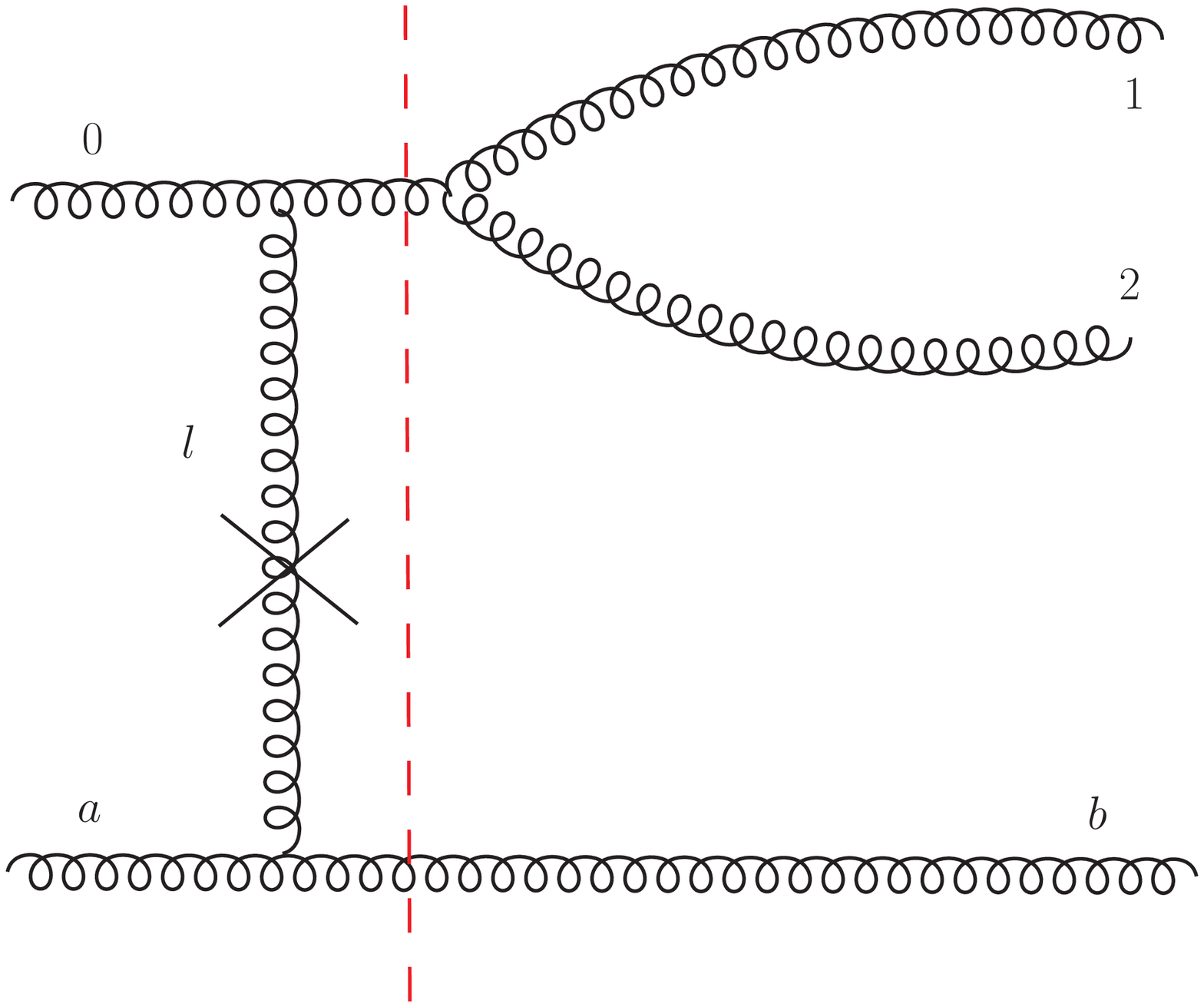,width=0.3\textwidth}\hspace*{1cm}\epsfig{file=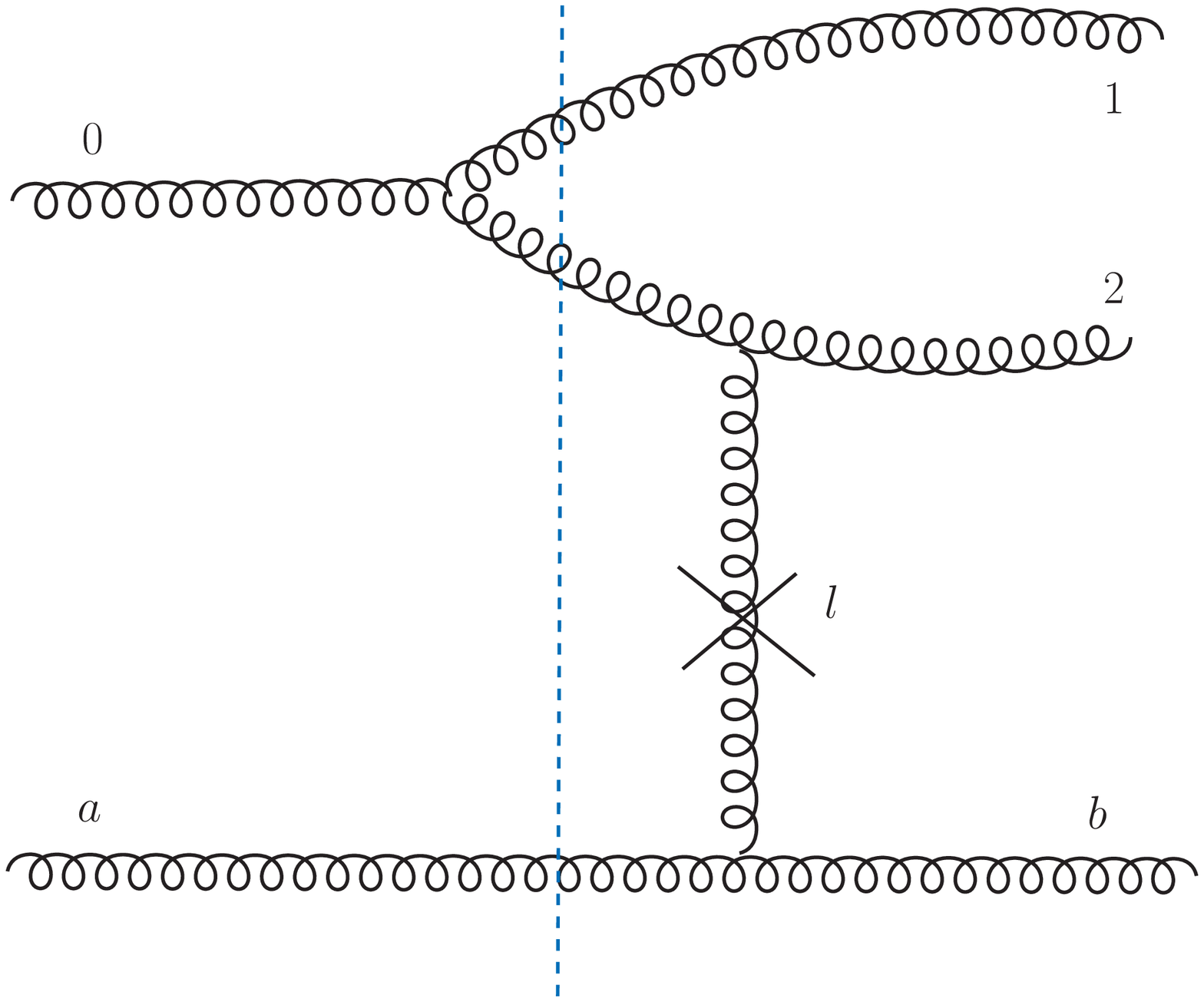,width=0.3\textwidth}}
\caption{Interaction of the wave function with 2 gluons with the target. The cross on the gluon line indicates that it is an instantaneous interaction. Vertical dashed lines denote the energy denominators; short dashes (blue lines), denominators in the initial radiation states; long dashed (red lines), denominators in the final state radiation. }
\label{fig:graph2}
\end{figure}

\begin{figure}
\centerline{\epsfig{file=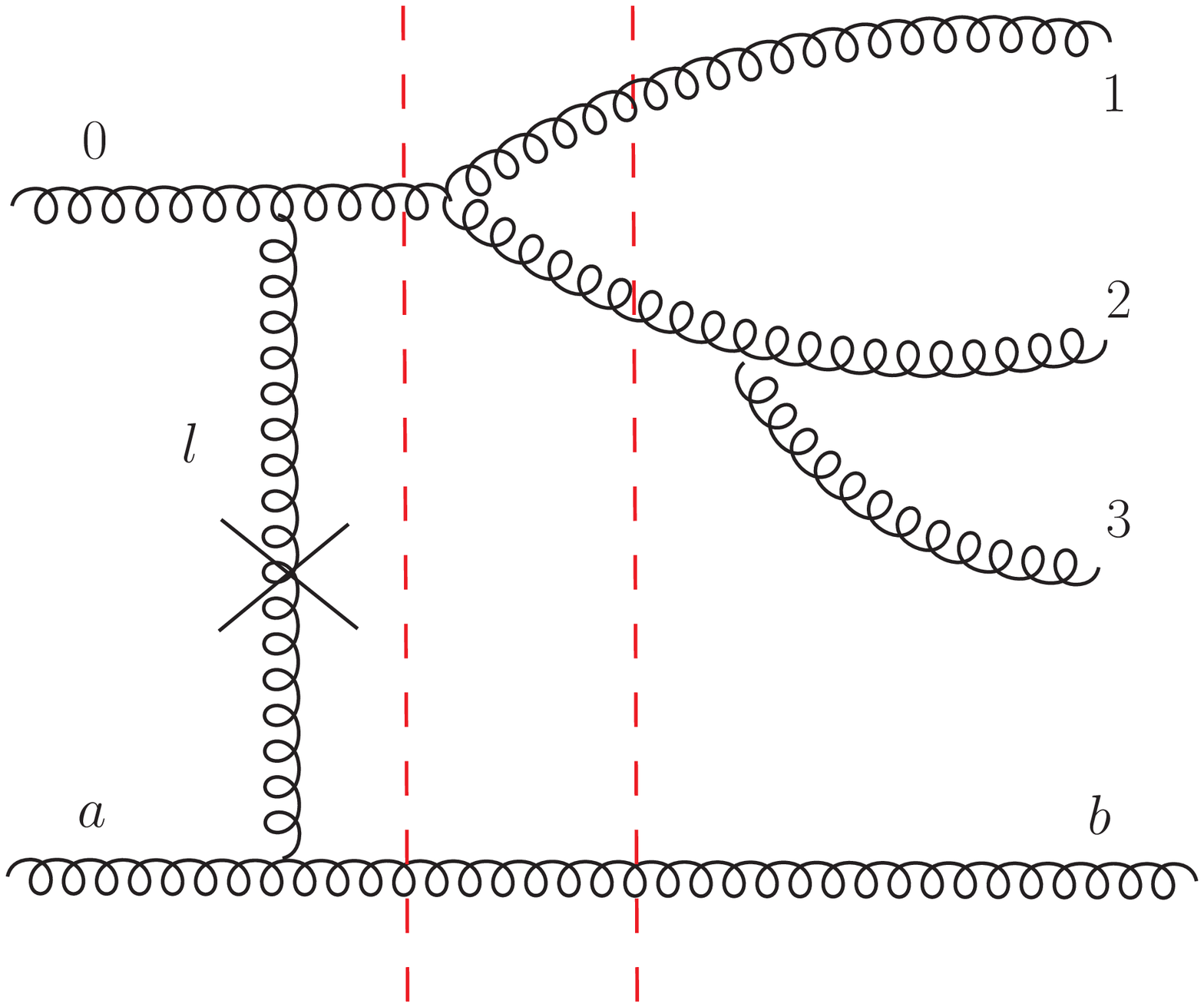,width=0.3\textwidth}\hspace*{0.4cm}\epsfig{file=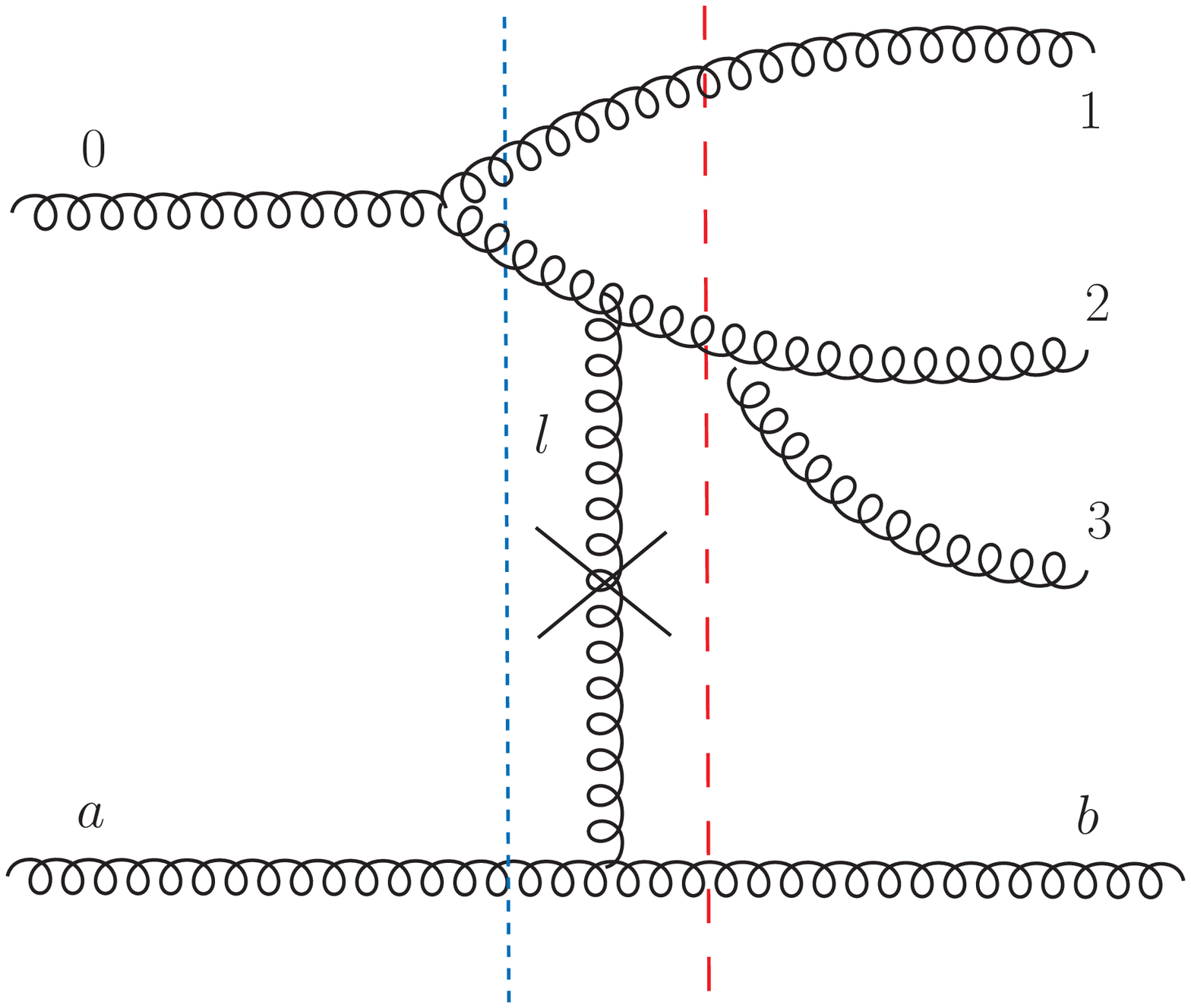,width=0.3\textwidth}\hspace*{0.4cm}\epsfig{file=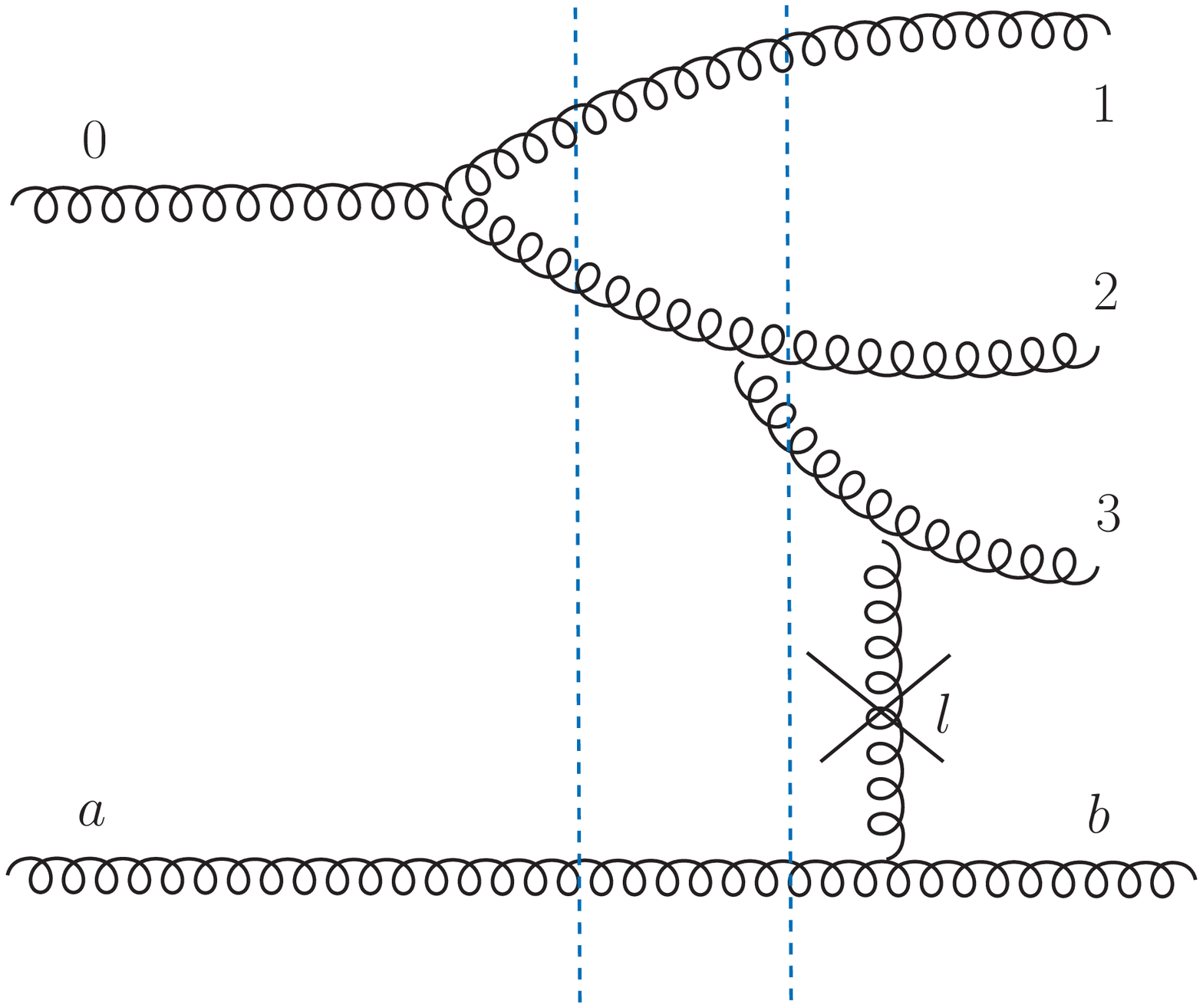,width=0.3\textwidth}}
\caption{Interaction of the wave function with 3 gluons with the target. The cross on the gluon line indicates that it is an instantaneous interaction. Vertical dashed lines denote the energy denominators; short dashes (blue lines), denominators in the initial radiation states; long dashed (red lines), denominators in the final state radiation.}
\label{fig:graph3}
\end{figure}

In order to demonstrate the equivalence with the MHV amplitudes 
it is essential to include the final state radiation in the wave function, which occurs after the interaction with the target has taken place.
 For example, in the case of the $2\rightarrow3$ amplitude,
we have to take into account graphs shown in Fig.~\ref{fig:graph2}. The gluon with the cross and labeled by the  momentum $l$ is the Coulomb gluon, and the gluon with momentum $P$ is the target gluon. The latter one has a large $P^-$ component. The dashed lines denote the energy denominators, which have to be taken before and after the interaction occurred.
Similarly for the $2\rightarrow4$ amplitude we will need to take into account graphs shown in Fig.~\ref{fig:graph3}. In general for this kinematics the number  of possible classes of diagrams
is $n-3$ with $n-4$ energy denominators both in the initial and final state.
Obviously within each of these classes one needs to sum over the  different possibilities of the 
evolution  of the wave function. For example in the case of graphs depicted in Fig.~\ref{fig:graph3}
there is a corresponding set of graphs where the last emitted gluon is coming from the splitting of the most upper gluon in the hadron wave function.

The dominance of the instantaneous (Coulomb) gluon exchange follows from the fact, that the other possible types of exchanges are suppressed w.r.t.\ the dominant exchange by large energy denominators. One can show easily that the relative suppression factor decrease with the total energy like $M^2 / s$, where $M^2$ is the invariant mass of the system produced by the projectile fragmentation. At fixed $M^2$, and for $s\to \infty$, the relative contributions of the non-Coulomb exchanges can be made arbitrarily small. Also, the contribution from scattering mediated by the four-gluon vertex is suppressed.
\subsection{Factorization of fragmentation tree amplitudes}

In constructing the full MHV amplitudes in the high energy limit we shall
use the results for the initial state evolution supplemented by the
analysis of fragmentation of the scattered states into on-shell final
state gluons.

The key feature of the fragmentation amplitudes is the independence
of fragmentation of the gluon trees originating from different parents.
We consider a (typically disconnected) fragmentation tree with a topology 
$\Theta = \Theta_1 \cup \ldots \cup \Theta_m$ where 
$\Theta_1,\ldots \Theta_m$ are topologies of the fragmentation trees 
of $m$~parent virtual gluons into $n$~on-shell gluons,
so that the first parent gluon, denoted by $(12\ldots n_1)$ fragments 
into gluons $(1,2,\ldots, n_1)$, the second parent gluon,  
$(n_1+1, n_1+2,\ldots, n_1)$ fragments into gluons $(n_1+1,n_1+2,\ldots, n_2)$,
and so on.
Then the amplitude of the fragmentation tree $\Theta$, denoted by
$T_{\Theta}$ may be factorized into fragmentation amplitudes of  
the parent gluons, $T_{\Theta_i}$ in the following way,
$T_{\Theta}[(1 \ldots n_1),(n_1+1\ldots n_2),\ldots,
(n_{m-1}\ldots n) \, \to \, 1,2,\ldots n]$, 
\begin{multline}
\label{eq:factm}
T_{\Theta}[(1 \ldots n_1),(n_1+1\ldots n_2),\ldots,
(n_{m-1}\ldots n) \, \to \, 1,2,\ldots n]
\; = \; 
T_{\Theta_1}[(1 \ldots n_1)\, \to \, 1,2,\ldots n_1]\, 
\\
\times \;
T_{\Theta_2}[(n_1+1 \ldots n_2)\, \to \, n_1+1,n_1+2,\ldots n_2]\,\times
\ldots\, \times\,
T_{\Theta_m}[(n_{m-1}+1 \ldots n)\, \to \, n_{m-1},n_{m-1}+1,\ldots n],
\end{multline}
provided that we sum over all possible light-cone time orderings of the
splittings within all connected trees, while preserving the topologies.
This property is not obvious in the light-cone formulation, as for
a given time ordering of the splittings, the variables related to 
different trees are mixed in the energy denominators, and the factorization
property (\ref{eq:factm}) does not hold. It only holds if the 
complete summation over all possible splitting orderings is performed.
In fact, the factorization property of independent tree amplitudes is quite 
intuitive. Nevertheless we shall provide a formal proof of this property 
within the light-cone field theory framework.

In order to prove (\ref{eq:factm}) we shall consider 
two sequences of splittings, $\,(a_i),\; i=0,1,\ldots,m\,$ and  
$\,(b_j),\; j=0,1,\ldots,n\,$, that occur in two topologically 
disconnected trees, $\Theta_a$ and $\Theta_b$, after the scattering. 
The internal topology of each tree is arbitrary, in particular each tree, 
$\Theta_a$ and $\Theta_b$, is not necessarily connected. 
The light-cone times of splittings $a_i$ and $b_j$ will be denoted 
by $\tau(a_i)$ and $\tau(b_j)$ respectively.
In the computation of the fragmentation amplitude of the tree 
$\Theta = \Theta_a \cup \Theta_b$ one sums over all possible 
time-orderings of splittings. We shall denote the splitting ordering
by ${\cal T}(\Theta)$.  This summation may be decomposed into
three nested summations: two independent external summations of the time 
orderings within each tree, $\Theta_a$ and $\Theta_b$, allowed by the 
topology, and the internal summation over all orderings of splittings 
between trees $\Theta_a$ and $\Theta_b$, with a fixed ordering within 
each tree,
\be
\sum_{{\cal T}(\Theta_a \cup \Theta_b)} \; = \; 
\sum_{{\cal T}(\Theta_a)} \; \sum_{{\cal T}(\Theta_b)} \; 
\sum_{{\cal T}(\Theta_a \cup \Theta_b) \;|\; {\cal T}(\Theta_a) \,\wedge\, {\cal T}(\Theta_b)}\,.
\ee
The two external summations are performed for each tree independently, so
they preserve the factorization property (\ref{eq:factm}).
Therefore, it is enough to prove (\ref{eq:factm}) for the arbitrary fixed 
splitting time orderings, ${\cal T}(\Theta_a)$ and  ${\cal T}(\Theta_b)$.
Thus, in what follows we fix the topologies $\Theta_a$ and  $\Theta_b$ and 
the splitting orderings ${\cal T}(\Theta_a)$ and  ${\cal T}(\Theta_b)$.
It is convenient to label the splittings according to the time order, 
$\,\tau(a_1)<\tau(a_{2})<\ldots<\tau(a_{m})\,$ and 
$\,\tau(b_1)<\tau(b_{2})<\ldots<\tau(b_{n})$.

When the topology of the tree $\Theta$ is fixed, then the contributions 
of splitting vertices to the amplitude does not depend on the time ordering
of the splittings. Furthermore, this contributions factorize into 
a product of independent contribution from trees $\Theta_a$ and $\Theta_b$. 
This is not true only for the energy denominators that, 
in general, mix the variables of both trees. So, in order to complete
the proof it is enough to show that the sum over relative splitting 
orderings, 
$\,{\cal T}(\Theta_a \cup \Theta_b) \;|\; 
{\cal T}(\Theta_a) \,\wedge\, {\cal T}(\Theta_b)\,$ 
of the products of the energy denominators of tree $\Theta$
factorizes between the  trees $\Theta_a$ and $\Theta_b$.
The energy denominators for tree $\Theta_a$, in the absence of tree 
$\Theta_b$, will be denoted by $\,A_1,A_2,\ldots,A_m\,$, where, in the
final state we choose to attribute the denominator to the line before
splitting. Analogously, the denominators for tree $\Theta_b$, in the absence 
of tree $\Theta_a$, will be denoted by $\,B_1,B_2,\ldots,B_n\,$. 
We shall prove the following factorization property for energy denominators:
\be
\label{eq:denfact}
\sum_{(i_p,p=1,\ldots,n+m)} \;\; 
\prod_{p=1} ^{n+m} \,{1 \over A_{i_p} + B_{p-i_p}} 
\;\; = \;\;  \prod_{i=1} ^{m} \, {1 \over A_{i}} \;\; 
\prod_{j=1} ^{n} \,{1 \over B_{j}},
\ee
where we used the additivity of the energy denominator of two evolving 
trees and introduced auxiliary symbols $A_0=B_0=0$. The series 
$\,(i_p,p=1,\ldots,n+m)\,$ represents the ordering of splittings
between the chains in the following way: $i_p$ is the number of splittings of 
$\Theta_a$ corresponding to $p$ splittings of $\Theta_a \cup \Theta_b$
assuming the time ordering $\,{\cal T}(\Theta_a \cup \Theta_b) \;|\; {\cal T}(\Theta_a) \,\wedge\, {\cal T}(\Theta_b)\,$, and $p-i_p$ is the corresponding 
number of splittings of $\Theta_b$.  

Equation  (\ref{eq:denfact}) may be proven by double induction, with respect
to~$m$ and~$n$. Let us first take $n=m=1$. Then we have:
\be
{1\over A_1}{1\over A_1+B_1} +  {1\over B_1}{1\over A_1+B_1} \; = \; 
{1 \over A_1 B_1},
\ee
in accord with (\ref{eq:denfact}). Next, for $n=1$ and an arbitrary 
$m \geq 2$ we consider the l.h.s.\ of (\ref{eq:denfact}):
\begin{multline}
\label{eq:mstep}
 \sum_{q=1} ^{m}  {1 \over A_1+B_1}\ldots {1 \over A_q +B_1} {1 \over A_q}\ldots {1\over A_m}
\; + \; {1 \over B_1}{1\over A_1 + B_1} \ldots {1\over A_m+B_1}
\\
= \; {1\over A_1+B_1} \, \left[\, 
\left( 
 \sum_{q=2} ^{m} {1 \over A_q} \ldots  {1 \over A_{m}} 
 {1 \over A_2+B_1}  \ldots {1\over A_{m} + B_1}\, 
\right)
\; + \; {1 \over B_1}{1\over A_2 + B_1} \ldots {1\over A_{m}+B_1}
\right. \\
\; + \; \left. {1 \over A_1}\ldots {1 \over A_m}\,\right]
\; = \;   {1\over A_1+B_1}\left(
{1 \over A_2}\ldots {1 \over A_{m}}{1\over B_1} \; + \;  
{1 \over A_1}\ldots {1 \over A_{m-1}}{1 \over A_{m}}\right)  \; = \;
 {1 \over A_1}\ldots {1 \over A_{m}}{1\over B_1}\,,
\end{multline}
where the first two terms in the square bracket in the middle line of 
(\ref{eq:mstep}) were evaluated using (\ref{eq:denfact}) for $m-1$.
This proves (\ref{eq:denfact}) for $n=1$.

In the second step we use the mathematical induction w.r.t.\ $n$, for 
a fixed $m$. The first step is to split the sum, 
$\,\sum_{(i_p,p=1,\ldots,n+m)}\,$, into contributions 
$\Sigma_1,\Sigma_2,\ldots \Sigma_m, \Sigma_{m+1}$ in  which
the first splitting in the $b$-sequence, $b_1$, occurs before $a_1$, 
between $a_1$ and $a_2$, \ldots, between  $a_{m-1}$ and $a_m$ and 
after $a_m$, respectively. 
In each of these contributions we sum over all the allowed orderings
of the later splittings, $b_2,\ldots,b_{n}$, w.r.t.\ the $a_i$ splittings.
Clearly, when $b_1$ occurs between $a_q$ and $a_{q+1}$ ($1\leq q \leq m-1$),
then all the later splittings,  $b_2,\ldots,b_{n}$, must occur after 
the splitting $a_{q}$. It implies that in $\Sigma_q$ it is necessary to 
perform the complete summation over  $b_2,\ldots,b_{n}$ orderings 
relative to the sequence $a_1,a_2,\ldots,a_q$. The splittings 
$\{a_1,a_2,\ldots,a_q,b_1,b_2,\ldots,b_{n-1}\}$ (with all possible orderings)
are then followed by the (ordered) sequence of splittings: 
$(b_n, a_{q+1}, \ldots, a_{m})$.
The summation of the products of energy denominators over all orderings
of the splittings $\{a_1,a_2,\ldots,a_q,b_1,b_2,\ldots,b_{n-1}\}$, preserving
the ordering of $(a_i)$ and $(b_j)$
in $\Sigma_q$ may be evaluated by applying (\ref{eq:denfact})
for the number of $a$-splittings equal $n-1$ and the number of 
$b$-splittings equal~$q$. The remaining denominators, corresponding
to $( b_n, a_{q+1}, \ldots, a_{m})$ read, $A_q+B_1$,
$A_{q+1}+B_1$, \ldots, $A_{m-1} + B_1$, $A_m+B_1$, respectively.
The final answer for $\Sigma_q$ reads,
\be
\Sigma_q \; = \; \left[\,{1\over B_2}\ldots {1\over B_{n}}\;
{1\over A_q}\ldots {1\over A_m}\,\right]\;  
{1\over A_q+B_1} {1\over A_{q-1} + B_1} \ldots {1\over A_1+ B_1}\,.
\ee
The l.h.s.\ of (\ref{eq:denfact}) is equal to sum over contributions 
$\Sigma_q$:
\begin{multline}
\label{eq:sumsig}
\Sigma_m + \Sigma_{m-1} \ldots  + \Sigma_{1} \; = \;
\left[\, {1\over B_2}{1\over B_2} \ldots {1\over B_{n}}\,\right] \;
{1\over B_1}{1 \over A_1 + B_1} {1\over A_2 + B_1}\ldots  {1\over A_m  + B_1} \\
+ \; \left[\,{1\over B_2}{1\over B_3} \ldots {1\over B_{n}}{1\over A_m}
\,\right]\;
{1 \over A_1 + B_1} {1\over A_2 + B_1}\ldots {1\over A_m+ B_1} 
 \\ + \; \ldots \; 
 + \; \left[{1\over B_2}{1\over B_3} \ldots {1\over B_{n}}
{1\over A_m}{1\over A_{m-1}}\ldots{1\over A_{2}}\,\right]\;
{1\over A_{2} + B_1}{1\over A_1+ B_1} \\
 + \;
\left[\, {1\over B_2}{1\over B_3} \ldots {1\over B_{n}}
{1\over A_1}{1\over A_2}\ldots{1\over A_{m}}\,\right]\;  
{1\over A_1+ B_1} \, ,
\end{multline}
where the contributions resummed using (\ref{eq:denfact}) are 
given in the square brackets. All the terms in the r.h.s.\ of
(\ref{eq:sumsig}) have a common prefactor $1/(B_2\ldots B_{n})$ that
can be factored out. The summation of remaining factors 
in the l.h.s.\ of (\ref{eq:sumsig}) is precisely the one evaluated in
(\ref{eq:mstep}),  and the result of this summation is 
given by $1/(B_1 A_1 \ldots A_m)$. This completes the proof of the induction 
step in $n$, with fixed $m$. Taking into account the obvious symmetry 
between $m$ and $n$, it proves also the induction step in $m$, with fixed $n$.
Therefore (\ref{eq:denfact}) holds for all $m$ and $n$. This, in turn, proves
(\ref{eq:factm}).

\subsection{Fragmentation of a single gluon}
\label{sec:fragm}

For the comparison of the amplitudes obtained in the light cone perturbation theory with the known expressions for the MHV amplitudes it is necessary to 
 compute both the initial and final  state emissions. This is essential,  as we need to take into a 
 account graphs which include both type of processes as depicted in  Figs.~\ref{fig:graph2} and \ref{fig:graph3}.  We consider here the fragmentation of a single, off-shell gluon (labeled by $(12\ldots n)$)  into 
the final state of $n$ on-shell gluons, $1,2,\ldots n$.
The $n$ final state gluons have transverse momenta  $\uvec{k}_1,\ldots,\uvec{k}_n$ and
the longitudinal momentum fractions $z_1,\ldots,z_n$. The initial gluon has transverse momentum $\uvec{k}_{(1\ldots n)}$ and the longitudinal fraction $z_{(1\ldots n)}$ where we again used shortcut notation $\uvec{k}_{(1\ldots n)}=\sum_{j=1}^{n} \uvec{k}_j$ and   $z_{(12\ldots n)} = \sum_{i=1} ^n z_i$ (see Sec.~\ref{sec:exactkinematics}).

\begin{figure}[ht]
\centerline{\epsfig{file=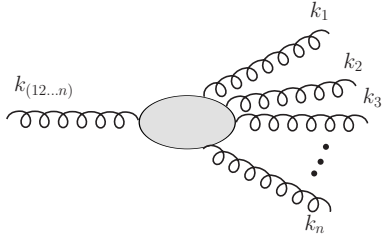,width=0.3\textwidth}}
\caption{Pictorial representation of the fragmentation amplitude $T[(12\ldots n) \to 1,2,\ldots, n]$
for the single off-shell initial gluon.}
\label{fig:fragsinglegluon}
\end{figure}

 We will denote the fragmentation part of the amplitude for 1 to n gluons as $T[(12\ldots n) \to 1,2,\ldots, n]$.
Let us consider the $n=2$ case. Using the computational methods developed in Sec.~\ref{sec:exactkinematics} we can write the expression for the fragmentation in the following way 
\be
T[(12) \to 1,2]  \; = \;  
\frac{g}{\tilde{D}_2}  {v^*_{12} \over \sqrt{\xi_{12}} } =
g \, {v^*_{12} \over \xi^{3/2} _{12} |v_{12}|^2}\,  = g \, \xi_{12}^{-3/2} \, \frac{1}{v_{12}} \; .  
\ee
 To obtain the above expression we have used the formula for the 3-gluon vertex (\ref{eq:vhel}) and the definition of the complex variable $v_{12}$ (\ref{eq:v23}). The energy denominator for the fragmentation $\tilde{D}_2$ was obtained  from expression (\ref{eq:dendifference2}). The calculation  for the  fragmentation of one to three gluons  reads
\begin{multline}
T[(123) \to 1,2,3] \; = \; \frac{1}{\tilde{D}_3} \, g^2 \, \left[\frac{v_{1(23)}^*}{\xi_{1(23)}^{1/2}\xi_{23}^{3/2}v_{23}} \,+\, \frac{v_{(12)3}^*}{\xi_{(12)3}^{1/2}\xi_{12}^{3/2} v_{12}} \, \right] \, = \\ =
  \frac{1}{\tilde{D}_3} \, g^2 \left(\frac{z_{(123)}}{z_1z_2z_3}\right)^{3/2} \,  \frac{v_{1(23)}^* \xi_{1(23)}v_{12}+v_{(12)3}^* \xi_{(12)3}v_{23}}{v_{12}v_{23}}  \; ,
  \label{eq:t123}
  \end{multline}
where the energy denominator for this case
\be
\tilde{D}_3= \sum_{i=1}^3 \frac{k_i^2}{z_i}-\frac{(k_{(123)})^2}{z_{(123)}} \;.
\label{eq:den123gen}
\ee
It is straightforward to verify that the numerator in the last line of (\ref{eq:t123})
is equal to the denominator $\tilde{D}_3$. Note, that this is a case of $n=2$ of the generalization of the formula Eq.~(\ref{eq:ndenominator}) derived in Sec.~\ref{sec:exactkinematics}, in the kinematics when the transverse momentum of the first  particle in the fragmentation chain is not vanishing $k_{(123)}\neq 0$ and when $z_{(12\dots n)}\neq 1$.  We also used the fact that
$$
\xi_{1(23)}\xi_{23} \, = \, \xi_{(12)3}\xi_{12} \, = \, \frac{z_{(123)}}{z_1z_2z_3} \; .
$$
Based on these two examples for $n=2$ and $n=3$ we postulate a general formula, that generalizes computations 
carried out explicitly for $n \leq 4$. The conjectured amplitude of 
fragmentation depicted in Fig.~\ref{fig:fragsinglegluon} reads
\be
\label{eq:frag1}
T[(12\ldots n) \to 1,2,\ldots, n] \; = \; 
g^{n-1} \left( {{z_{(12\ldots n)}} \over {z_1 z_2 \ldots z_n}} \right)^{3/2} \;
{1\over v_{12} v_{23} \ldots v_{n-1\, n}} \; .
\ee
It is interesting to note that the form of the above amplitude for the fragmentation is dual to that of the gluon  wave function with $n$-components.
Namely, upon the replacement of the gluon light-cone velocities $v_{ij}$ in (\ref{eq:frag1}) by the  gluon transverse positions $r_{ij}$,
we obtain the formula (\ref{eq:onshelltrcoowf}) for the on-shell gluon wave function in the coordinate space (modulo the overall factors with $z_i$ fractions).

In order to prove the above formula (\ref{eq:frag1})
 we will need first to show that the relation (\ref{eq:den123gen})
is valid for arbitrary values of $n$. 
The generalization of Eq.~(\ref{eq:ndenominator}) can be easily shown by noting
that in this case we have 
\be
v_{(1\ldots i)(i+1 \ldots n)} \xi _{(1\ldots i)(i+1 \ldots n)}
\; = 
 \sum_{j=1}^{i} k_j \, - \, \frac{z_{(1\dots i)}}{z_{(1\dots n)}}\, \sum_{j=1}^{n} k_j \, = \,  k_{(1\ldots i)} -\frac{z_{(1\dots i)}}{z_{(1\dots n)}}k_{(1\ldots n)}\;,
 \label{eq:vxi_sumk_gen}
\ee
which is also a generalization of relation (\ref{eq:vxi_sumk}) for the case when  $\sum_{j=1}^{n} k_j\neq 0$.
The analogous proof as in (\ref{eq:ndenominator}) can be shown to proceed as follows
\begin{multline}
\sum_{i=1}^{n} v_{i\,i+1}\xi_{(1\dots i)(i+1\dots n+1)} v_{(1\dots i)(i+1\dots n+1)} = \sum_{i=1}^{n} v_{i\,i+1} \left(\sum_{j=1}^{i} k_j \, - \, \frac{z_{(1\dots i)}}{z_{(1\dots n+1)}}\, \sum_{j=1}^{n+1} k_j\right)= \\
=\, \sum_{i=1}^{n} \bigg(\frac{k_i}{z_i}-\frac{k_{i+1}}{z_{i+1}}\bigg) \left(\sum_{j=1}^{i} k_j \, - \, \frac{z_{(1\dots i)}}{z_{(1\dots n+1)}}\, \sum_{j=1}^{n+1} k_j \right) \\
 = \sum_{i=1}^{n} \frac{k_i^2}{z_i}+\frac{k_{n+1}^2}{z_{n+1}} -\frac{k_{n+1}}{z_{n+1}}\sum_{j=1}^{n+1} k_j - \frac{\sum_{j=1}^{n+1} k_j}{\sum_{j=1}^{n+1}z_j}\left[\sum_{i=1}^n \frac{k_i}{z_i} z_i+\sum_{i=1}^n \frac{k_i}{z_i} \sum_{j=1}^{i-1} z_j -\frac{k_{n+1}}{z_{n+1}}\sum_{j=1}^n z_j -\sum_{i=1}^{n-1}\frac{k_{i+1}}{z_{i+1}}\sum_{j=1}^i z_j \right]\; .
\label{eq:proof_denom_gen1}
\end{multline}
Again, the second and fourth terms in the $[\ldots]$ cancel and we are left with
\begin{multline}
\label{eq:proof_denom_gen2}
 \sum_{i=1}^{n+1} \frac{k_i^2}{z_i}-\frac{k_{n+1}}{z_{n+1}}\sum_{j=1}^{n+1} k_j - \frac{\sum_{j=1}^{n+1} k_j}{\sum_{j=1}^{n+1}z_j}\left[ \sum_{j=1}^n k_j-\frac{k_{n+1}}{z_{n+1}} \left(\sum_{j=1}^{n+1}z_j-z_{n+1} \right)\right] = \\
=\,  \sum_{i=1}^{n+1} \frac{k_i^2}{z_i} - \frac{\left(\sum_{j=1}^{n+1} k_j\right)^2}{\sum_{j=1}^{n+1}z_j}=
 \sum_{i=1}^{n+1} \frac{k_i^2}{z_i} -\frac{(k_{(1\ldots n+1)})^2}{z_{(1\ldots n+1)}} = \tilde{D}_{n+1} \; .
\end{multline}

\begin{figure}[ht]
\centerline{\epsfig{file=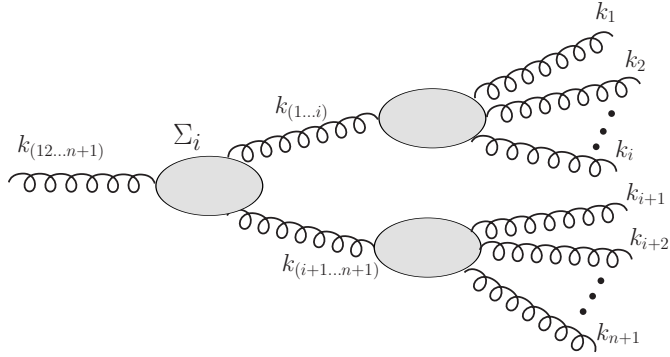,width=0.5\textwidth}}
\caption{Pictorial representation of one step in the fragmentation, Eq.~(\ref{eq:fragonestep}).}
\label{fig:fragmaster}
\end{figure}

The proof of the conjectured amplitude (\ref{eq:frag1}) can now be performed by mathematical induction, using the
factorization property (\ref{eq:factm}) and the relation (\ref{eq:proof_denom_gen2}). 
Let us take the fragmentation into $n+1$ gluons denoted by $T[(1,2,\ldots, n+1) \to 1,2,\ldots, n+1]$ 
and represent it by lower fragmentation factors  
$T[ (1\ldots i)  \to 1,\ldots,i\,]$ and 
$T[ (i+1\ldots n+1)\to i+1,\ldots,n+1]$ 
by separating out the first splitting (in all possible realizations). 
We obtain
\begin{multline}
T[(12 \ldots n+1) \to 1,2,\ldots,n+1] \; = \; {g\over \tilde D_{n+1}} \; 
\sum _{i=1} ^n  \, \left\{\,
{v^*_{(1\ldots i)(i+1\ldots n+1)} 
\over \sqrt{\xi_{(1\ldots i)(i+1\ldots n+1)}}} \; \right. \\
\left. \times \;\rule{0em}{1.8em}
T[(1\ldots i) \to 1,\ldots,i\,] \; T[(i+1\ldots n+1) \to i+1,\ldots,n+1]\,
\right\}\, . 
\label{eq:fragonestep}
\end{multline}
This expression is the final state analog of formula (\ref{eq:recurrence1}) for the iteration of the wave function and it is schematically depicted in Fig.~\ref{fig:fragmaster}.
%
This may be rewritten using formula (\ref{eq:frag1}) for lower number
of gluons:
\begin{multline}
T[(12\ldots n+1) \to 1,2,\ldots,n+1] \; = \; g^{n}{1\over \tilde D_{n+1}} \; 
\sum _{i=1} ^n  \,
\left(
{ z_{(1\ldots i)}z_{(i+1\ldots n+1)} \over
z_1\ldots z_i\, z_{i+1}\ldots z_{n+1}}
\right)^{3/2} \;
{ 1 \over 
\xi^{3/2}_{(1\ldots i)(i+1\ldots n+1)}}\\
\times \;
{\xi_{(1\ldots i)(i+1\ldots n+1)}\, v^*_{(1\ldots i)(i+1\ldots n+1)}
\over  
(v_{12} \ldots v_{i-1\, i})\; (v_{i+1\, i+2}\ldots v_{n\, n+1})} \; .
\end{multline}
The expression 
$$
\left( { z_{(1\ldots i)}z_{(i+1\ldots n+1)} \over
z_1\ldots z_i\, z_{i+1}\ldots z_{n+1}}\right)^{3/2}{ 1 \over 
\xi^{3/2}_{(1\ldots i)(i+1\ldots n+1)}} = 
\left( {{z_{(12\ldots n+1)}} \over {z_1 z_2 \ldots z_{n+1}}} \right)^{3/2}  \; ,
$$
is independent of $i$ and can  be factored out in front of the sum.
Taking the common denominator we finally arrive at
\begin{multline}
T[(12\ldots n+1) \to 1,2,\ldots,n+1] \; = 
 \, \left( {{z_{(12\ldots n+1)}} \over {z_1 z_2 \ldots z_{n+1}}} \right)^{3/2} \;
{g^{n} \over \tilde D_{n+1}} \; 
{ \sum _{i=1} ^n 
\xi_{(1\ldots i)(i+1\ldots n+1)}\, v^*_{(1\ldots i)(i+1\ldots n+1)}
 \, v_{i\,i+1} \over v_{12} v_{23} \ldots v_{n\, n+1}} \, =\\
 g^{n} \left( {{z_{(12\ldots n+1)}} \over {z_1 z_2 \ldots z_{n+1}}} \right)^{3/2} 
{1\over v_{12} v_{23} \ldots v_{n\, n+1}} \;,
\end{multline}
where to get the last line we used the proof  in Eqs.~(\ref{eq:proof_denom_gen1}, \ref{eq:proof_denom_gen2}) that the sum in the numerator in the first line is equal to the energy denominator $\tilde{D}_{n+1}$.  The above expression is identical to (\ref{eq:frag1}) for $n+1$ which completes the  proof.
\subsection{MHV amplitudes from the LCPT}

\begin{figure}[ht]
\centerline{\epsfig{file=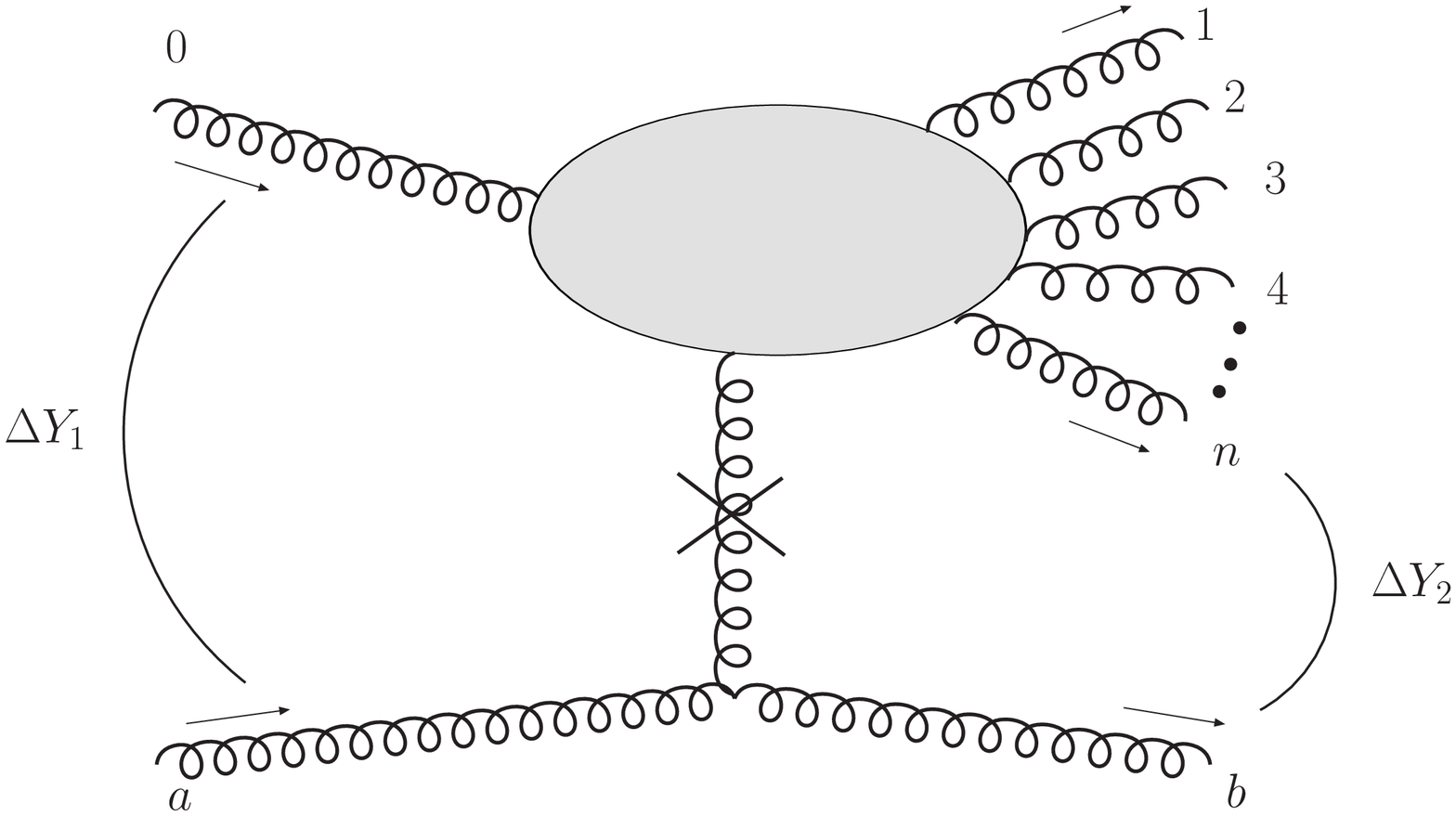,width=0.5\textwidth}}
\caption{The $2\rightarrow n+1$ on-shell gluon amplitude in the high energy limit. The gluon $0$ dissociates into the gluon cascade (indicated by a blob) which interacts via high energy gluon (with a cross) with the gluon $a \to b$. The large rapidity difference $\Delta Y_1 \sim \Delta Y_2$ between the light cone cascade and the lower gluon is taken. The arrows indicate the momentum flow: the gluons $0,a$ are incoming and $1,\dots,n,b$ are outgoing. All the gluons have  $+$ helicity and it  is conserved.}
\label{fig:mhvlightcone}
\end{figure}

It is interesting to see how the MHV amplitudes emerge in the LCPT framework.
We have derived the exact expressions for the light cone wave functions in Sec.~\ref{sec:exactkinematics} and the fragmentation amplitudes of the virtual gluon in Sec.~\ref{sec:fragm}. In this section we compute the $2\to n+1$ gluon scattering amplitude in the limit when a projectile gluon produces $n$~gluons off a single gluon target, and the produced states is  far in rapidity from the  target gluon. The situation is depicted schematically in Fig.~\ref{fig:mhvlightcone}. The result will be shown to reproduce the known analytic form of the MHV amplitudes in the adopted kinematic limit. 
In fact, due to this kinematic limitation we do not provide here the exact derivation of the MHV amplitudes in the general case. Note, however, that the high energy approximation is made only for the coupling of the incoming and outgoing target gluon, and no other approximations are made. Therefore, the comparison of our results to the MHV amplitudes provides a thorough cross check of our approach and the calculations. 

We label the gluons in the way shown in Fig.~\ref{fig:mhvlightcone}. 
The incoming gluon labeled $0$, develops into a cascade of gluons $1,\dots,n$ and scatters on a target gluon $a \to b$. For the high energy case the dominant contributions are given  by the instantaneous exchange of the Coulomb gluon. The exchanged gluon can be attached to any gluon in the cascade. Therefore we include both initial and final state emissions. However, since we are considering color ordered amplitudes it will be sufficient to take into account the attachment to the gluons in the form of the diagrams shown in Figs.~\ref{fig:graph1} and \ref{fig:graph2}.  The gluons $0$ and $a$ are incoming with helicities $+$, and all gluons through the whole cascade down to the final state, carry the positive helicity.

In the simplest case  when the upper part of the diagram is just  single gluon $0 \to 1$ we have  only $2\rightarrow 2$ scattering and the  helicity is conserved. The MHV amplitude  for $2\rightarrow 2$ scattering (\ref{eq:ptmhv}) in this case
 \be
|M(a,0 \to 1,b)| = g^2 \, \left|\frac{\langle a 0 \rangle^4}{\langle a 0 \rangle \langle 0 1 \rangle \langle 1 b \rangle \langle b a \rangle}\right| \;  = \; g^2 \, \frac{s}{|t|} \;,
\label{eq:twototwo}
\ee
where we have used the fact that the spinor products can be regarded as complex square roots of the Mandelstam invariants, see formulae (\ref{eq:ijs}), and in this case they are equal to
$$
|\, \langle a 0 \rangle\, | = |\, \langle 1 b\rangle \, | =\sqrt{s} \, , \hspace*{2cm}  | \, \langle 0 1 \rangle \, | = | \,\langle b a \rangle \, |=\sqrt{|t|} \;.  
$$
The explicit computation of the Coulomb gluon exchange in the LCPT in the high energy limit gives 
\be
|\widetilde M(a,0 \to 1,b)| \; = \; {g^2 \over \sqrt{z_a z_b z_0 z_1}} \, 
\frac{s}{|t|} \;,
\label{eq:twototwo2}
\ee
Obviously results (\ref{eq:twototwo}) and  (\ref{eq:twototwo2}) agree up to the conventional factor  $1/\sqrt{z_a z_b z_0 z_1}$, used in the LCPT amplitudes \cite{Lepage:1980fj,Brodsky:1997de}.

\begin{figure}[ht]
\centerline{\epsfig{file=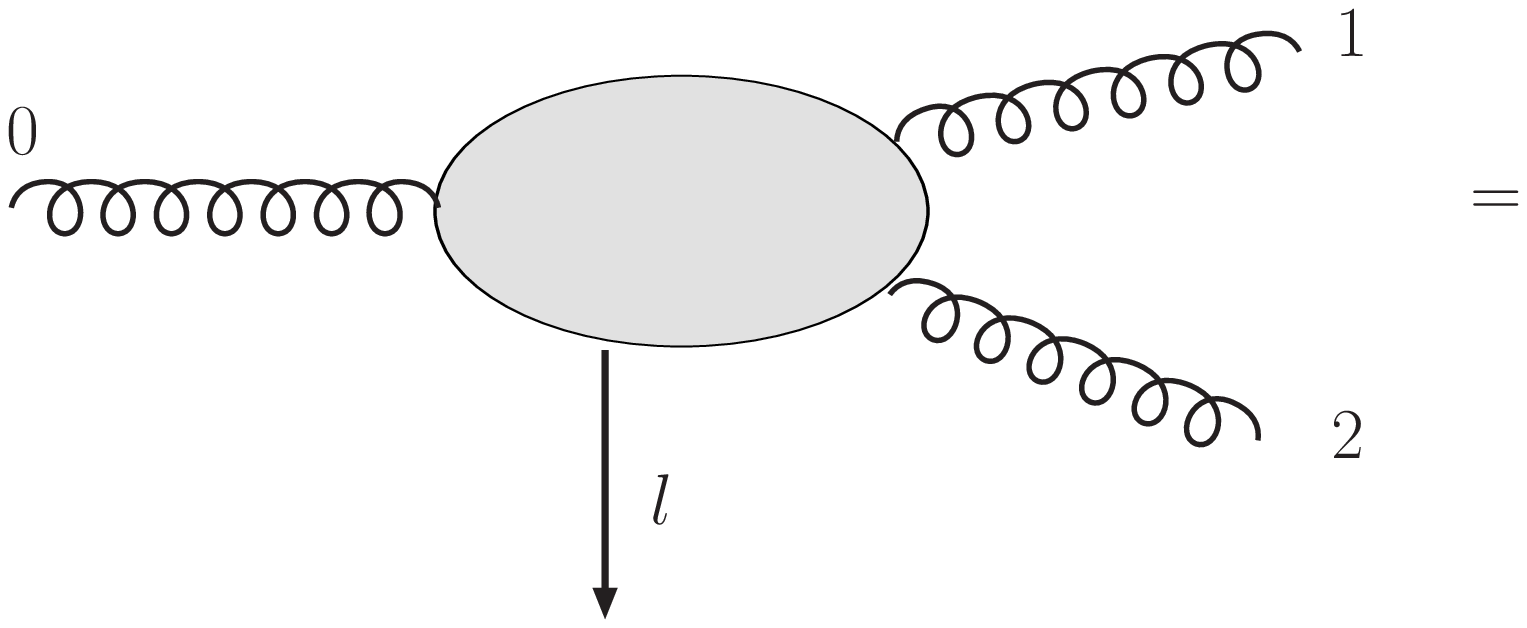,width=0.3\textwidth}\hspace*{0.3cm}\epsfig{file=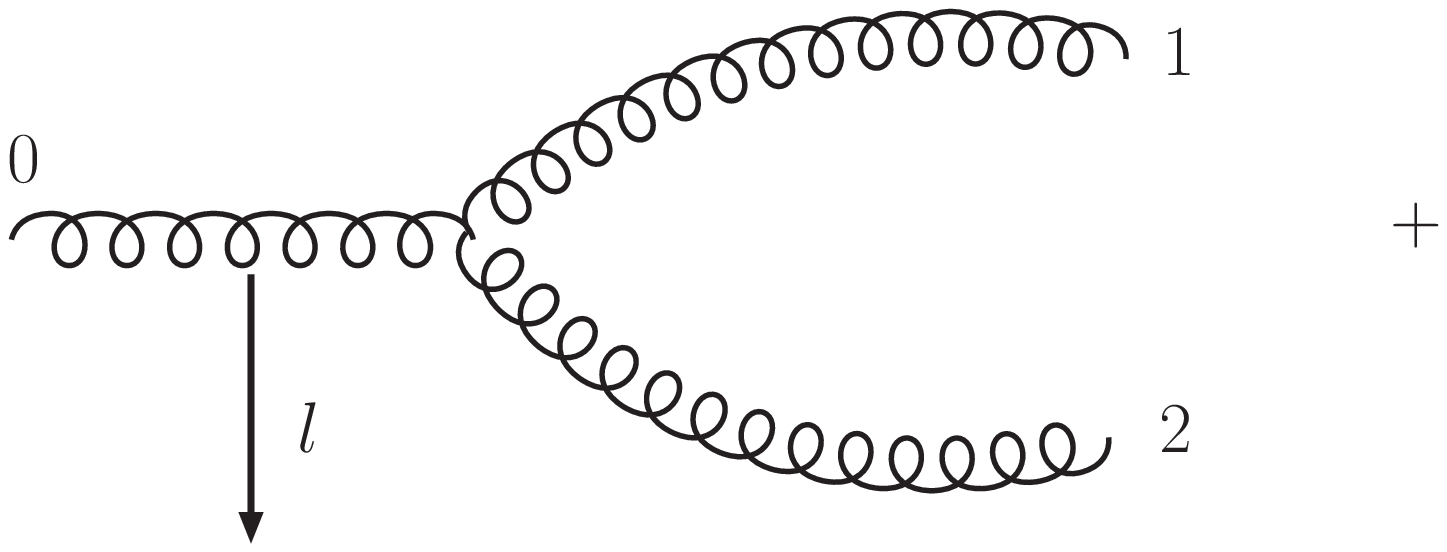,width=0.28\textwidth}\hspace*{0.5cm}\epsfig{file=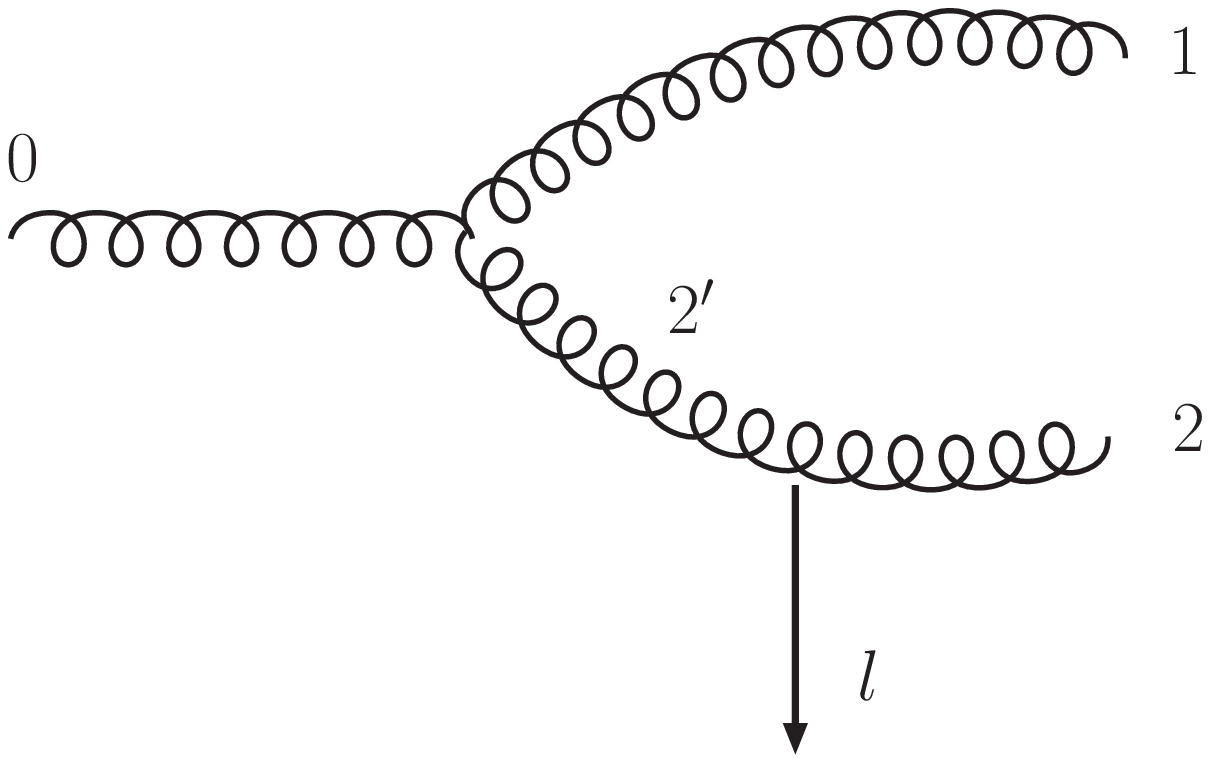,width=0.23\textwidth}}
\caption{Pictorial representation of Eq.~(\ref{eq:psi2fact}). The down-pointing arrows indicate the momentum transfer caused by the exchange of the $t$ channel gluon.}
\label{fig:graph2psitilde}
\end{figure}

In what follows, we shall factorize the amplitude of the gluon exchange, 
$\sim g^2 s / |t|\,$, from the amplitude of the projectile gluon evolution down to the $n$-gluon final state, $\tilde \Psi_n$. Thus, we write:
\be
\widetilde M(a,0 \to b,1,2,\ldots n) \, = \, {1\over \sqrt{z_a z_b}}
g^2 {s\over |t|} \;\, \tilde \Psi_n(1,2,\ldots n), 
\label{eq:m_psi}
\ee
where the prefactor $1/\sqrt{z_a z_b}$ is, again, a result of the convention
adopted in the LCPT, and as usual the momentum conservation $\delta$-functions represented by $\Delta^{(n+1)}$ (recall that this factor is implicit in the wave functions) are factored out from the amplitude. In general, we have,
\be
M(a,0 \to b,1,2,\ldots n) \; = \; \sqrt{z_a z_b z_0 z_1 \ldots z_n}\,
\widetilde M(a,0 \to b,1,2,\ldots n)
\label{eq:mmt}
\ee
Let us now consider the $2\to 3$ process depicted in Fig.~\ref{fig:graph1}. 
Using the light cone framework we can rewrite the upper part of both graphs in the following way
\be
\tilde{\Psi}_2(1,2) \; = \;   \Psi_1(12') T[(12)\rightarrow 1,2] \, + \,  \Psi_2(1,2') \, ,
\label{eq:psi2fact}
\ee
where the first term on the right hand side of the above equation (fragmentation in the final state) corresponds to the upper portion of left graph in Fig.~\ref{fig:graph1} and the second term (the initial wave function) to the analogous part in the graph on the right hand side. To be more precise we represent Eq.~(\ref{eq:psi2fact})
graphically in Fig.~\ref{fig:graph2psitilde}. Note, that the prefactor $z_2$, related to the Coulomb gluon coupling to gluons 2 and $2'$, cancels against the factors of $1/\sqrt{z_2}$, coming from the vertices connected by gluon $2'$. 
\begin{figure}[ht]
\centerline{\epsfig{file=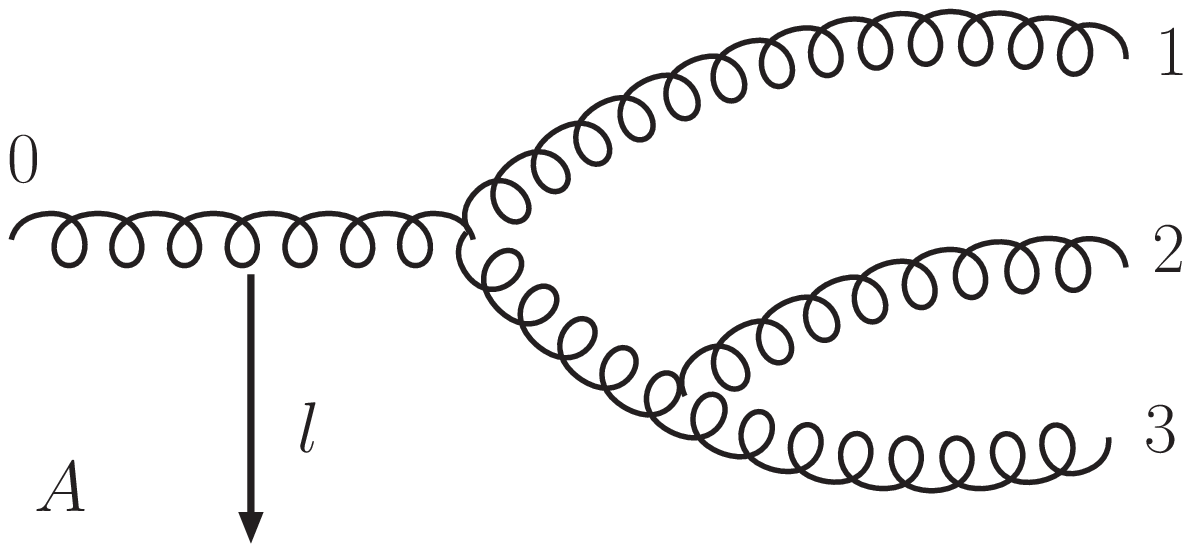,width=0.25\textwidth}\hspace*{0.9cm}\epsfig{file=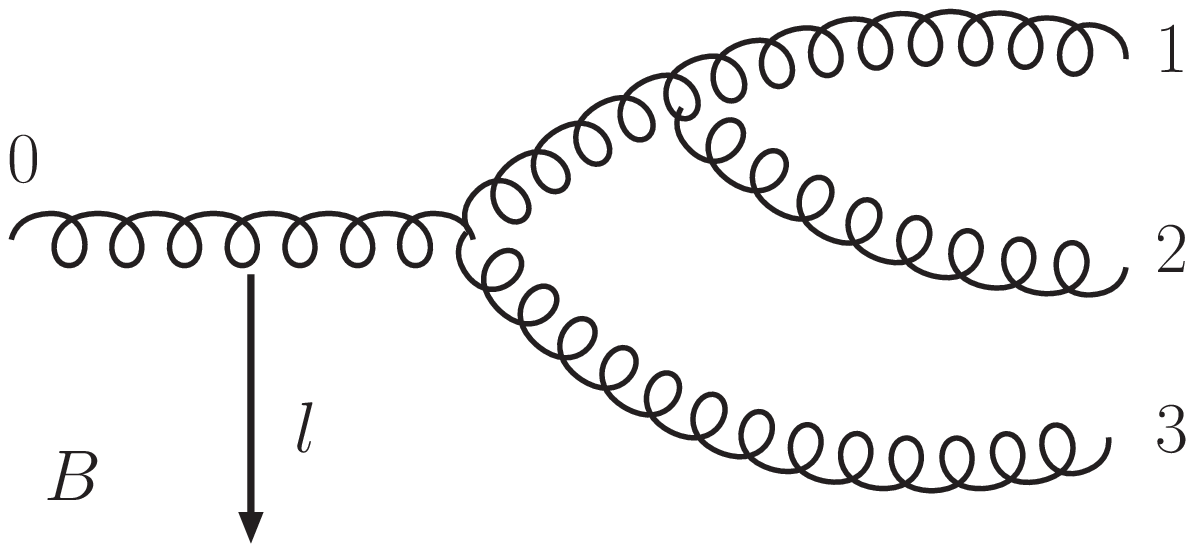,width=0.25\textwidth}\hspace*{0.9cm}\epsfig{file=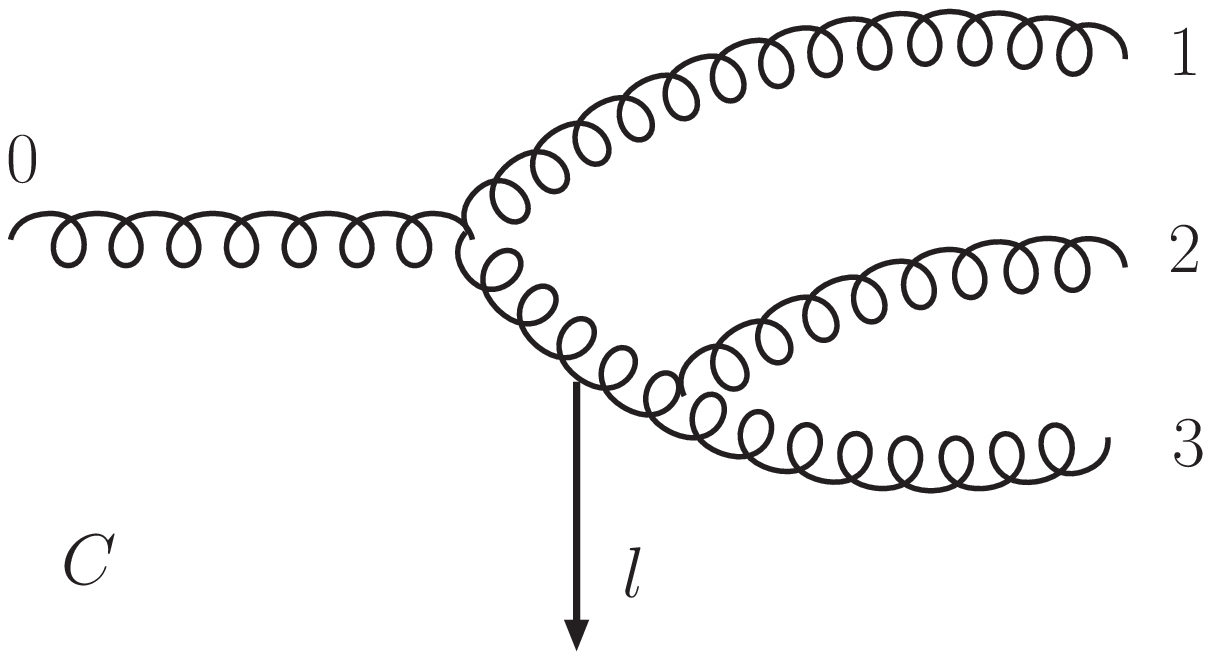,width=0.25\textwidth}}
\centerline{\epsfig{file=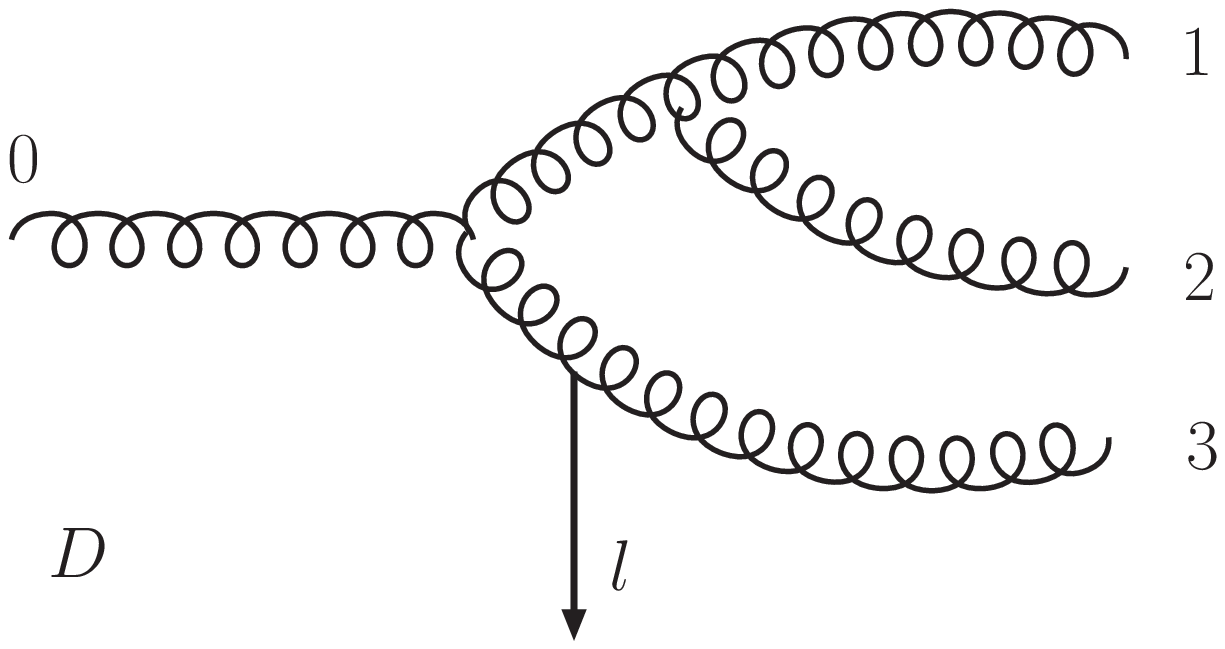,width=0.25\textwidth}\hspace*{0.9cm}\epsfig{file=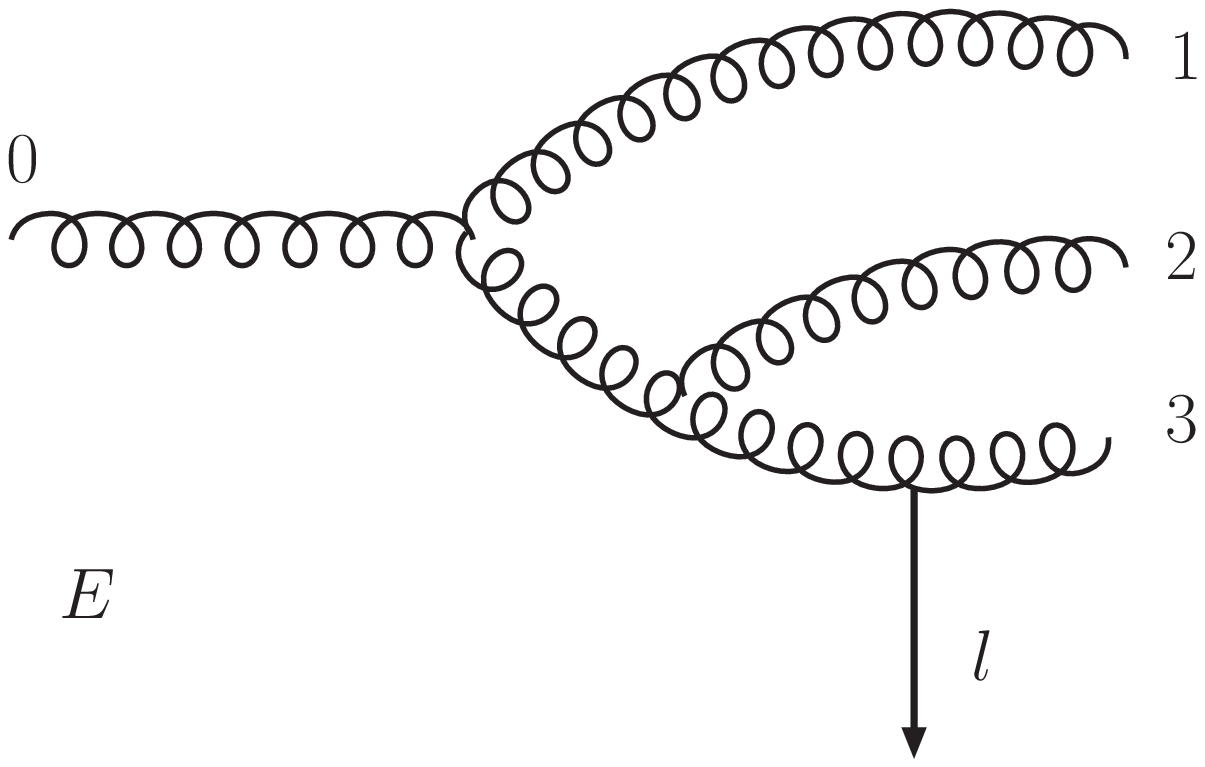,width=0.25\textwidth}\hspace*{0.9cm}\epsfig{file=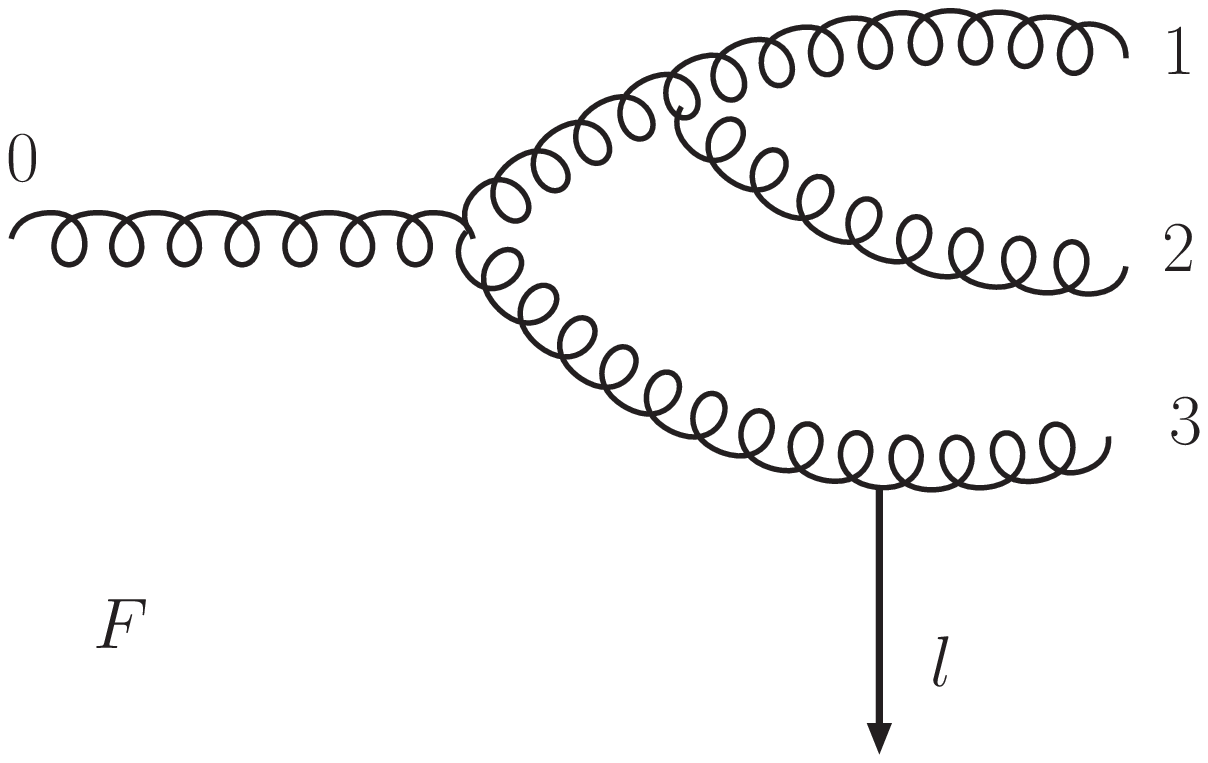,width=0.25\textwidth}}
\caption{Diagrams contributing to the $2 \to 4$ amplitude. The down-pointing arrows indicate the momentum transfer caused by the exchange of the $t$ channel gluon.}
\label{fig:graph3psitilde}
\end{figure}
The notation $2'$ indicates that we need to take into account the transverse momentum carried by the exchanged gluon (the longitudinal fraction of the gluon momentum is neglected). It is equal to $l=-k_{(12)}$.
For the process depicted in the last graph in Fig.~\ref{fig:graph2psitilde} we have $k_{2'} = k_2+ k_{(12)} =-k_1$.
Putting in the explicit expressions for the wave function $\Psi_2(1,2')$ Eq.~(\ref{eq:psinfact}) and $T[(12)\rightarrow 1,2]$ Eq.~(\ref{eq:frag1}) we obtain
\be
\tilde{\Psi}_2(1,2) \; = \; g\, \bigg( \, \frac{1}{(z_1 z_2)^{3/2}} \frac{1}{v_{12}} -\frac{1}{(z_1 z_2)^{3/2}} \frac{1}{v_{12'}} \,\bigg) \; = \; \frac{g}{(z_1 \, z_2)^{3/2}} \frac{v_{12'}-v_{12}}{v_{12} v_{12'}} \, = \,\frac{g}{(z_1 \, z_2)^{3/2}} \frac{z_1\,k_{(12)}}{k_1\, v_{12}} \; . 
\label{eq:tildepsi2}
\ee
For the case of the $2\rightarrow 4 $ process  we have the following contributions
\be
\label{eq:psi3fact}
\tilde{\Psi}_3(1,2,3) \, = \, \Psi_1(123')T[(123)\rightarrow 1,2,3] \, + \, \Psi_2(1,(23)')\, T[(23)\rightarrow 2,3] \, + \, \Psi_2(12,3')\, T[(12)\rightarrow 1,2] \, + \,\Psi_3(1,2,3') \; .
\ee
The first term on the right hand side corresponds to the sum of the  diagrams A and B in Fig.~\ref{fig:graph3psitilde}, the second and the third terms correspond to the diagrams C and D, and the last term is the sum of diagrams E and F. 
Again, the explicit calculation using (\ref{eq:psinfact}) and (\ref{eq:frag1}) yields
\begin{multline}
\tilde{\Psi}_3(1,2,3) \, = \, g^2\,\left[\frac{1}{(z_1 z_2 z_3)^{3/2}}\,\frac{1}{v_{12}v_{23}} \, - \, \frac{1}{\xi_{1(23)}^{3/2}}
\frac{1}{v_{1(23)'}} \left(\frac{1}{\xi_{23}}\right)^{3/2}\,\frac{1}{v_{23}} \right.\\
\left.-  \frac{1}{\xi_{(12)3}^{3/2}}
\frac{1}{v_{(12)3'}} \left(\frac{1}{\xi_{12}}\right)^{3/2}\,\frac{1}{v_{12}} + \frac{1}{\sqrt{z_1 z_2 z_3}}
\frac{1}{\xi_{1(23)}\xi_{(12)3}} \frac{1}{v_{(12)3'} v_{1(23)'}} \right] \, = \,
\frac{g^2}{{(z_1 z_2 z_3})^{3/2}} \; \frac{k_{(123)} z_1 }{k_1 \,  v_{12}\, v_{23}} \; .
\label{eq:tildepsi3}
\end{multline}
In deriving (\ref{eq:tildepsi3}) we used the following relations 
$$
v_{1(23)'} = k_1 \frac{1}{z_1 (z_2+z_3)} = \frac{k_1}{\xi_{1(23)}}, \; \; \; v_{(12)3'} = k_{(12)} \frac{1}{z_3 (z_1+z_2)} = \frac{k_{(12)}}{\xi_{(12)3}} \; ,
$$
as well as  $\xi_{1(23)}\xi_{23}=\xi_{(12)3}\xi_{12}=z_1 z_2 z_3$. The case for $2\rightarrow 5$ proceeds in analogy. The corresponding  expression for $\tilde{\Psi}_4$ can be found by taking the following sum
\begin{multline}
\tilde{\Psi}_4 =
 \Psi_4(1,2,3,4')+\\
\Psi_3(1,2,(34)') T[(34)\rightarrow 3,4]+\Psi_3(1,23,4') T[(23)\rightarrow 2,3]+\Psi_3(12,3,4')T[(12)\rightarrow 1,2]+\\
 \Psi_2(1,(234)')T[(234)\rightarrow 2,3,4]+\Psi_2(1,(234)')T[(234)\rightarrow 2,3,4]+\\
 \Psi_2((12),(34)') T[(12)\rightarrow1,2]T[(34)\rightarrow 3,4]+\Psi_1(1234')T[(1234)\rightarrow 1,2,3,4] \; ,
 \label{eq:psi4fact}
\end{multline}
where we have considered  all possible attachments of the Coulomb gluon.
The explicit calculations are straightforward but lengthy, with the result being
\begin{equation}
\tilde{\Psi}_4 = g^3 \frac{1}{(z_1 z_2 z_3 z_4)^{3/2}} \, \frac{z_1 \, k_{(1234)}}{k_1 \, v_{12} v_{23} v_{34}} \; .
\label{eq:tildepsi4}
\end{equation}
\begin{figure}[ht]
\centerline{\epsfig{file=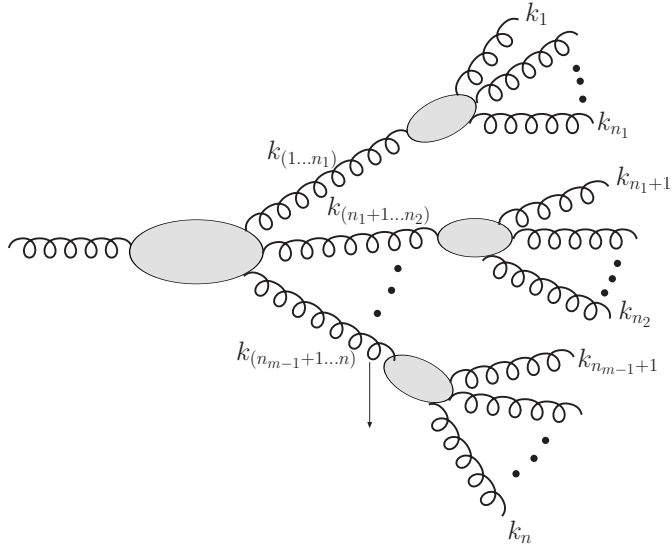,width=0.5\textwidth}}
\caption{Schematic representation of formula (\ref{eq:psifinmaster}). The down-pointing arrow indicates the momentum transfer due to the exchange of the Coulomb gluon. The part of the graph to the left of this arrow (before the scattering) is the wave function $\Psi_m$ and the part of the graph to the right are the fragmentation amplitudes $T$, see Eq.~(\ref{eq:psifinmaster}).}
\label{fig:master}
\end{figure}
The expression for the general $n$ can be found from the following formula
\begin{multline}
\label{eq:psifinmaster}
\tilde \Psi_n(1,2,\ldots ,n) \; = \;
\sum_{m=1} ^n \, \sum_{(1\leq n_1<n_2<\ldots<n_{m-1}\leq n)}\;
\Psi_m((1\ldots n_1)(n_1+1\ldots n_2)\ldots(n_{m-1}+1\ldots n))\\
\times\;
T[(1\ldots n_1)\to 1,\ldots, n_1]\;T[(n_1+1\ldots n_2)\to n_1+1,\ldots, n_2]\,
\ldots \,T[(n_{m-1}+1\ldots n)\to n_{m-1}+1,\ldots, n] \;.
\end{multline}
This formula, depicted in Fig.~\ref{fig:master} is obtained by taking into account all possible attachments of the exchanged gluon in the cascade. 
The additional assumption is the 
 factorization of fragmentation of virtual
gluon (\ref{eq:factm}).
For the complete proof one has to use the explicit expressions 
 for the fragmentation amplitude  $T$ given by (\ref{eq:frag1}) and the initial state 
wave function $\Psi_n$ given by (\ref{eq:psinfact}).  The general proof is quite lengthy  therefore we give it in the appendix A.
The final result is the following general form for the $\tilde{\Psi}_n$  for an arbitrary number of emitted gluons
\be
\label{eq:psifin}
\tilde\Psi_n(1,2,\ldots n) \; = \; g^{n-1} \, {k_{(1\ldots n)} \over k_1 / z_1}
\; 
{1 \over \sqrt{z_1 z_2 \ldots z_n}} \; {1 \over z_1 z_2 \ldots z_n} \; 
{1 \over v_{12} v_{23} \ldots v_{n-1\, n}} \; . 
\ee
We see that it is a generalization of the case for $n=2,3,4$ given in 
 (\ref{eq:tildepsi2}), 
(\ref{eq:tildepsi3}) and (\ref{eq:tildepsi4}).
Note that,  $v_{01} = -\frac{k_1}{z_1}$ (as we have chosen  the transverse momentum of particle 0 to vanish $\uvec{k}_0=0$ ).
\begin{figure}[ht]
\centerline{\epsfig{file=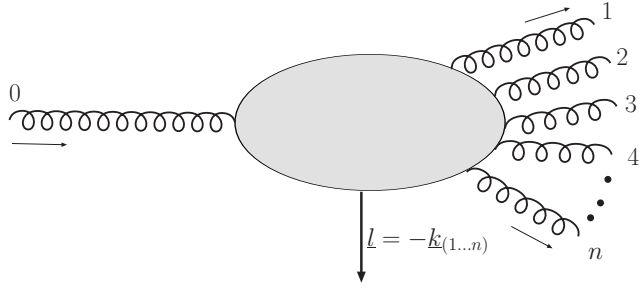,width=0.5\textwidth}}
\caption{Pictorial representation of $\tilde{\Psi}_n$. The down-pointing arrow indicates the momentum transfer caused by the exchange of the $t$ channel gluon. The sum over all possible attachments to the gluons in the cascade is performed. }
\label{fig:tildepsin}
\end{figure}
This object is graphically represented in Fig.~\ref{fig:tildepsin}.
 In order to facilitate the comparison with the MHV amplitude we use  the relation $\langle  i\, i+1 \rangle  = \sqrt{z_i z_{i+1}}\,  v_{i \, i+1}$  to get
\be
\label{eq:psifin_mhvform}
\tilde\Psi_n(1,2,\ldots n) \; = \; g^{n-1} \,
\; 
{1 \over \sqrt{z_1 z_2 \ldots z_n}} \; {1 \over \sqrt{ z_n}} \; 
{k_{(1\dots n)} \over \langle 01 \rangle \langle 12 \rangle  \langle 23 \rangle  \ldots  \langle n-1\, n \rangle} \; . 
\ee
We also need to include the Coulomb gluon exchange between gluon
$a\rightarrow b$ and the gluons in the cascade.  
Let us rewrite the $s/|t|$ part of the prefactor in (\ref{eq:m_psi}),
\be
{s \over |t|} \, = \, 
\sqrt{z_n}\;\left|
{\langle a0 \rangle ^4 \over 
\langle a0 \rangle\, \langle n b \rangle  \langle b a \rangle 
k_{(1\ldots n)} } \right| \, + \, {\cal O}(s^0) \;. 
\label{eq:stex}
\ee
Collecting together 
(\ref{eq:m_psi}), (\ref{eq:mmt}), (\ref{eq:psifin_mhvform}), and (\ref{eq:stex}), we obtain (up to a complex phase factor) the amplitude for $2\rightarrow n+1$ scattering
\be
M(0;a \to 1,\ldots,n;b) \; \simeq \; 
g^{n+1} {\langle a0 \rangle ^4 \over 
\langle a0 \rangle\, \langle 01 \rangle \, \langle 12 \rangle \,
\langle n-1\, n \rangle \, \langle n b \rangle  \langle b a \rangle}\,
, 
\ee
which is equivalent to the MHV amplitude (\ref{eq:ptmhv}).
The approximation sign ``$\simeq$'' is used because  the exchange between the gluon cascade and gluon $a$ was computed in the high energy limit. 

\section*{Summary}
\label{sec:summary}
In this paper we have analyzed multi-gluon cascades keeping the exact representation of the gluon kinematics using the light cone perturbation theory. For the choice of gluon helicities corresponding to the MHV amplitudes, we considered the components of the incoming gluon light cone wave function with an arbitrary number of virtual gluons. We found a hierarchy of recurrence relation between the multi-gluon components that holds at the tree level. The hierarchy was solved exactly in the case of the on-shell incoming gluon. A compact exact form of the real gluon wave function is presented. Interestingly enough, the natural variables which appear in the computation are closely related to the spinor products which appear in the computation of the maximally helicity violating amplitudes. 

We also improved the color dipole evolution equation at small $x$ by relaxing partially the soft gluon approximation. Thus, we included into analysis an  entanglement between the longitudinal and transverse degrees of freedom, as it follows from the form of the energy denominators of the light cone perturbation theory. We obtained a modified kernel of the color dipole evolution in which the emission of large dipoles is exponentially suppressed above the size dependent on the longitudinal momentum fraction of the softer emission in the dipole / gluon splitting. The new kernel contains modified Bessel function of the second kind and it is no longer conformally invariant in two transverse dimensions. It also leads to a much slower diffusion in transverse space than the original LL dipole kernel. 

Furthermore, we showed that the modified dipole evolution kernel gives at the NLL accuracy the most singular pieces at the collinear limit of the exact NLL BFKL kernel, related to double logarithmic terms in the collinear limit. 
Notably, these are the terms of the NLL~BFKL kernel which violate the conformal invariance in the two transverse dimensions both in the case of the ${\cal N}=4$~SYM theory and in the case of QCD case. This supports a conjecture that these terms are related rather to universal kinematical effects than to dynamical details of the theory. 

We have also investigated the scattering of the gluon, described by the light cone wave function on a target gluon and the final state emissions of gluons.
We showed explicitly within LCPT that fragmentation amplitudes of different virtual gluons emerging from the scattering into real gluons are independent, despite an apparent entanglement. The fragmentation amplitude has been computed exactly of a single virtual gluon with the positive helicity into an arbitrary number of real gluons with positive helicities. Interestingly, we found a duality between the initial state wave function of the real gluon and the fragmentation amplitude of the virtual gluon. This result may hint on a deeper duality between the initial and final state evolution in the gluon cascade. 

Finally, we have constructed  the $2 \to n+1$ gluon amplitudes in the MHV helicity configuration by scattering the wave function on the target gluon, followed by the fragmentation of the scattered state. We chose the kinematical situation in which the projectile gluon produces $n$~gluons in arbitrarily large rapidity separation from the target gluon. In this limit we reproduced the Parke--Taylor MHV amplitudes.

\section*{Acknowledgments}
We especially  thank Ted Rogers for discussions.
We also thank Jochen Bartels, Francois Gelis, Al Mueller, Gregory Soyez and  Raju Venugopalan for comments and discussions. This work
was supported by the  Polish Ministry of Education grant No.\ N202 249235.
L.~M.\ gratefully acknowledges the support of the DFG grant No.\ SFB 676.

\section*{Appendix A}
We provide here the proof of the explicit expression for  the upper portion of the scattering amplitudes (\ref{eq:psifin}) which will be done via recursive induction.
Let us  begin with the formula (\ref{eq:psifinmaster}) for the function $\tilde{\Psi}_n$ introduced in Sec.~\ref{sec:mhv} 
\begin{multline}
\label{eq:psifinmastera}
\tilde \Psi_n(1,2,\ldots ,n) \; = \;
\sum_{m=1} ^n \, \sum_{(1\leq n_1<n_2<\ldots<n_{m-1}\leq n)}\;
\Psi_m((1\ldots n_1)(n_1+1\ldots n_2)\ldots(n_{m-1}+1\ldots n))\\
\times\;
T[(1\ldots n_1)\to 1,\ldots, n_1]\;T[(n_1+1\ldots n_2)\to n_1+1,\ldots, n_2]\,
\ldots \,T[(n_{m-1}+1\ldots n)\to n_{m-1}+1,\ldots, n] \; .
\end{multline}
This formula arises when we take into account all possible attachments of the exchanged gluon between the cascade of gluons and the target gluon. Therefore we need to combine the wave function (initial state) with the fragmentation function (final state radiation) in all possible ways. This is accounted for by performing the summations in (\ref{eq:psifinmastera}).
The above sum was used to compute the explicit expressions for $\tilde{\Psi}_2,\tilde{\Psi}_3,\tilde{\Psi}_4$ which are given by the Eqs.~(\ref{eq:tildepsi2},\ref{eq:tildepsi3},\ref{eq:tildepsi4}) by using the 
 formulae for the wave functions and the fragmentation functions.
Here we prove that using the Eq.~(\ref{eq:psifinmastera}) we get the following explicit  result  for  $\tilde{\Psi}_n$   (\ref{eq:psifin}) in case of the arbitrary $n$ number of gluons 
\be
\label{eq:psifina}
\tilde\Psi_n(1,2,\ldots n) \; = \; g^{n-1} \, {k_{(1\ldots n)} \over k_1 / z_1}
\; 
{1 \over \sqrt{z_1 z_2 \ldots z_n}} \; {1 \over z_1 z_2 \ldots z_n} \; 
{1 \over v_{12} v_{23} \ldots v_{n-1\, n}} \; . 
\ee
As a first step we need to take into account the momentum transfer from the exchanged gluon in the explicit expressions for the wave functions. For example,  the wave function $\Psi_n$  with $n=m$  (with the momentum
transfer included) which appears in the sum (\ref{eq:psifinmastera}) has the following form

\[
\Psi_n(1,2,\ldots,n') \; = \; (-1)^{n-1}g^{n-1}\, {1 \over \sqrt{z_1 z_2 \ldots z_n}}\,
{1\over\xi_{(12\ldots n-1)n}\,\xi_{(12\ldots n-2)(n-1\,n )} \,
\ldots \, \xi_{1(2\ldots n)}}
\]
\be
\times\;
{1\over v_{(12\ldots n-1)n'}\, v_{(12\ldots n-2)(n-1\,n )'} \,
\ldots \, v_{1(2\ldots n)'}} \; ,
\label{eq:psinfactprime}
\ee
where primes indicate the momenta of the wave function before the momentum transfer has happened as discussed in Sec.~\ref{sec:mhv} and shown in Figs.~\ref{fig:graph2psitilde}  and \ref{fig:graph3psitilde}.
From the momentum conservation (the transverse momentum of the initial particle is zero) we get following relations between the primed and unprimed 
momenta
$$
k_{n'} = -k_{(1\dots n-1)}, \; k_{(n-1 n)'} = -k_{(1\dots n-2)}, \; \dots \; , k_{(2\dots n)'} = -k_1 \; .
$$
Therefore in general we have 
$$
v_{(1\dots j)(j+1\dots n)'}= \frac{k_{(1\dots j)}}{z_{(1\dots j)}}-\frac{k_{(j+1\dots n)'}}{z_{(j+1\dots n)}}=
\frac{k_{(1\dots j)}}{\xi_{(1\dots j)(j+1\dots n)}} \; .
$$
These relations are true for any wave function appearing in expression (\ref{eq:psifinmastera}).
Using these relations we can rewrite the wave function (\ref{eq:psinfactprime}) as
\be
\Psi_n(1,2,\ldots,n') \; = \; (-1)^{n-1}g^{n-1}\, {1 \over \sqrt{z_1 z_2 \ldots z_n}}\,
{1\over k_1 \, k_{(12)} \,
\ldots \, k_{(1\dots n-1)}} \; ,
\label{eq:psinfactprimek}
\ee
which is equivalent to (\ref{eq:psinki}). For each term in the sum (\ref{eq:psifinmastera}) we have the factor which depends on the longitudinal degrees of freedom in the following way
\begin{multline}
\frac{1}{\sqrt{z_{(1\dots n_1)}z_{(n_1+1 \dots n_2)} \dots z_{(n_{m-1} +1\dots n)} }} \left(\frac{z_{(1\dots n_1)}z_{(n_1+1 \dots n_2)} \dots z_{(n_{m-1} +1\dots n)}   }{z_1 \dots z_n}\right)^{3/2} \; = \\
=\; \frac{1}{\sqrt{z_1 \dots z_n}}  \left(\frac{z_{(1\dots n_1)}z_{(n_1+1 \dots n_2)} \dots z_{(n_{m-1} +1\dots n)}   }{z_1 \dots z_n}\right) \; .
\end{multline}
Therefore we can factor out the term $1/\sqrt{z_1\dots z_n}$ in front
of the whole expression. This factor is purely conventional as it comes from our choice of the normalization of the momenta. 
The common factor $1/z_1 z_2\dots z_n$ in the parenthesis  $()$ will be left under the sum for the time being.
It will prove to be useful to  use another  notation for the momenta. We introduce auxiliary variables $p_{i i+1}$ defined as follows
$$
p_{12}=k_1, \; p_{23}=k_{(12)}, \; p_{34}=k_{(123)}, \; \ldots, \;  p_{n-1 n}=k_{(1\dots n-1)} \; .
$$

Using (\ref{eq:psifinmastera}), the new variables, explicit expressions for wave function (\ref{eq:psinfactprimek}) and fragmentation amplitude  we get for the general case $n$
\begin{multline}
\tilde \Psi_n(1,2,\ldots ,n) \; = \; \frac{g^{n-1}}{\sqrt{z_1 z_2 \dots z_n}} \left
\{ \; \frac{z_{(1\dots n)}}{z_1\dots z_n \,v_{12}\dots v_{n-1 n}} \;- \right .\\
\left[  \, \frac{z_{(23\dots n)}}{z_2\dots z_n}\frac{1}{p_{12} v_{23}\dots v_{n-1 n}} \, + \, \frac{z_{(12)}z_{(3\dots n)}}{z_1 z_2\dots z_n}\frac{1}{v_{12} p_{23}  v_{34}\dots v_{n-1 n}}\, +\, \dots \, +\, \frac{z_{(1\dots n-1)}}{z_1 z_2\dots z_{n-1}}\frac{1}{v_{12}\dots v_{n-2 n-1} p_{n-1 n} } \, \right] \; +\\
\sum_{1\le<i_1<i_2\le n-1} \frac{z_{(1 \dots i_1)} z_{(i_1+1 \dots i_2)} z_{(i_2+1 \dots n)}}{z_1 \dots z_n}\frac{1}{v_{12}\dots p_{i_1 i_1+1}\dots p_{i_2 i_2+1}  \dots  v_{n-1 n}} \; -\\
\sum_{1\le<i_1<i_2<i_3\le n-1} \frac{z_{(1 \dots i_1)} z_{(i_1+1 \dots i_2)} z_{(i_2+1 \dots i_3)} z_{(i_3+1 \dots n)} }{z_1 \dots z_n}\frac{1}{v_{12}\dots p_{i_1 i_1+1}\dots p_{i_2 i_2+1}  \dots p_{i_3 i_3+1}  \dots  v_{n-1 n}} \; +\\
+ \; \ldots\ldots\ldots \; +\\
(-1)^{n-1} \left[\, \frac{z_{(12)}}{z_1 z_2}\frac{1}{v_{12} p_{23}\dots p_{n-1 n}} \, + \, \frac{z_{(23)}}{z_2 z_3}\frac{1}{p_{12} v_{23}  p_{34}\dots p_{n-1 n}}\, + \, \dots \, +\, \frac{z_{(n-1 n)}}{ z_{n-1} z_n}\frac{1}{p_{12}\dots p_{n-2 n-1} v_{n-1 n} } \, \right] \; +\\
\left. (-1)^{n} \frac{1}{p_{12}p_{23}\dots p_{n-1 n}} \; \right\}\; ,
\label{eq:tildepsin_a}
\end{multline}
where $\dots\dots\dots$ means that we need to take higher order sums with $1\le i_1< i_2 \dots < i_m \le n-1$.
We see that in the expression above, each term has a form of 
$$
\frac{z_{(1 \ldots i_1)} z_{(i_1+1 \ldots i_2)}  \ldots z_{(i_m+1 \ldots n)} }{z_1 \ldots z_n}\frac{1}{v_{12}\ldots p_{i_1 i_1+1}\ldots p_{i_2 i_2+1}  \ldots\ldots p_{i_m i_m+1}  \ldots  v_{n-1 n}} \; ,
$$
with $1\le i_1< i_2 \dots < i_m \le n-1$, where $m$ can range from $0$ to $n-1$.
In other words we can regard (\ref{eq:tildepsin_a}) as a sum over the different possible insertions of two kinds of elements $p_{i_k i_{k}+1}$  and $v_{j_l j_{l}+1}$ into the chain with $n-1$  sites. We write down the above formula in the form with the common denominator, which is equal to
\be
{\cal D}_{n} = z_1 z_2 \dots z_n \, v_{12} v_{23} \dots v_{n-1 n} \, p_{12} p_{23}\dots p_{n-1 n} \; .
\label{eq:commondenominator}
\ee
The numerator  is then equal to (we omitted the $g^{n-1}$ factor here)
\begin{multline}
{\cal N}_{n} \; =\; z_{(1\dots n)} p_{12}p_{23}\dots p_{n-1 n} \; -\\
\left[ z_1 z_{(2\dots n)} v_{12} p_{23}\dots p_{n-1 n} + z_{(12)} z_{(3\dots n)} p_{12} v_{23} p_{34} \dots p_{n-1 n} +\dots + 
z_{(1\dots n-1)}z_n p_{12} \dots p_{n-2 n-1} v_{n-1 n} \right]+\\
\sum_{1 \le i_1 < i_2 \le n-1} z_{(1 \dots i_1)} z_{(i_1+1 \dots i_2)} z_{(i_2+1 \dots n)} \, p_{12}\dots v_{i_1 i_1+1}\dots v_{i_2 i_2+1}  \dots  p_{n-1 n} \; - \\
\sum_{1\le<i_1<i_2<i_3\le n-1} \frac{z_{(1 \dots i_1)} z_{(i_1+1 \dots i_2)} z_{(i_2+1 \dots i_3)} z_{(i_3+1 \dots n)} }{z_1 \dots z_n}{p_{12}\dots v_{i_1 i_1+1}\dots v_{i_2 i_2+1}  \dots v_{i_3 i_3+1}  \dots  p_{n-1 n}} \; \\
+ \; \dots\dots\dots \; + \; \\
(-1)^{n-1} \left[ {z_{(12)}z_3\ldots z_n}{p_{12} v_{23}\dots v_{n-1 n}}  +  {z_1z_{(23)}\ldots z_n} {v_{12} p_{23}  v_{34}\dots v_{n-1 n}}\, + \, \dots \, + {z_1 \ldots z_{(n-1 n)}}{v_{12}\dots v_{n-2 n-1} p_{n-1 n} } \right] \\
+\;(-1)^n \, z_1 z_2 \dots z_n v_{12}\dots v_{n-1 n} \; .
\label{eq:numerator}
\end{multline}
We assume that the formula (\ref{eq:psifina}) is true for $n$. In that case
\be
{\cal N}_{n} = z_1 p_{23} p_{34} \dots p_{n-1 n} k_{(1\dots n)} \; .
\label{eq:Nn}
\ee
Let us take the case of $n+1$ which reads
\begin{multline}
{\cal N}_{n+1} \; =\; z_{(1\dots n+1)} p_{12}p_{23}\dots p_{n-1 n} p_{n n+1}\; -\\
\left[ z_1 z_{(2\dots n+1)} v_{12} p_{23}\dots p_{n n+1} + z_{(12)} z_{(3\dots n+1)} p_{12} v_{23} p_{34} \dots p_{n n+1} +\dots + 
z_{(1\dots n)}z_{n+1} p_{12} \dots p_{n-1 n} v_{n n+1} \right]+\\
\sum_{1 \le i_1 < i_2 \le n} z_{(1 \dots i_1)} z_{(i_1+1 \dots i_2)} z_{(i_2+1 \dots n+1)} \, p_{12}\dots v_{i_1 i_1+1}\dots v_{i_2 i_2+1}  \dots  p_{n n+1}  \\
+ \; \dots\dots\dots \; + \; 
(-1)^{n+1} \, z_1 z_2 \dots z_n z_{n+1} v_{12}\dots v_{n-1 n}  v_{n n+1}\; .
\label{eq:numeratorn+1}
\end{multline}
Obviously in  every term in ${\cal N}_{n+1}$    there can be either  $p_{n n+1}$ or $v_{n n+1}$ present.
 We collect these terms and obtain
\begin{multline}
{\cal N}_{n+1} = p_{n n+1} \left( \ldots\ldots \right) \; + \;  
z_{n+1} v_{n n+1} \, \left[ -z_{(1\dots n)} p_{12} \dots p_{n-1 n} \; +  \right. \\
 \left. z_1 z_{(2\dots n)} v_{12} p_{23}\dots p_{n-1 n} + z_{(12)} z_{(3\dots n)} p_{12} v_{23} p_{34} \dots p_{n-1 n} +\dots + 
z_{(1\dots n-1)}z_n p_{12} \dots p_{n-2 n-1} v_{n-1 n} \; \right. \\
-\sum_{1 \le i_1 < i_2 \le n-1} z_{(1 \dots i_1)} z_{(i_1+1 \dots i_2)} z_{(i_2+1 \dots n)} \, p_{12}\dots v_{i_1 i_1+1}\dots v_{i_2 i_2+1}  \dots  p_{n-1 n} \;+ \\
\left. \; +  \; \dots\dots\dots \; + \;  (-1)^{n+1} \,  z_1 z_2 \dots z_n v_{12}\dots v_{n-1 n} \; \right] \;.
\end{multline}
The terms in the squared parenthesis $[\ldots]$ combine to $-{\cal N}_{n}$ of (\ref{eq:numerator}).
We make the inductive step and use (\ref{eq:Nn}) for this expression. Using $p_{n n+1}=k_{(1\dots n)}$ the numerator ${\cal N}_{n+1}$ reads then
\begin{multline}
{\cal N}_{n+1} = p_{n n+1} \left\{ z_{(1\dots n+1)} p_{12}p_{23}\dots p_{n-1 n} \right.-\\
\left[ z_1 z_{(2\dots n+1)} v_{12} p_{23}\dots p_{n-1 n} + z_{(12)} z_{(3\dots n+1)} p_{12} v_{23} p_{34} \dots p_{n-1 n} +\dots + 
z_{(1\dots n-1)}z_{(n n+1)} p_{12} \dots p_{n-2 n-1} v_{n-1 n} \right]
\; + \; \\
\sum_{1 \le i_1 < i_2 \le n-1} z_{(1 \dots i_1)} z_{(i_1+1 \dots i_2)} z_{(i_2+1 \dots n+1)} \, p_{12}\dots v_{i_1 i_1+1}\dots v_{i_2 i_2+1}  \dots  p_{n-1 n} \; - \\
  \; \dots\dots\dots \;  \\
\left.-z_{n+1} v_{n n+1} z_1 p_{23} p_{34} \dots p_{n-1 n} \right\} \; ,
\end{multline}
where the last term comes from making the inductive step as described above and we have factored out  $p_{n n+1}$.
Now we combine terms in the parenthesis $\{ \ldots \}$ which contain either $p_{n-1 n}$ or $v_{n-1 n}$. We then observe that
the coefficient in front of $z_{(n n+1)} v_{n-1 n}$ will combine to ${\cal N}_{n-1}$ and we can again use (\ref{eq:Nn}) for $n-1$.
Therefore the term proportional to   $v_{n-1 n}$  in $\{ \ldots \}$  reads
$$
-z_{(n n+1)} v_{n-1 n} z_1 p_{23} p_{34} \dots p_{n-1 n} \; .
$$
Therefore all the terms now contain  $p_{n-1 n}$ and it can be again factored out.
We see that we need to apply this procedure  recursively $n-1$ times, and each time we pick up a term which will be proportional to $z_{(j \ldots n+1)} v_{j j+1}$ (where $j$ decreases with each step). Making the inductive step we see that such term will contain  $p_{j j+1}$ which can be factored out. After performing all these steps the result reads then
\be
{\cal N}_{n+1} = p_{23}\dots p_{n-1 n} p_{n n+1} \left[ z_{(1\dots n+1)} p_{12} - z_1 z_{(2\dots n+1)}v_{12} - z_1 z_{(3\dots n+1)}v_{23} - \dots - \, z_1z_{(n n+1)} v_{n-1 n}  - z_1 z_{n+1} v_{n n+1}  \right] \; ,
\label{eq:prefinalNn+1}
\ee
where all $p_{j j+1}$ terms with $2\le j \le n$ have been factored out.
We can use 
\be
-z_1 z_{(k\dots n+1)}v_{1k} - z_1 z_{(k+1\dots n+1)}v_{k k+1} = - z_1 z_k v_{1k} -z_1 z_{(k+1\dots n+1)} (v_{1k}+v_{k k+1})= - z_1 z_k v_{1k} -z_1 z_{(k+1 \dots n+1)} v_{1 k+1} \; ,
\ee
to rewrite the numerator as
\be
{\cal N}_{n+1} = p_{23}\dots p_{n-1 n} p_{n n+1} \left[ z_{(1\dots n+1)} p_{12} - z_1 z_2 v_{12} - z_1 z_3  v_{13} - \dots - \, z_1z_{n} v_{1 n}  - z_1 z_{n+1} v_{1 n+1}  \right] \; .
\ee
Using $p_{12}=k_1$ and $z_1 z_j v_{1 j } = z_j k_1 - z_1 k_j$ we  find that the terms in the $[\ldots]$ combine to
$z_1 k_{(1\ldots n+1)}$ which gives
\be
{\cal N}_{n+1} = z_1 p_{23}\dots p_{n-1 n} p_{n n+1} k_{(1\ldots n+1)} \;,
\label{eq:finalNn+1}
\ee
and this  completes the proof for $\tilde{\Psi}_{n+1}$. Let us finally note that in the relations (\ref{eq:prefinalNn+1})
to (\ref{eq:finalNn+1}) were also used in the explicit calculations done for $n=2,3,4$. It is important to note that
these relations hold also in the cases where $z_{(1\dots n+1)} \neq 0$, which is important as in the proof we recursively
move to lower $n$ where the partial sums over longitudinal $z$ and transverse momenta $k$ can take different values.


\begin{thebibliography}{999}

\bibitem{AlMueller}   A.~H.~Mueller,
Nucl.\ Phys.\  B {\bf 415}, 373 (1994).

\bibitem{Chen:1995pa}
Z.~Chen and A.~H.~Mueller,
Nucl.\ Phys.\  B {\bf 451}, 579 (1995).

\bibitem{Weinberg:1966jm}
S.~Weinberg,
Phys.\ Rev.\  {\bf 150}, 1313 (1966).
  
\bibitem{Bjorken:1970ah}
  J.~D.~Bjorken, J.~B.~Kogut and D.~E.~Soper,
  Phys.\ Rev.\  D {\bf 3}, 1382 (1971).

\bibitem{Lepage:1980fj}
  G.~P.~Lepage and S.~J.~Brodsky,
  Phys.\ Rev.\  D {\bf 22}, 2157 (1980).

\bibitem{Brodsky:1997de}
S.~J.~Brodsky, H.~C.~Pauli and S.~S.~Pinsky,
Phys.\ Rept.\  {\bf 301}, 299 (1998)
[arXiv:hep-ph/9705477].

\bibitem{BFKL} E.A.\ Kuraev, L.N.\ Lipatov and V.S.\ Fadin, {
Sh.\ Eksp.\ Teor.\
Fiz.} {\bf 72} (1977) 373, ({ Sov.\ Phys.\ JETP} {\bf  45}
(1977) 199); \\Ya.\ Ya.\
Balitzkij and L.N.\ Lipatov, { Yad.\ Fiz.} {\bf 28} (1978)
1597 ({ Sov.\ J.\
Nucl.\ Phys.} {\bf 28} (1978) 822), \\J.B.\ Bronzan and R.L.\
Sugar, { Phys.\ Rev.}
{\bf D17} (1978) 585; \\T.\ Jaroszewicz, {Acta.\ Phys.\ Polon.}
{\bf B11} 
(1980) 965.

\bibitem{Parke:1986gb}
  S.~J.~Parke and T.~R.~Taylor,
  Phys.\ Rev.\ Lett.\  {\bf 56}, 2459 (1986).

\bibitem{Mangano:1990by}
  M.~L.~Mangano and S.~J.~Parke,
  Phys.\ Rept.\  {\bf 200} (1991) 301
  [arXiv:hep-th/0509223].

\bibitem{Balitsky:2008zz}
  I.~Balitsky and G.~A.~Chirilli,
  Phys.\ Rev.\  D {\bf 77}, 014019 (2008)
  [arXiv:0710.4330 [hep-ph]].
  
\bibitem{Fadin:2007xy}
  V.~S.~Fadin and R.~Fiore,
  Phys.\ Lett.\  B {\bf 661}, 139 (2008)
  [arXiv:0712.3901 [hep-ph]].

\bibitem{Fadin:2007de}
  V.~S.~Fadin, R.~Fiore, A.~V.~Grabovsky and A.~Papa,
  Nucl.\ Phys.\  B {\bf 784}, 49 (2007)
  [arXiv:0705.1885 [hep-ph]].

\bibitem{Fadin:2007ee}
  V.~S.~Fadin, R.~Fiore and A.~Papa,
  Phys.\ Lett.\  B {\bf 647}, 179 (2007)
  [arXiv:hep-ph/0701075].
\bibitem{Fadin:2006ha}
  V.~S.~Fadin, R.~Fiore and A.~Papa,
  Nucl.\ Phys.\  B {\bf 769}, 108 (2007)
  [arXiv:hep-ph/0612284].

\bibitem{Salam:1995uy}
  G.~P.~Salam,
  Nucl.\ Phys.\  B {\bf 461}, 512 (1996)
  [arXiv:hep-ph/9509353].

\bibitem{Mueller:1996te}
  A.~H.~Mueller and G.~P.~Salam,
  Nucl.\ Phys.\  B {\bf 475}, 293 (1996)
  [arXiv:hep-ph/9605302].
  
  \bibitem{Kovchegov:1999yj}
  Yu.~V.~Kovchegov,
  Phys.\ Rev.\  D {\bf 60}, 034008 (1999)
  [arXiv:hep-ph/9901281].
  
  \bibitem{KOV2} Yu. V. Kovchegov, { Phys. Rev.} D {\bf 61} (2000) 074018 [arXiv:hep-ph/9905214].
  
  \bibitem{Balitsky:1995ub}
  I.~Balitsky,
  Nucl.\ Phys.\  B {\bf 463}, 99 (1996)
  [arXiv:hep-ph/9509348].
  
 \bibitem{BAL2} I.~Balitsky, Phys.\ Rev.\ Lett.\  {\bf 81}, 2024 (1998)
  [arXiv:hep-ph/9807434];\\
              Phys.\ Rev.\  D {\bf 60}, 014020 (1999)
  [arXiv:hep-ph/9812311]. 
 
\bibitem{BAL3} I.~Balitsky,    Phys.\ Lett.\  B {\bf 518}, 235 (2001)
  [arXiv:hep-ph/0105334].

\bibitem{Balitsky:2001mr}
  I.~Balitsky and A.~V.~Belitsky,
  Nucl.\ Phys.\  B {\bf 629}, 290 (2002)
  [arXiv:hep-ph/0110158].

\bibitem{Balitsky:2004rr}
  I.~Balitsky,
  Phys.\ Rev.\  D {\bf 70}, 114030 (2004)
  [arXiv:hep-ph/0409314].

\bibitem{Balitsky:2006pf}
  I.~Balitsky,
  Nucl.\ Phys.\ Proc.\ Suppl.\  {\bf 152}, 275 (2006).

\bibitem{Balitsky:2005we}
  I.~Balitsky,
  Phys.\ Rev.\  D {\bf 72}, 074027 (2005)
  [arXiv:hep-ph/0507237].

  \bibitem{MCLERVEN} L. McLerran and  R. Venugopalan,  
                   {\em Phys. Rev.} {\bf D49} (1994) 2233;  
                   {\em Phys. Rev.} {\bf D49} (1994) 3352;\\  
                   {\em Phys. Rev.} {\bf D50} (1994) 2225;
                   R. Venugopalan,  {\it Acta Phys. Polon.} {\bf B30} (199) 3731; \\ 
                   E. Iancu, A. Leonidov and L. McLerran, 
                   {\em Nucl.Phys.} {\bf A692} (2001) 583;\\ 
                   E. Ferreiro, E. Iancu, A. Leonidov and  L. McLerran, 
                   {\tt hep-ph/0109115}. 
                   
                     
\bibitem{CGC1}    E. Iancu and L. McLerran,  
                  {\it Phys. Lett.} {\bf B510} (2001) 133.  

  
\bibitem{JAL}  J. Jalilian-Marian, A. Kovner, L. McLerran and H. Weigert,  
                    {\it Nucl. Phys.} {\bf B504} (1997) 415;  \\
                    {\it Phys. Rev.} {\bf D59} (1999) 014014; 
                    {\it Phys. Rev.} {\bf D59} (1999) 034007.  
  
\bibitem{Iancu:2003xm}
  E.~Iancu and R.~Venugopalan,
  arXiv:hep-ph/0303204.


\bibitem{Fadin:1996nw}
  V.~S.~Fadin and L.~N.~Lipatov,
  Nucl.\ Phys.\  B {\bf 477}, 767 (1996)
  [arXiv:hep-ph/9602287].

\bibitem{Fadin:1998py}
  V.~S.~Fadin and L.~N.~Lipatov,
  Phys.\ Lett.\  B {\bf 429}, 127 (1998)
  [arXiv:hep-ph/9802290].

\bibitem{Camici:1997ij}
  G.~Camici and M.~Ciafaloni,
  Phys.\ Lett.\  B {\bf 412}, 396 (1997)
  [Erratum-ibid.\  B {\bf 417}, 390 (1998)]
  [arXiv:hep-ph/9707390].
  
\bibitem{Ciafaloni:1998gs}
  M.~Ciafaloni and G.~Camici,
  Phys.\ Lett.\  B {\bf 430}, 349 (1998)
  [arXiv:hep-ph/9803389].
  
\bibitem{Fadin:2005zj}
  V.~S.~Fadin and R.~Fiore,
  Phys.\ Rev.\  D {\bf 72}, 014018 (2005)
  [arXiv:hep-ph/0502045].
  
  
\bibitem{Fadin:2004zq}
  V.~S.~Fadin and R.~Fiore,
  Phys.\ Lett.\  B {\bf 610}, 61 (2005)
  [Erratum-ibid.\  B {\bf 621}, 61 (2005)]
  [arXiv:hep-ph/0412386].
  
\bibitem{Balitsky:2006wa}
  I.~Balitsky,
  Phys.\ Rev.\  D {\bf 75}, 014001 (2007)
  [arXiv:hep-ph/0609105].
  
  
  \bibitem{Kovchegov:2006vj}
  Yu.~V.~Kovchegov and H.~Weigert,
  Nucl.\ Phys.\  A {\bf 784}, 188 (2007)
  [arXiv:hep-ph/0609090].
  
   \bibitem{Andersson:1995jt}
  B.~Andersson, G.~Gustafson, H.~Kharraziha and J.~Samuelsson,
  Z.\ Phys.\  C {\bf 71}, 613 (1996).
 
\bibitem{Andersson:1995ju}
  B.~Andersson, G.~Gustafson and J.~Samuelsson,
  Nucl.\ Phys.\  B {\bf 467}, 443 (1996).
  
\bibitem{Kwiecinski:1996td}
  J.~Kwiecinski, A.~D.~Martin and P.~J.~Sutton,
  Z.\ Phys.\  C {\bf 71}, 585 (1996)
  [arXiv:hep-ph/9602320].
  
\bibitem{Salam:1998tj}
  G.~P.~Salam,
  JHEP {\bf 9807}, 019 (1998)
  [arXiv:hep-ph/9806482].

\bibitem{Ross:1998xw}
  D.~A.~Ross,
  Phys.\ Lett.\  B {\bf 431}, 161 (1998)
  [arXiv:hep-ph/9804332].
   
\bibitem{Kwiecinski:1997ee}
  J.~Kwiecinski, A.~D.~Martin and A.~M.~Stasto,
  Phys.\ Rev.\  D {\bf 56}, 3991 (1997)
  [arXiv:hep-ph/9703445].
  
\bibitem{Avsar:2005iz}
  E.~Avsar, G.~Gustafson and L.~Lonnblad,
  JHEP {\bf 0507}, 062 (2005)
  [arXiv:hep-ph/0503181].
  
\bibitem{Avsar:2006jy}
  E.~Avsar, G.~Gustafson and L.~Lonnblad,
  JHEP {\bf 0701}, 012 (2007)
  [arXiv:hep-ph/0610157].

\bibitem{Salam:1999cn}
  G.~P.~Salam,
  Acta Phys.\ Polon.\  B {\bf 30}, 3679 (1999)
  [arXiv:hep-ph/9910492].

\bibitem{Ciafaloni:1998iv}
  M.~Ciafaloni and D.~Colferai,
  Phys.\ Lett.\  B {\bf 452}, 372 (1999)
  [arXiv:hep-ph/9812366].
  
\bibitem{Ciafaloni:1999yw}
  M.~Ciafaloni, D.~Colferai and G.~P.~Salam,
  Phys.\ Rev.\  D {\bf 60}, 114036 (1999)
  [arXiv:hep-ph/9905566].
  
\bibitem{Ciafaloni:1999au}
  M.~Ciafaloni, D.~Colferai and G.~P.~Salam,
  JHEP {\bf 9910}, 017 (1999)
  [arXiv:hep-ph/9907409].

\bibitem{Ciafaloni:2003ek}
  M.~Ciafaloni, D.~Colferai,  G.~P.~Salam and A.~M.~Stasto,
  Phys.\ Lett.\  B {\bf 576}, 143 (2003)
  [arXiv:hep-ph/0305254].
  
\bibitem{Ciafaloni:2003rd}
  M.~Ciafaloni, D.~Colferai, G.~P.~Salam and A.~M.~Stasto,
  Phys.\ Rev.\  D {\bf 68}, 114003 (2003)
  [arXiv:hep-ph/0307188].
  
\bibitem{Ciafaloni:2007gf}
  M.~Ciafaloni, D.~Colferai, G.~P.~Salam and A.~M.~Stasto,
  JHEP {\bf 0708}, 046 (2007)
  [arXiv:0707.1453 [hep-ph]].
  
\bibitem{Altarelli:1999vw}
G.~Altarelli, R.~D.~Ball and S.~Forte,
Nucl.\ Phys.\ B {\bf 575},  (2000) 313;

\bibitem{Altarelli:2000mh}
G.~Altarelli, R.~D.~Ball and S.~Forte,
Nucl.\ Phys.\ B {\bf 599},  (2001) 383.

  \bibitem{Altarelli:2001ap}
    G.~Altarelli, R.~D.~Ball and S.~Forte,
    arXiv:hep-ph/0104246.

\bibitem{ABF_improved}
G.~Altarelli, R.~D.~Ball and S.~Forte,
Nucl.\ Phys.\ B {\bf 621},  (2002) 359;

\bibitem{Altarelli:2003hk}
G.~Altarelli, R.~D.~Ball and S.~Forte,
Nucl.\ Phys.\ B {\bf 674},  (2003) 459;

\bibitem{Altarelli:2005ni}
G.~Altarelli, R.~D.~Ball and S.~Forte,
Nucl.\ Phys.\ B {\bf 742},  (2006) 1.

\bibitem{THORNE} R.S.~Thorne,  Phys. Rev. {\bf D64} (2001) 074005; Phys. Lett. {\bf 474} (2000) 372.

\bibitem{ThWh06}
  C.~D.~White and R.~S.~Thorne,
  Phys.\ Rev.\  D {\bf 75},  (2007) 034005.

\bibitem{Manohar:1998xv}
  A.~V.~Manohar,
  arXiv:hep-ph/9802419.


\bibitem{GradRyzh} I.S.~Gradshteyn and I.M.~Ryzhik, {\it Table of Integrals, Series and Products}, VIth edition, Academic Press, 2000.

\bibitem{Dokshitzer:2005bf}
  Yu.~L.~Dokshitzer, G.~Marchesini and G.~P.~Salam,
  Phys.\ Lett.\  B {\bf 634}, 504 (2006)
  [arXiv:hep-ph/0511302].
  
\bibitem{DM}
  Yu.~L.~Dokshitzer and G.~Marchesini,
  Phys.\ Lett.\  B {\bf 646}, 189 (2007)
  [arXiv:hep-th/0612248].

\bibitem{Marchesini:2006ax}
  G.~Marchesini, Talk given at Workshop on Future Prospects in QCD at High Energies, Brookhaven, Upton, New York, 17-22 Jul 2006; 
  arXiv:hep-ph/0605262.
  
\bibitem{Bartels:1992ym}
  J.~Bartels,
  Phys.\ Lett.\  B {\bf 298}, 204 (1993).
  
\bibitem{Bartels:1993ih}
  J.~Bartels,
  Z.\ Phys.\  C {\bf 60}, 471 (1993).
  
  \bibitem{KGBAS} 
  K.~J.~Golec-Biernat and A.~M.~Stasto,
  Nucl.\ Phys.\  B {\bf 668}, 345 (2003)
  [arXiv:hep-ph/0306279].

\bibitem{Froissart} M.~Froissart, Phys. \ Rev. {\bf 123} (1961) 1053; A.~Martin, Phys. Rev. {\bf 129} (1963) 1432.

  \bibitem{KW}  A.~Kovner and U.~A.~Wiedemann,
  Phys.\ Rev.\  D {\bf 66}, 034031 (2002)
  [arXiv:hep-ph/0204277].
  
  \bibitem{Kovner:2001bh}
  A.~Kovner and U.~A.~Wiedemann,
  Phys.\ Rev.\  D {\bf 66}, 051502 (2002)
  [arXiv:hep-ph/0112140].
  
\bibitem{Kovner:2002yt}
  A.~Kovner and U.~A.~Wiedemann,
  Phys.\ Lett.\  B {\bf 551}, 311 (2003)
  [arXiv:hep-ph/0207335].
  
  \bibitem{Heisenberg}  W.~Heisenberg, Z. \ Phys. {\bf 133}, (1952) 65.
 
\bibitem{DelDuca:1995zy}
  V.~Del Duca,
  Phys.\ Rev.\  D {\bf 52}, 1527 (1995)
  [arXiv:hep-ph/9503340].
  
\bibitem{DelDuca:1993pp}
  V.~Del Duca,
  Phys.\ Rev.\  D {\bf 48}, 5133 (1993)
  [arXiv:hep-ph/9304259].



\end{thebibliography}
\end{document}